\documentclass[graybox]{svmult}

\usepackage{type1cm}        
\usepackage{makeidx}         
\usepackage{graphicx}        
\usepackage{multicol}        
\usepackage[bottom]{footmisc}

\usepackage{newtxtext}       %
\usepackage{newtxmath}       

\usepackage{amsmath,amsfonts,mathtools,url,soul,verbatim,lineno}
\modulolinenumbers[1]
\usepackage{tabulary}
\newcolumntype{C}[1]{>{\centering\arraybackslash}p{#1}}

\theoremstyle{definition}
\newtheorem{dfn}{Definition}
\newtheorem{pro}[dfn]{Problem}
\newtheorem{thm}[dfn]{Theorem}

\newtheorem{cor}[dfn]{Corollary}
\newtheorem{lem}[dfn]{Lemma}
\newtheorem{exa}[dfn]{Example}

\newcommand{\R}{\mathbb R}
\newcommand{\Z}{\mathbb Z}
\newcommand{\al}{\alpha}
\newcommand{\be}{\beta}

\newcommand{\Ga}{\Gamma}
\newcommand{\de}{\delta}
\newcommand{\la}{\lambda}
\newcommand{\La}{\Lambda}
\newcommand{\ti}{\tilde}
\newcommand{\si}{\sigma}
\newcommand{\ep}{\varepsilon}

\newcommand{\IT}{\mathrm{IT}}
\newcommand{\CSC}{\mathrm{CSC}}
\newcommand{\iso}{\mathrm{Iso}}

\newcommand{\Or}{\mathrm{O}}

\newcommand{\GL}{\mathrm{GL}}

\newcommand{\AMD}{\mathrm{AMD}}
\newcommand{\EMD}{\mathrm{EMD}}
\newcommand{\sym}{\mathrm{Sym}}
\newcommand{\vol}{\mathrm{Vol}}
\newcommand{\bs}{\hfill $\blacksquare$}
\newcommand{\bd}{\partial}
\newcommand{\vl}{\,:\,}

\makeindex             

\begin{document}

\title*{Introduction to Periodic Geometry and Topology}
\author{Olga Anosova and Vitaliy Kurlin}
\institute{Olga Anosova \at University of Liverpool, Liverpool L69 3BX, UK \email{oanosova@liv.ac.uk}
\and Vitaliy Kurlin \at University of Liverpool, Liverpool L69 3BX, UK \email{vkurlin@liv.ac.uk}}
%
%
\maketitle

\vspace*{-2cm}

\abstract{
This monograph introduces key concepts and problems in the
new research area of Periodic Geometry and Topology for materials applications.
Periodic structures such as solid crystalline materials or textiles were
previously classified in discrete and coarse ways that depend on manual choices or are unstable under perturbations.
Since crystal structures are determined in a rigid form, their finest natural equivalence is defined by rigid motion or isometry, which preserves inter-point distances.
Due to atomic vibrations, isometry classes of periodic point sets form a continuous space whose geometry and topology were unknown. 
The key new problem in Periodic Geometry is to unambiguously parameterize this space of isometry classes by continuous coordinates that allow a complete reconstruction of any crystal.
The major part of this manuscript reviews the recently developed isometry invariants to resolve the above problem: (1) density functions computed from higher order Voronoi zones, (2) distance-based invariants that allow ultra-fast visualizations of huge crystal datasets, and (3) the complete invariant isoset (a DNA-type code) with a first continuous metric on all periodic crystals.
The main goal of Periodic Topology is classify textiles up to periodic isotopy, which is a continuous deformation of a thickened plane without a fixed lattice basis.
This practical problem substantially differs from past research focused on links in a fixed thickened torus.
}

\begin{figure}[h!]
\includegraphics[height=30mm]{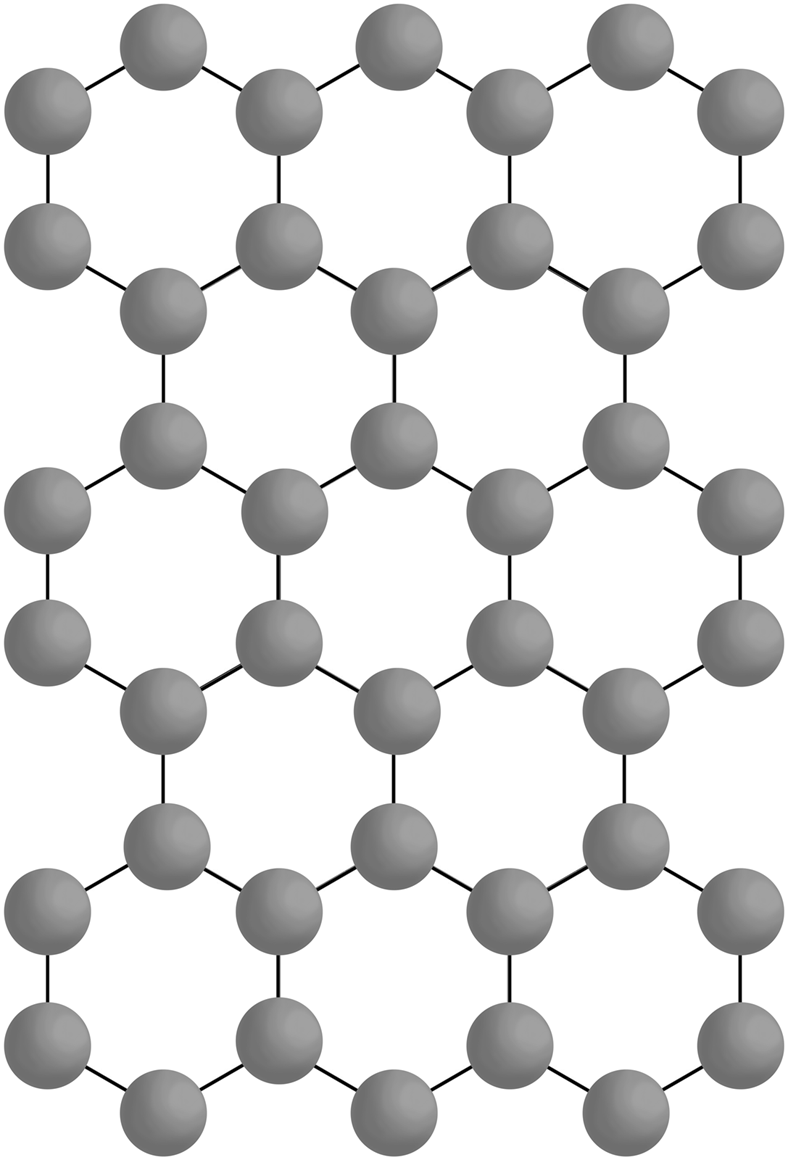}
\hspace*{1pt}
\includegraphics[height=30mm]{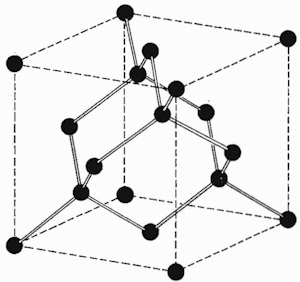}
\hspace*{1pt}
\includegraphics[height=30mm]{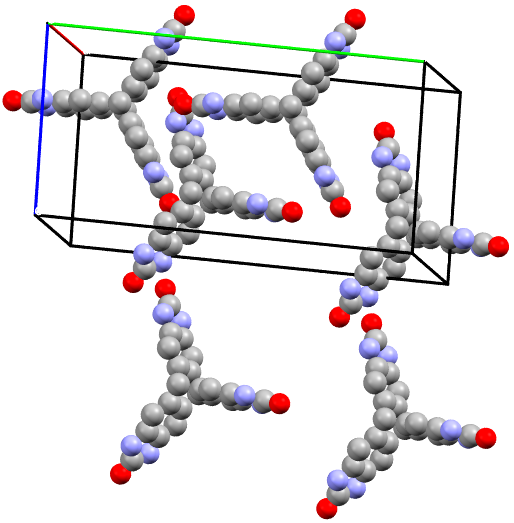}
\hspace*{1pt}
\includegraphics[height=30mm]{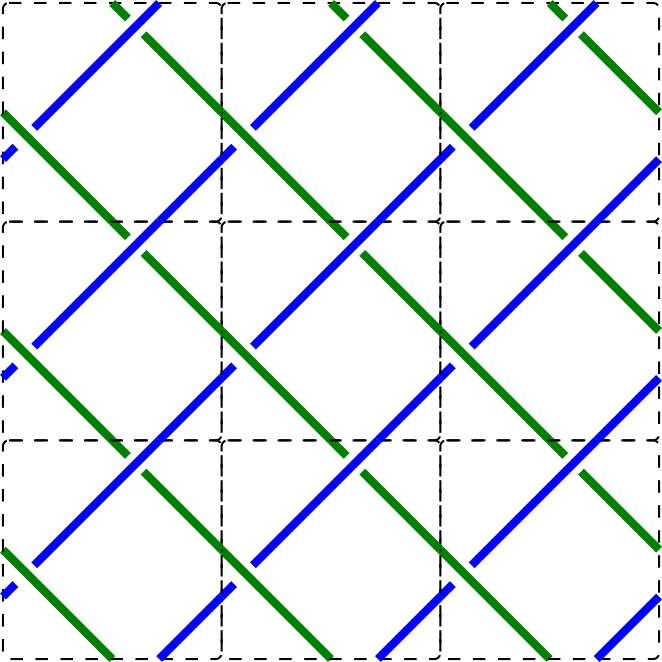}
\caption{
Examples of periodic structures in nature.
\textbf{1st}: graphene consists of carbon atoms arranged in a honeycomb way.
\textbf{2nd}: diamond consists of carbon atoms arranged in a tetrahedral way.
\textbf{3rd}: a molecular crystal with a rectangular unit cell. 
\textbf{4th}: a periodic textile based on a square pattern.}
\label{fig:crystals+textile}      
\end{figure}


\section{Introduction: motivations for research on periodic structures}
\label{sec:intro}

Periodic structures are common in nature.
All solid crystalline material (briefly, \emph{crystals}), including natural minerals such as diamond, or synthesized materials, such as graphene, have structures of periodically repeated unit cells in Fig.~\ref{fig:crystals+textile}.
\medskip

Soft clothing materials (briefly, \emph{textiles}) also have an underlying periodic structure, but their periodicity is usually two-directional in practice, while most crystals are periodic in three directions.
One great exception is a class of 2-dimensional materials, including
graphene consisting of carbons atoms arranged as in Fig.~\ref{fig:crystals+textile}. 
\medskip

The initial part of the paper discusses geometric problems for structures  that are periodic in $n$ independent directions in $\R^n$.
Our motivation comes from the practical case $n=3$ of periodic crystals.
Since atoms are much better defined physical objects than inter-atomic bonds, we represent any crystal by a periodic set of points at all atomic centers.
Each point can be labeled by a chemical element and other physical properties, such as an electric charge for ions.
For simplicity, we introduce all key concepts and state main problems in the hardest case of indistinguishable points.
All stated results can be easily extended to labeled points and even periodic graphs.
\smallskip

A periodic point set $S$ is obtained from a finite \emph{motif} $M$ of points in a \emph{unit cell} $U$ (parallelepiped) by translations along all integer linear combinations of vectors along edges of $U$. 
Points in a motif $M\subset U$ are usually given by the coordinates in the basis of $U$.
These coordinates are numbers in $[0,1]$ and are called \emph{fractional}.
\smallskip

A typical representation of $S$ by a pair $(U,M)$ is highly ambiguous, because $S$ can be obtained from infinitely many different unit cells (or bases) with suitable motifs.
\smallskip

\begin{figure}[ht]
\includegraphics[width=\textwidth]{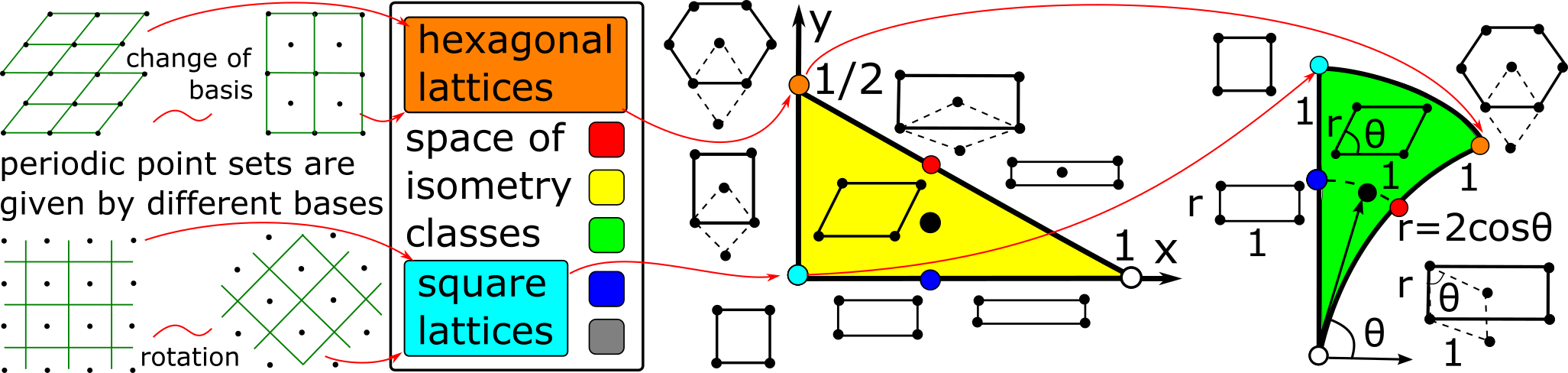}
\caption{The space of all isometry classes of lattices has explicit parameterizations only for $n=2,3$.}
\label{fig:space_isometry_classes}      
\end{figure}

The two hexagonal lattices in the top left corner of Fig.~\ref{fig:space_isometry_classes} are represented in two ways: as the parallelogram $U$ with the basis $(1,0),(\frac{1}{2},\frac{\sqrt{3}}{2})$ and the single point $(0,0)\in M$, and as the larger rectangle $U$ with the basis $(\sqrt{3},0),(0,\frac{\sqrt{3}}{2})$ and the two points $(0,0),(\frac{1}{2},\frac{1}{2})\in M$.
So one can change a pair $(U,M)$ and keep all points of $S$.

The crucial complication comes from the fact that crystal structures are determined in a rigid form, hence should be considered \emph{equivalent up to rigid motions}, which are compositions of translations and rotations in $\R^3$, see the bottom left corner of Fig.~\ref{fig:space_isometry_classes}
We consider \emph{isometries}, which preserve Euclidean distances and include also reflections, because distinguishing orientations of isometric sets is easier.
\medskip

Taking into account rigid motions makes pairs $(U,M)$ even more ambiguous for representing a periodic point set $S$. 
Even if we keep the unit cell $U$ (or a basis) fixed, one can shift all points $p$ in a motif $M\subset U$ by the same vector, which changes all fractional coordinates of $p$, but produces an isometric set. 
We should use only isometry invariants, not cell parameters of $U$ or fractional coordinates of $p\in M$.
\medskip

Rising up to the next level of abstraction, the middle picture in Fig.~\ref{fig:space_isometry_classes} shows the space of all isometry classes of lattices (periodic sets with 1-point motifs).
Continuity of this space was largely ignored in the past, because discrete invariants such as symmetry groups break down under perturbations of points.
Continuous parameterizations were explicitly constructed for lattices in dimension $n=2$.
\medskip

Any 2D lattice can be associated to a quadratic form $(1+x)u^2+2yuv+(1-x)v^2$, where $x,y$ parameterize the yellow triangle in Fig.~\ref{fig:space_isometry_classes}.
This triangle is the fundamental domain of the $\GL_2(\Z)$ action on positive quadratic forms \cite[section~6.2]{zhilinskii2016introduction}.
\medskip

The difficulties for $n=3$ \cite[section~6.3]{zhilinskii2016introduction} show that a new approach is needed for the general case of periodic point sets. 
The curved triangle in Fig.~\ref{fig:space_isometry_classes} is parameterized by alternative parameters for $n=2$ motivated by the new \emph{isoset} in section~\ref{sec:isoset_complete}.
\medskip

\begin{figure}[ht]
\includegraphics[height=42mm]{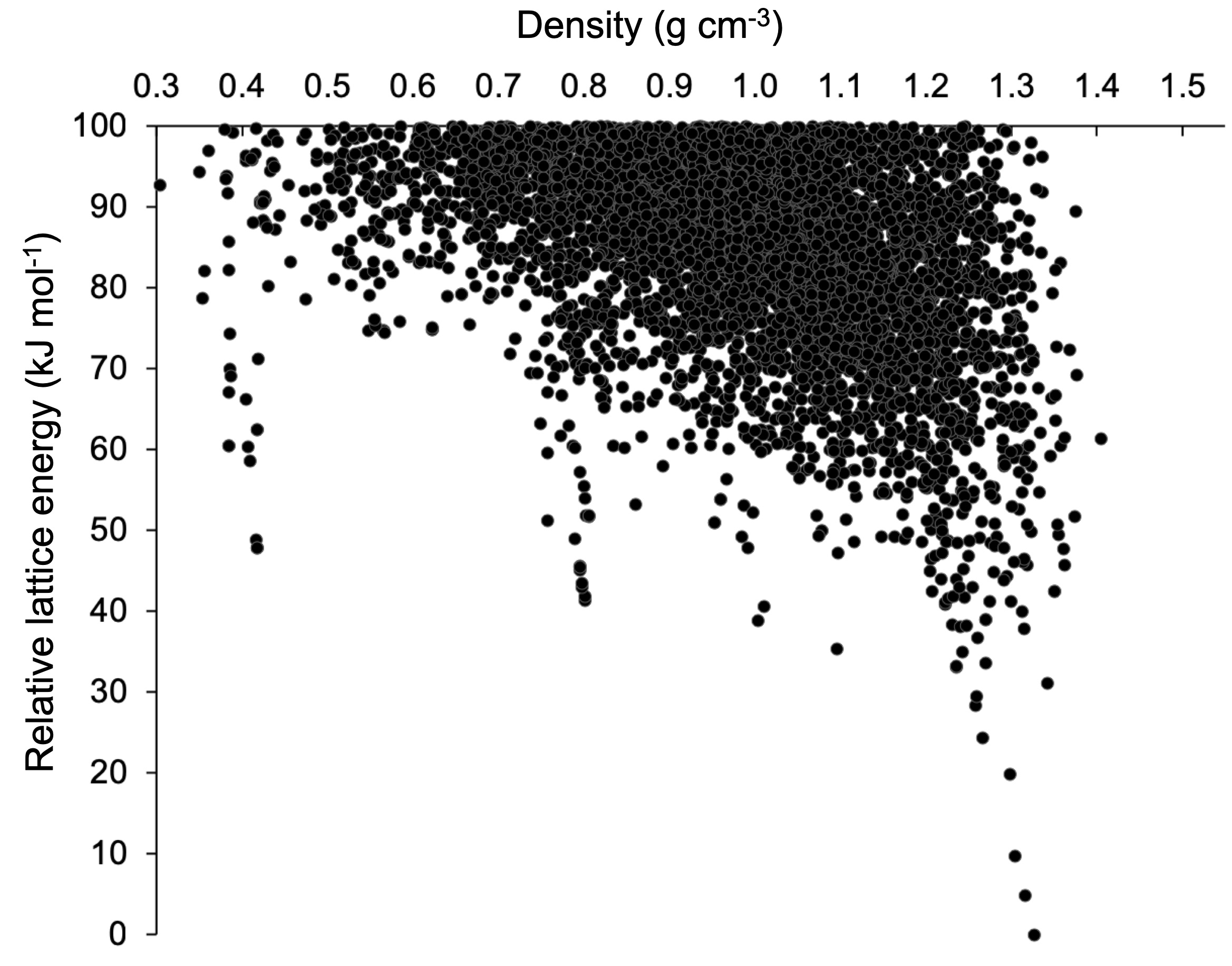}
\includegraphics[height=42mm]{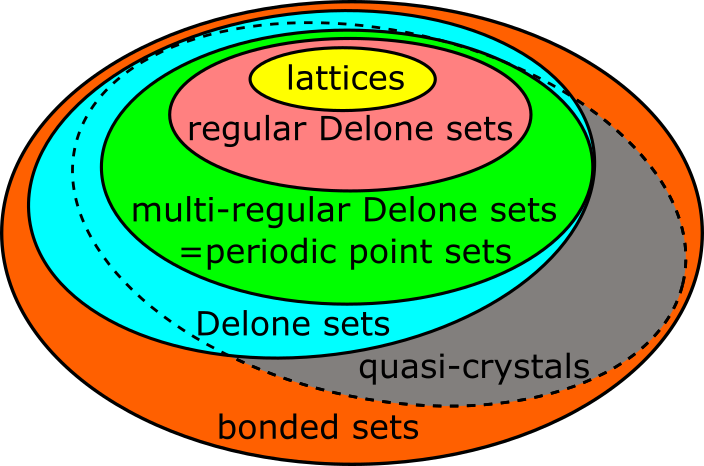}
\caption{\textbf{Left}: 5679 simulated crystals on the left are a discrete sample from a continuous but not-yet-parameterized space of all periodic point sets.
\textbf{Right}: hierarchy of materials models.}
\label{fig:CSP+hierarchy}      
\end{figure}

The practical motivation for continuous isometry invariants comes from Crystal Structure Prediction (CSP).
To predict new crystals, any CSP software starts from almost random positions of atoms or molecules in a random unit cell and iteratively optimizes a complicated energy function (\emph{energy}), whose lower values indicate a potential thermodynamic stability.
A typical CSP outputs the energy-vs-density landscape in the left picture of Fig.~\ref{fig:CSP+hierarchy} shows 5679 predicted crystals based on the T2 molecule from the 3rd picture in Fig.~\ref{fig:crystals+textile}.
This 12-week supercomputer work \cite{pulido2017functional} is an \emph{embarrassment of over-prediction} as coined by Prof Sally Price \cite{price2018zeroth}, because only five crystals were actually synthesized. 
Each point in the CSP landscape represents one simulated crystal by its density $\rho$ and energy, where $\rho$ is the weight of atoms in a unit cell $U$ divided by the cell volume $\vol[U]$.
Many of these 5679 crystals are nearly identical, which was impossible to automatically recognize by past tools. 

The hierarchy of models for classes of materials in the right hand side picture of Fig.~\ref{fig:CSP+hierarchy} shows that lattices are the simplest periodic structures, while Delone sets can model non-periodic materials.  
The International Union of Crystallography defines a crystal as any material that `has essentially a sharp diffraction pattern'  \cite{crystal}.
This description includes quasi-crystals, whose discovery by Shechtman \cite{shechtman1984metallic} was rewarded by the Nobel prize in 2011.  
Since there is no theoretical model equivalent to a description of a `sharp diffraction pattern', the gray area of quasi-crystals in Fig.~\ref{fig:CSP+hierarchy} is bounded by a dashed curve.
Bonded sets \cite{dolbilin2019regular} model amorphous materials.
\medskip

The new area of \emph{Periodic Geometry} studies point-based periodic structures up to isometries in a new continuous way, which differs from past discrete approaches.
\medskip

Section~\ref{sec:problem} formally introduces all concepts and states the first key problem of a continuous isometry classification.
Section~\ref{sec:past} discusses the relevant past work.
Section~\ref{sec:AMD} introduces the first and fastest continuous isometry invariants: the infinite sequence of Average Minimum Distances $\AMD_k$ \cite{widdowson2020average}.
Section~\ref{sec:densities} reviews another infinite family of isometry invariants: continuous density functions $\psi_k(t)$.
New Theorem~\ref{thm:densities1D} explicitly describes $\psi_k(t)$ for periodic 1D sets. 
Section~\ref{sec:isoset_complete} describes the \emph{isoset} to completely classify all periodic point sets up to isometry.
Section~\ref{sec:isoset_continuous} uses isosets to define the first metric on periodic point sets whose continuity under perturbations is proved in Theorem~\ref{thm:continuity}.
Section~\ref{sec:algorithms} justifies polynomial time algorithms to compute and compare isosets, and to approximate the continuous metric above.

\begin{figure}[ht]
\includegraphics[width=\textwidth]{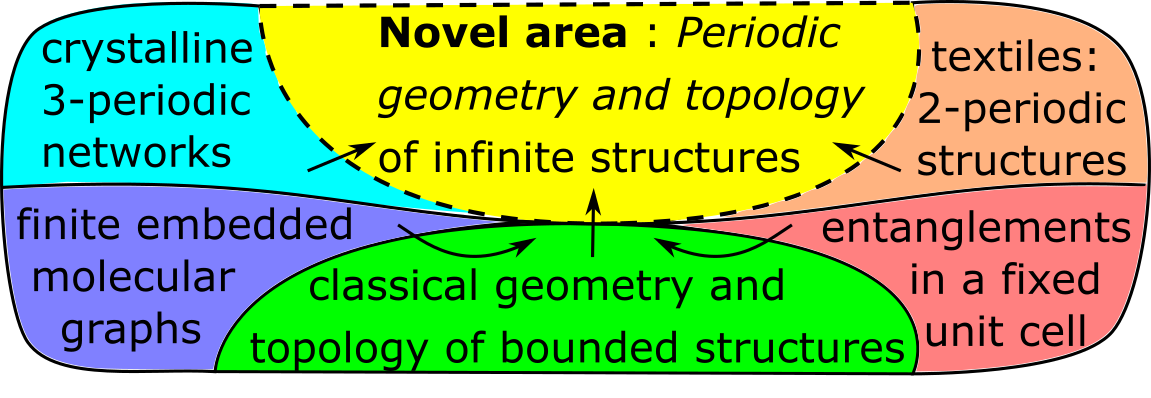}
\caption{Relations of Periodic Geometry and Topology with previously established research areas.}
\label{fig:PGT}      
\end{figure}

Section~\ref{sec:textiles} introduces \emph{Periodic Topology} as a new extension of knot theory to infinite periodic structures.
Almost all past research studied periodic knots or graphs in a fixed (usually cubic) cell with boundary conditions.
Due to a fixed unit cell, this traditional approach is equivalent to classifying knots in a thickened torus $T^2\times[0,1]$ or a 3-dimensional torus $T^3$.
The key difference is the new equivalence: a \emph{periodic isotopy} is a continuous deformation via periodic structures without fixing a cell.
\medskip

When results or proofs are taken from past work, detailed references are given.
The first author contributed the results and proofs from \cite{anosova2021isometry} about completeness of isosets.
The second author contributed all remaining results in the manuscript.
We thank Herbert Edelsbrunner and all our collaborators for the fruitful discussions. 

\section{Key concepts and problems of Periodic Geometry for point sets}
\label{sec:problem}

In the Euclidean space $\R^n$, any point $p\in\R^n$ is represented by the vector $ p$ from the origin of $\R^n$ to $p$.
The \emph{Euclidean} distance between points $p,q\in\R^n$ is denoted by $|p-q|$.
Throughout the paper the word \emph{crystal} refers to a periodic set of points only in $\R^3$, while a \emph{periodic point set} is used in general for any dimension $n\geq 1$.
  
\begin{dfn}[a lattice $\La$, a unit cell $U$, a motif $M$, a periodic set $S=M+\La$]
\label{dfn:crystal}
For a linear basis $ v_1,\dots, v_n$ in $\R^n$, a {\em lattice} is $\La= p+\{\sum\limits_{i=1}^n c_i v_i : c_i\in\Z\}$.
The \emph{unit cell} $U( v_1,\dots, v_n)=\left\{ \sum\limits_{i=1}^n c_i v_i \vl c_i\in[0,1) \right\}$ is the parallelepiped spanned by the basis.
A \emph{motif} $M$ is any finite set of points $p_1,\dots,p_m\in U$.
A \emph{periodic point set} is the Minkowski sum $S=M+\La=\{ u+ v :  u\in M,  v\in \La\}$.
If $M$ consists of a single point $p$, then $p+\La$ is a translated lattice, which will be also called a lattice for brevity.
A unit cell $U$ of a periodic set $S=M+\La$ is \emph{primitive} if any vector $ v$ that translates $S$ to itself is an integer linear combination of the basis of the cell $U$, i.e. $ v\in\La$.
\bs
\end{dfn}

\begin{figure}[ht]
\includegraphics[width=\textwidth]{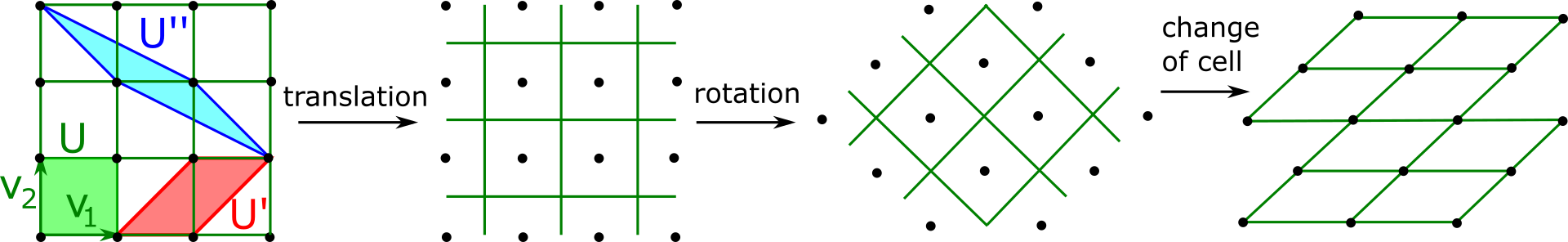}
\caption{
\cite[Fig.~2]{anosova2021isometry}
\textbf{Left}: three primitive cells $U,U',U''$ of the square lattice $S$. Other pictures show periodic sets $\La+M$ with different bases and motifs, which are all isometric to the lattice $S$.}
\label{fig:square_lattices}
\end{figure}

A primitive unit cell $U$ of any lattice has a motif of one point counted with weights as follows.
Points strictly inside $U$ have weight 1, points inside the faces of $U$ have weight $\frac{1}{2}$, corners of $U\subset\R^2$ have weight $\frac{1}{4}$ and so on. 
All unit cells in Fig.~\ref{fig:square_lattices} are primitive, because the four corners of a square are counted as one point in $U$.
\medskip

Any crystal is usually given in the form of a Crystallographic Information File (CIF), which contains parameters of a unit cell $U$ and fractional coordinates of points in a motif $M$.
Fig.~\ref{fig:square_lattices} shows that this representation as a pair $(U,M)$ is highly ambiguous if we try to compare periodic point sets up to translations and rotations.
\medskip

Since most solid crystalline materials are considered as rigid bodies, Periodic Geometry studies periodic point sets up to rigid motion or \emph{isometry} preserving distances.
Since all atoms vibrate at finite temperature (above the absolute zero), 
 any periodic set and its periodic perturbation are slightly different isometry classes.
\medskip
 
Hence thermal vibrations of atoms motivate to study the continuous space formed by all isometry classes of periodic point sets.
A continuous parameterization of this space would give us a `geographic' map containing all known crystals.
Even more importantly, many unexplored points in this map are potential new materials.
\medskip
 
Any phase transition between different forms of the same crystalline material such as auxetic \cite{borcea2018periodic} can be represented by a continuous path in the above space.
The bottleneck distance $d_B$ naturally measures a maximum displacements of atoms.
 
\begin{dfn}[bottleneck distance]
\label{dfn:bottleneck_distance} 
For a fixed bijection $g:S\to Q$ between periodic point sets $S,Q\subset\R^n$, the \emph{maximum deviation} is the supremum $\sup\limits_{p\in S}|p-g(p)|$ of Euclidean distances over all points $p\in S$.
The \emph{bottleneck distance} is the infimum $d_B(S,Q)=\inf\limits_{g:S\to Q}\;\sup\limits_{p\in S}|p-g(p)|$ over all bijections $g$ between infinite sets.
\bs
\end{dfn}

\begin{figure}[ht]
\includegraphics[width=\textwidth]{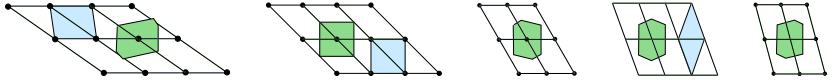}
\caption{
All lattices can be continuously deformed to each other (modified from \cite[Fig.~6.7]{zhilinskii2016introduction}).}
\label{fig:lattice_deformations}
\end{figure}

The bottleneck distance can be formally converted into a metric on the space of isometry classes by taking another infimum $d_B([S],[Q])=\inf_f d_B(S,f(Q))$ over all isometries $f$ of $\R^n$. 
Since it is impractical to compute $d_B$, it makes sense to look for other metrics on the continuous space $\CSC$ of isometry classes of periodic points sets. 
Fig.~\ref{fig:lattice_deformations} illustrates the path-connectivity of this space for 2D lattices.

For now we treat all points as identical.
When new invariants are defined in sections~\ref{sec:AMD}, \ref{sec:densities}, \ref{sec:isoset_complete}, we will describe how to enrich these invariants by labels of points. 
\medskip

An isometry \emph{invariant} is any function or property $I(S)$ preserved by any isometry applied to a periodic point set $S$.
In simplest cases, the values of $I$ are numerical, for example $\AMD(S)$ from section~\ref{sec:AMD} is a sequence of real numbers.
Section~\ref{sec:densities} reviews the density fingerprint $\Psi(S)$, which is a sequence of continuous functions $\psi_k(t):[0,+\infty)\to[0,1]$. 
Section~\ref{sec:isoset_complete} defines the isoset $I(S;\al)$ as a collection of isometry classes of finite $\al$-clusters. 
For practical applications, it should be easier to compare two values of any invariant than comparing original periodic sets.
\medskip

\begin{figure}[h!]
\includegraphics[height=37mm]{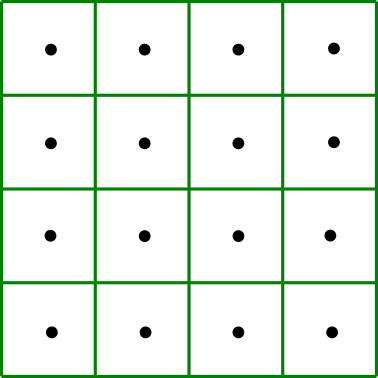}
\hspace*{1mm}
\includegraphics[height=37mm]{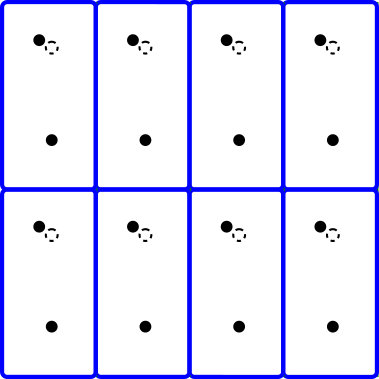}
\hspace*{1mm}
\includegraphics[height=37mm]{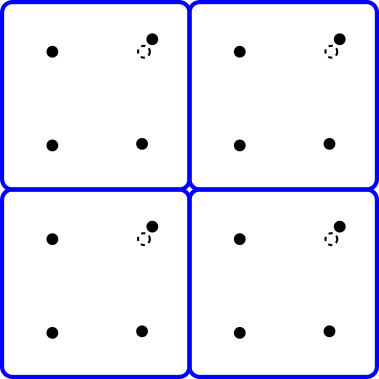}
\caption{
A continuous invariant should take close values on these slightly different periodic sets.}
\label{fig:lattice_perturbations}
\end{figure}

The novel condition in Problem~\ref{pro:isometry_classification} below is the Lipschitz continuity under perturbations, which can be measured in the bottleneck distance $d_B$ between point sets.
This continuity is justified by the example of close periodic sets in Fig.~\ref{fig:lattice_perturbations} showing that the volume of a primitive unit cell discontinuously changes under perturbations.
Similarly, all discrete invariants such as symmetry groups are also discontinuous.
\medskip

The traditional classification of crystals by symmetries cuts the space of isometry classes into disjoint pieces.
The continuity requirement will allow us to quantify small differences between nearly identical crystals, which appear as slightly different approximations to the same local minimum in Crystal Structure Prediction.

\begin{pro}[isometry classification]
\label{pro:isometry_classification}
Find a complete and continuous isometry invariant $I$ of periodic point sets $S\subset\R^n$ satisfying the following conditions:
\smallskip

\noindent
(a) \emph{invariance} : if periodic sets $S,Q$ are isometric, then $I(S)=I(Q)$;
\smallskip

\noindent
(b) \emph{computability} : a distance between values of $I$ is computable fast, for example in a polynomial time in the number of points in a motif $M$ of a periodic set $S$;
\smallskip

\noindent
(c) \emph{continuity} : $d(I(S),I(Q))\leq \la d_B(S,Q)$ for a suitable distance $d$ between invariant values, see condition (b), and a factor $C$ ideally independent of $S,Q$;
\smallskip

\noindent
(d) \emph{completeness} : if $I(S)=I(Q)$, then the periodic point sets $S,Q$ are isometric;
\smallskip

\noindent
(e) \emph{inverse design} : a periodic set $S$ can be explicitly reconstructed from $I(S)$;
\smallskip

\noindent
(f) \emph{parameterization} : the space of all isometry classes of periodic sets is parameterized by $I$, for example all realizable values of $I$ should be described.
\bs
\end{pro}

Condition~(\ref{pro:isometry_classification}a) is needed for any reliable comparison of crystals.
Many crystal descriptors include cell parameters or fractional coordinates, neither of which are isometry invariants. 
If a non-invariant takes different values on two crystals, these crystals can still be isometric.
Hence non-invariants can not justifiably distinguish crystals or predict crystal properties.
In condition~(\ref{pro:isometry_classification}b) a distance $d$ should satisfy all metric axioms, most importantly, if $d(I(S),I(Q))=0$, then $S,Q$ are isometric.
\medskip

Among all geometry-based invariants of crystals, only the physical density $\al$ seems to theoretically satisfy condition~(\ref{pro:isometry_classification}c), not discrete invariants such as symmetry groups and not even the volume of a primitive unit cell as shown in Fig.~\ref{fig:lattice_perturbations}.
\medskip

Condition~(\ref{pro:isometry_classification}d) of completeness allows us to uniquely identify any crystal $S$ by its complete invariant.
Condition~(\ref{pro:isometry_classification}e) is the most recent requirement to convert a trial-and-error materials discovery into a guided exploration by trying new values of a complete invariant $I$.
Condition~(\ref{pro:isometry_classification}f) provides 
will enable an active exploration of the space instead of the current random sampling in Crystal Structure Prediction.
\medskip

Sections~\ref{sec:AMD}, \ref{sec:densities}, \ref{sec:isoset_complete} review invariants satisfying some or all conditions (\ref{pro:isometry_classification}abcde).
Condition~(\ref{pro:isometry_classification}f) may need another invariant, which should be easier than the isoset.
\medskip

For 3-point sets (triangles), an example invariant satisfying all the conditions is the triple of edge-lengths $a\geq b\geq c>0$ satisfying the triangle inequality $a<b+c$.
So the space of all isometry classes of 3-point sets is continuously parameterized by $a,b,c$. 
We could parameterize all triangles by two edges and the angle between them.
Periodic point sets may be also classified by different complete invariants.

\section{Review of the relevant past work on isometry classifications}
\label{sec:past}

Since the first papers in Periodic Geometry \cite{mosca2020voronoi}, \cite{edels2021}, \cite{widdowson2020average} have reviewed many past methods in crystallography, we briefly mention only the most relevant ones.
\medskip

The COMPACK algorithm \cite{chisholm2005compack} compares crystals by trying to match finite portions, which depends on tolerances and outputs irregular numbers in \cite[Table~1]{widdowson2020average}.
We mention 230 crystallographic groups in $\R^3$ and focus on continuous invariants below.
The discontinuity in comparisons of crystals and even lattices has been known since 1980 \cite{andrews1980perturbation}.
Two provably continuous metrics on lattices were defined in \cite{mosca2020voronoi}. 
\medskip

The pair distribution function (PDF) is based on inter-atomic distances \cite{toby1992accuracy}, but is smoothed and computed with cut-off parameters.
The Average Minimum Distances \cite{widdowson2020average} in section~\ref{sec:AMD} can be considered as invariant analogues of PDF.
The key advantages of AMD over other invariants are the fast running time and easy interpretability.  
\medskip

The Crystal Structure Prediction always visualizes simulated crystals as an energy-vs-density landscape. 
This single-value density $\rho$ is substantially extended to density functions \cite{edels2021} in section~\ref{sec:densities}.
The resulting \emph{densigram} is provably complete for periodic sets in a general position, but is slow to compute.
Though it is still unclear if one can reconstruct a periodic point set from its AMD sequence or densigram, these invariants are theoretically justified as reliable inputs for machine learning predictions.
\medskip

The most recent invariant \emph{isoset} reduces the isometry problem for infinite periodic sets to a finite collection of local $\al$-clusters. 
Completeness of the isoset \cite{anosova2021isometry} is proved in section~\ref{sec:isoset_complete}.
Continuity of the isoset is the new result proved in section~\ref{sec:isoset_continuous}.
\medskip

Though the isoset allows a full reconstruction of a periodic point set,
 the isoset $I(S;\al)$ grows in a radius $\al$ and its components ($\al$-clusters) should be compared up to orthogonal maps.
Condition~(\ref{pro:isometry_classification}f) requires an easier but complete invariant.
\medskip

The concept of the isoset emerged from the theory of Delone sets defined below.

\begin{dfn}[Delone sets]
\label{dfn:Delone}
An infinite set of points $S\subset\R^n$ is called a {\em Delone} set if the following conditions hold, see the hierarchy of materials models in Fig.~\ref{fig:CSP+hierarchy}:
\medskip

\noindent
(\ref{dfn:Delone}a) {\em packing} :
$S$ is {\em uniformly discrete}, i.e. there is a maximum \emph{packing radius} $r(S)$ such that all open balls $B(a;r(S))$ with radius $r(S)$ and centers $a\in S$ are disjoint;	
\medskip

\noindent
(\ref{dfn:Delone}b) {\em covering} :
$S$ is {\em relatively dense}, i.e. there is a minimum \emph{covering radius}  $R(S)$ such that all closed balls $\bar B(a;R(S))$ centered at all points $a\in S$ cover $\R^n$.
\bs
\end{dfn}

\begin{figure}[h!]
\includegraphics[height=55mm]{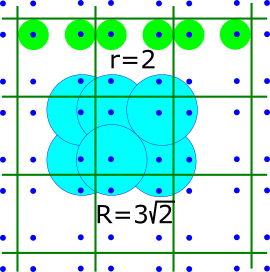}
\hspace*{2mm}
\includegraphics[height=55mm]{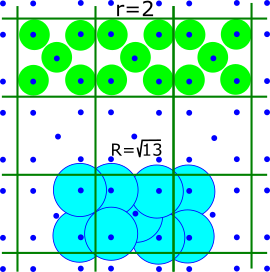}
\caption{
\textbf{Left}: the periodic point set $S_1$ has the points $(2,2),(2,8),(8,2),(8,8)$ in the square cell $[0,10)^2$, the packing radius $r=2$ and covering radius $R=3\sqrt{2}$.  
\textbf{Right}: $S_2$ has the extra point $(5,5)$ in the center of $[0,10)^2$, the packing radius $r=2$ and covering radius $R=\sqrt{13}$, see Definition~\ref{dfn:Delone}. }
\label{fig:covering+packing}
\end{figure}

The isometry problems for finite (non-periodic) point sets are thoroughly  reviewed in the book on Euclidean Distance Geometry \cite{liberti2017euclidean}.
Another related area is Rigidity Theory \cite{borcea2011minimally} whose key object is a finite or periodic graph with given lengths of edges.
The main problem is to find necessary and sufficient conditions for such a graph to have an embedding with straight-line edges, which is unique up to isometry \cite{kaszanitzky2021global}.
\medskip

The key novelty of Periodic Geometry is the aim to study the continuous space of all isometry classes of periodic structures, not isolated structures as in the past. 
Related results from \cite{dolbilin1998multiregular, bouniaev2017regular, dolbilin2019regular} and \cite{grishanov2009topological, morton2009doubly} will be reviewed in sections~\ref{sec:isoset_complete} and \ref{sec:textiles}.

\section{Average Minimum Distances (AMD) of a periodic point set}
\label{sec:AMD}

This section reminds the key results of \cite{widdowson2020average} and discusses new Examples~\ref{exa:SQ15} and~\ref{exa:SQ32}.

\begin{dfn}[Average Minimum Distances]
\label{dfn:AMD}
Let a periodic set $S=\La+M\subset\R^n$ have points $p_1,\dots,p_m$ in a motif.
Fix an integer $k\geq 1$.
For $j=1,\dots,k$, let $d_{ij}$ be the distance from the point $p_i\in M$
to its $j$-th nearest neighbor in the infinite set $S$.
The \emph{Average Minimum Distance} 
$\AMD_j(S)=\dfrac{1}{m}\sum\limits_{i=1}^m d_{ij}$ is the average of distances to $j$-th neighbors over points in the motif.
Set $\AMD^{(k)}=(\AMD_1,\dots,\AMD_k)$.
\bs
\end{dfn}

\newcommand{\amdh}{40mm}
\begin{figure}[h!]
\includegraphics[height=\amdh]{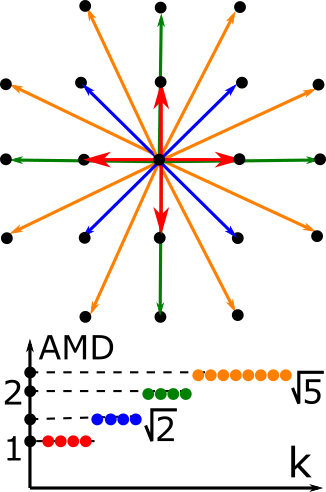}
\hspace*{3mm}
\includegraphics[height=\amdh]{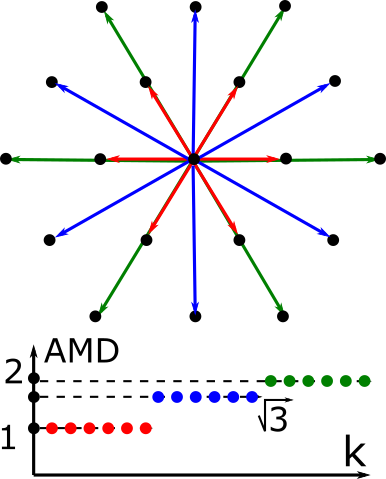}
\hspace*{3mm}
\includegraphics[height=\amdh]{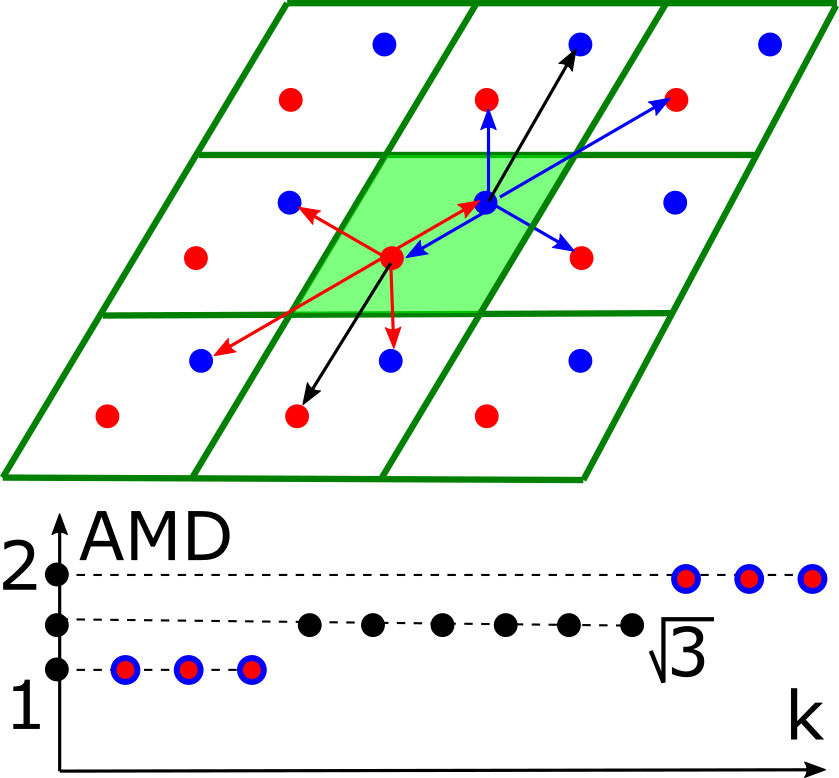}
\caption{\cite[Fig.~5]{widdowson2020average}. \textbf{Left}: in the square lattice, the $k$-th neighbors of the origin and corresponding $\AMD_k$ have the same colour, e.g. $\AMD_1=\dots=\AMD_4=1$ in red.
\textbf{Middle}: in the hexagonal lattice, the first 6 values are in red: $\AMD_1=\dots=\AMD_6=1$.
\textbf{Right}: the motif of two points (red and blue) within a green unit cell has the first three distances equal: $\AMD_1=\AMD_2=\AMD_3=1$.}
\label{fig:AMD}
\end{figure}

If $S$ is a lattice $\La$ containing the origin $0$, then $\AMD_k$ is the distance from $0$ to its $k$-th nearest neighbor in $\La$.
The averaging over points in a motif $M$ removes any dependence on point ordering within $M$ and also on a unit cell $U$.
\medskip

Notice that $k$ is not an input parameter that may change the AMD invariant.
Here $k$ is only the length of the sequence $\AMD^{(k)}(S)$.
The full $\AMD(S)$ is infinite.
The public codes for the AMD invariants are available in Python \cite{AMD_DW} and C++ \cite{AMD_MM}.
\medskip

The main proved results about the AMD invariants in \cite{widdowson2020average} are the following.
\smallskip

\noindent
\textbf{Isometry Theorem~4}: 
 invariance of the Average Minimum Distances $\AMD(S)$.
\smallskip

\noindent
\textbf{Continuity Theorem~9}: 
$|\AMD_k(S)-\AMD_k(Q)|\leq 2d_B(S,Q)$ for close $S,Q$.
\smallskip

\noindent
\textbf{Asymptotic Theorem~13}: for any periodic point set $S\subset\R^n$, if $k\to+\infty$, then $\AMD_k(S)$ approaches $c(S)\sqrt[n]{k}$. 
Here $c(S)$ is a point-based analogue of the density.
\smallskip

\noindent
\textbf{Complexity Theorem~14}: one can compute the vector $\AMD^{(k)}(S)$ in a time near linear in $km$, where $m$ is the number of points in a motif of a periodic set $S\subset\R^n$.
\medskip

\begin{exa}
\label{exa:SQ15}
\cite[Appendix~B]{widdowson2020average} discusses homometric periodic sets that can be distinguished by AMD and not by inter-point distance distributions or by powder diffraction patterns.
One of the challenging examples is the pair of 1D sets
$S_{15} = \{0,1,3,4,5,7,9,10,12\}+15\Z$ and
$Q_{15} = \{0,1,3,4,6,8,9,12,14\}+15\Z$, which have the period interval or the unit cell $[0,15]$ shown as a circle in Fig.~\ref{fig:SQ15}.
\medskip

\begin{figure}[ht]
\includegraphics[width=\textwidth]{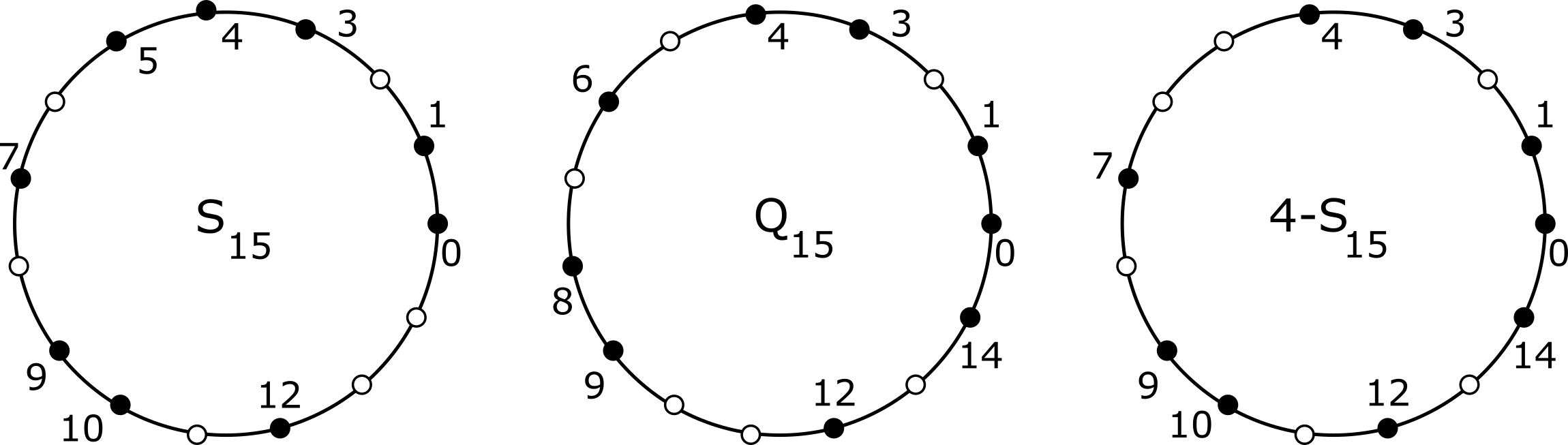}
\caption{Circular versions of the periodic sets $S_{15},Q_{15}$.
Distances are measured along round arcs.}
\label{fig:SQ15}      
\end{figure}

The periodic sets in Fig.~\ref{fig:SQ15} are obtained as Minkowski sums $S_{15}=U+V+15\Z$ and $Q_{15}=U-V+15\Z$ for 
$U = \{0, 4, 9\}$ and $V = \{0, 1, 3\}$.
The last picture in Fig.~\ref{fig:SQ15} shows the periodic set $4-S_{15}$ isometric to $S_{15}$.
Now the difference between $Q_{15}$ and $4-S_{15}$ is better visible: points $0,1,3,4,5,12,14$ are common, but points $6,8,9\in Q_{15}$ are shifted to $7,9,10\in 4-S_{15}$.
These sets cannot be distinguished by the more sophisticated density functions, see the beginning of section 5 in \cite{edels2021}.
\medskip

The highlighted cells in Tables~\ref{tab:S15} and~\ref{tab:Q15} show that $S_{15}$, $Q_{15}$ are distinguished by $\AMD_k$ for $k=3,4$.
Hence $\AMD$ is stronger than density functions in this case.
\bs
\end{exa}

\noindent
\begin{table}[h!]
\caption{\textbf{First row}: 9 points from the motif $M\subset S_{15}$ in Fig.~\ref{fig:SQ15}.
\textbf{Further rows}: distance from each point $p\in M$ to its $k$-th nearest neighbor in $S_{15}$.
\textbf{Last column}: $\AMD_k$ from Definition~\ref{dfn:AMD}. }
\label{tab:S15}
\begin{tabular}{C{10mm}|C{9mm}C{9mm}C{9mm}C{9mm}C{9mm}C{9mm}C{9mm}C{9mm}C{9mm}|C{10mm}}
\hline\noalign{\smallskip}
$S_{15}$ & 0 & 1 & 3 & 4 & 5 & 7 & 9 & 10 & 12 & $\AMD_k$ \\
\noalign{\smallskip}\svhline\noalign{\smallskip}
$k=1$      & 1 & 1 & 1 & 1 & 1 & 2 & 1 &  1 &  2 & 11/9 \\
\noalign{\smallskip}\noalign{\smallskip}
$k=2$      & 3 & 2 & 2 & 1 & 2 & 2 & 2 &  2 &  3 & 19/9 \\
\noalign{\smallskip}\noalign{\smallskip}
$k=3$      & 3 & 3 & 2 & 3 & 2 & 3 & 3 &  3 &  3 & \hl{25/9} \\
\noalign{\smallskip}\noalign{\smallskip}
$k=4$      & 4 & 4 & 3 & 3 & 4 & 3 & 4 &  5 &  4 & \hl{34/9} 
\end{tabular}
\end{table}

\begin{table}[h!]
\caption{\textbf{First row}: 9 points from the motif $M\subset Q_{15}$ in Fig.~\ref{fig:SQ15}.
\textbf{Further rows}: distance from each point $p\in M$ to its $k$-th nearest neighbor in $Q_{15}$.
\textbf{Last column}: $\AMD_k$ from Definition~\ref{dfn:AMD}. }
\label{tab:Q15}
\begin{tabular}{C{10mm}|C{9mm}C{9mm}C{9mm}C{9mm}C{9mm}C{9mm}C{9mm}C{9mm}C{9mm}|C{10mm}}
\hline\noalign{\smallskip}
$Q_{15}$ & 0 & 1 & 3 & 4 & 6 & 8 & 9 & 12 & 14 & $\AMD_k$ \\
\noalign{\smallskip}\svhline\noalign{\smallskip}
$k=1$      & 1 & 1 & 1 & 1 & 2 & 1 & 1 &  2  & 1 & 11/9 \\
\noalign{\smallskip}\noalign{\smallskip}
$k=2$      & 1 & 2 & 2 & 2 & 2 & 2 & 3 &  3  & 2 & 19/9 \\
\noalign{\smallskip}\noalign{\smallskip}
$k=3$      & 3 & 2 & 3 & 3 & 3 & 4 & 3 &  3  & 2 & \hl{26/9} \\
\noalign{\smallskip}\noalign{\smallskip}
$k=4$      & 3 & 3 & 3 & 4 & 3 & 4 & 5 &  4  & 4 & \hl{33/9}
\end{tabular}
\end{table}

\begin{figure}[ht]
\includegraphics[width=\textwidth]{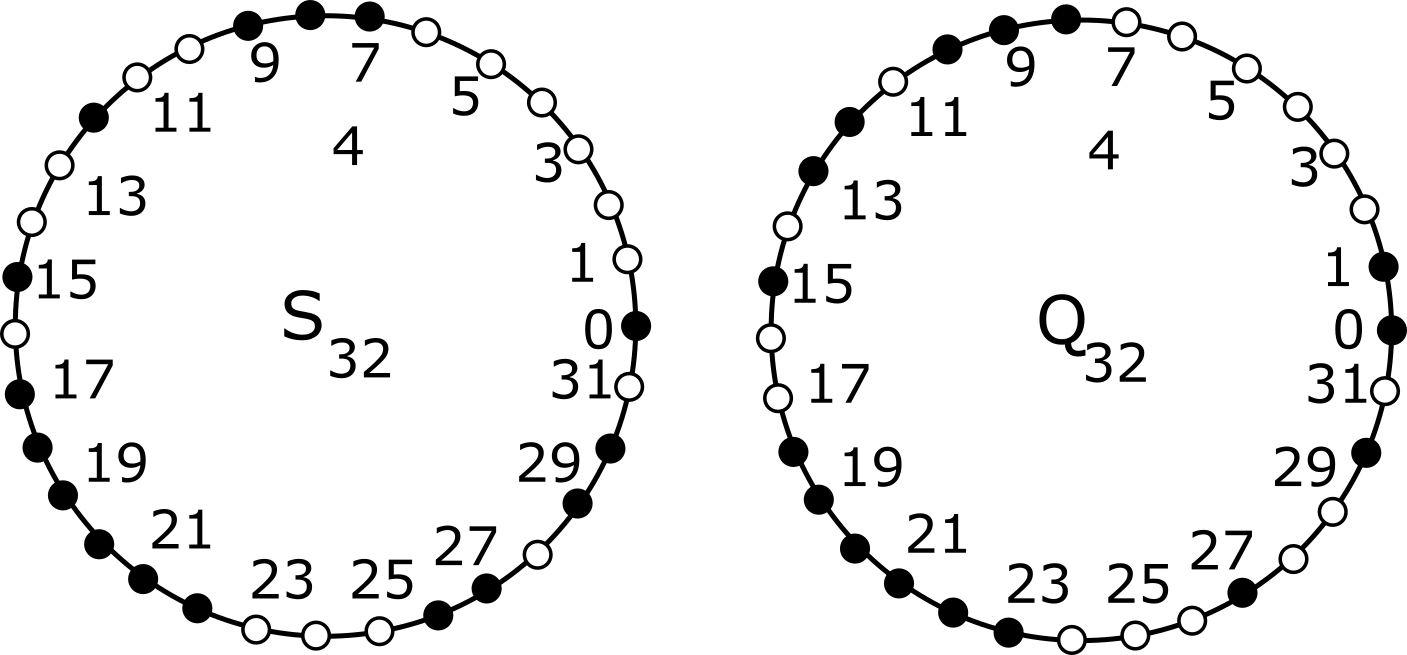}
\caption{Circular versions of the periodic sets $S_{32},Q_{32}$.
Distances are measured along round arcs.}
\label{fig:SQ32}      
\end{figure}

\begin{exa}
\label{exa:SQ32}
Another challenging pair of 1D periodic sets $S_{32},Q_{32}$ in Fig.~\ref{fig:SQ32} with period 32 was experimentally distinguished by density functions in \cite[Example 5.2]{edels2021}.
Tables~\ref{tab:S32},~\ref{tab:Q32} show that these sets are distinguished by $\AMD_k$ for $k=2,3$.
\bs
\end{exa}

\begin{table}[h!]
\caption{\textbf{First row}: 16 points from the motif $M\subset S_{32}$. 
\textbf{Further rows}: distance from each point $p\in M$ to its $k$-th nearest neighbor in $S_{32}$.
\textbf{Last column}: $\AMD_k$ from Definition~\ref{dfn:AMD}. }
\label{tab:S32}
\begin{tabular}{C{10mm}|C{5mm}C{5mm}C{5mm}C{5mm}C{5mm}C{5mm}C{5mm}C{5mm}C{5mm}C{5mm}C{5mm}C{5mm}C{5mm}C{5mm}C{5mm}C{5mm}|C{10mm}}
\hline\noalign{\smallskip}
$S_{32}$ & 0 & 7 & 8 & 9 & 12 & 15 & 17 & 18 & 19 & 20 & 21 & 22 & 26 & 27 & 29 & 30 & $\AMD_k$ \\
\noalign{\smallskip}\svhline\noalign{\smallskip}
$k=1$     & 2 & 1 & 1 & 1 &  3  &  2  &  1  &  1 &   1  &  1  &  1  &  1 &  1  &   1 &   1 &  1 & 20/16 \\
\noalign{\smallskip}\noalign{\smallskip}
$k=2$     & 3 & 2 & 1 & 2 &  3  &  3  &  2  &  1 &   1  &  1  &  1  &  2 &  3  &   2 &   2 &  2 & \hl{31/16} \\
\noalign{\smallskip}\noalign{\smallskip}
$k=3$     & 5 & 5 & 4 & 3 &  4  &  3  &  2  &  2 &   2  &  2  &  2  &  3 &  4  &   3 &   3 &  3 & \hl{50/16}
\end{tabular}
\end{table}

\begin{table}[h!]
\caption{\textbf{First row}: 16 points from the motif $M\subset Q_{32}$. 
\textbf{Further rows}: distance from each point $p\in M$ to its $k$-th nearest neighbor in $Q_{32}$.
\textbf{Last column}: $\AMD_k$ from Definition~\ref{dfn:AMD}. }
\label{tab:Q32}
\begin{tabular}{C{10mm}|C{5mm}C{5mm}C{5mm}C{5mm}C{5mm}C{5mm}C{5mm}C{5mm}C{5mm}C{5mm}C{5mm}C{5mm}C{5mm}C{5mm}C{5mm}C{5mm}|C{10mm}}
\hline\noalign{\smallskip}
$Q_{32}$ & 0 & 1 & 8 & 9 & 10 & 12 & 13 & 15 & 18 & 19 & 20 & 21 & 22 & 23 & 27 & 30 & $\AMD_k$ \\
\noalign{\smallskip}\svhline\noalign{\smallskip}
$k=1$      & 1 & 1 & 1 & 1 &  1  &  1  &  1  &  2 &   1  &  1  &  1  &  1 &  1  &   1 &   3 &  2 & 20/16 \\
\noalign{\smallskip}\noalign{\smallskip}
$k=2$      & 2 & 3 & 2 & 1 &  2  &  2  &  2  &  3 &   2  &  1  &  1  &  1 &  1  &   2 &   4 &  3 & \hl{32/16} \\
\noalign{\smallskip}\noalign{\smallskip}
$k=3$      & 5 & 6 & 4 & 3 &  2  &  3  &  3  &  3 &   3  &  2  &  2  &  2 &  2  &   3 &   5 &  3 & \hl{51/16} \\
\end{tabular}
\end{table}

\section{Density functions (densigrams) extend the single-value density}
\label{sec:densities}

This section reviews the key results of \cite{edels2021} and then proves new Theorem~\ref{thm:densities1D}.

\begin{dfn}[density functions]
\label{dfn:densities}
Let a periodic set $S=\La+M\subset\R^n$ have a unit cell $U$.
Fix an integer $i\geq 0$.
Let the subregion $U_k(t)\subset U$ be covered by $k$ closed balls with a radius $t>0$ and centers at all points of $S$.
The \emph{density function} is $\psi_k(t)=\vol[U_k(t)]/\vol[U]$.
The \emph{density fingerprint} is the sequence $\Psi[S]=\{\psi_k(t)\}_{k=0}^{+\infty}$.
Fig.~\ref{fig:densities_hex+sq} shows
the \emph{densigram} of accumulated functions $\sum\limits_{i=1}^k\psi_i(t)$.
\bs
\end{dfn}

\newcommand{\cheight}{24mm}
\newcommand{\bwidth}{50mm}
\newcommand{\dwidth}{64mm}
\begin{figure}[h!]
  \parbox{\bwidth}{
  \includegraphics[height=\cheight]{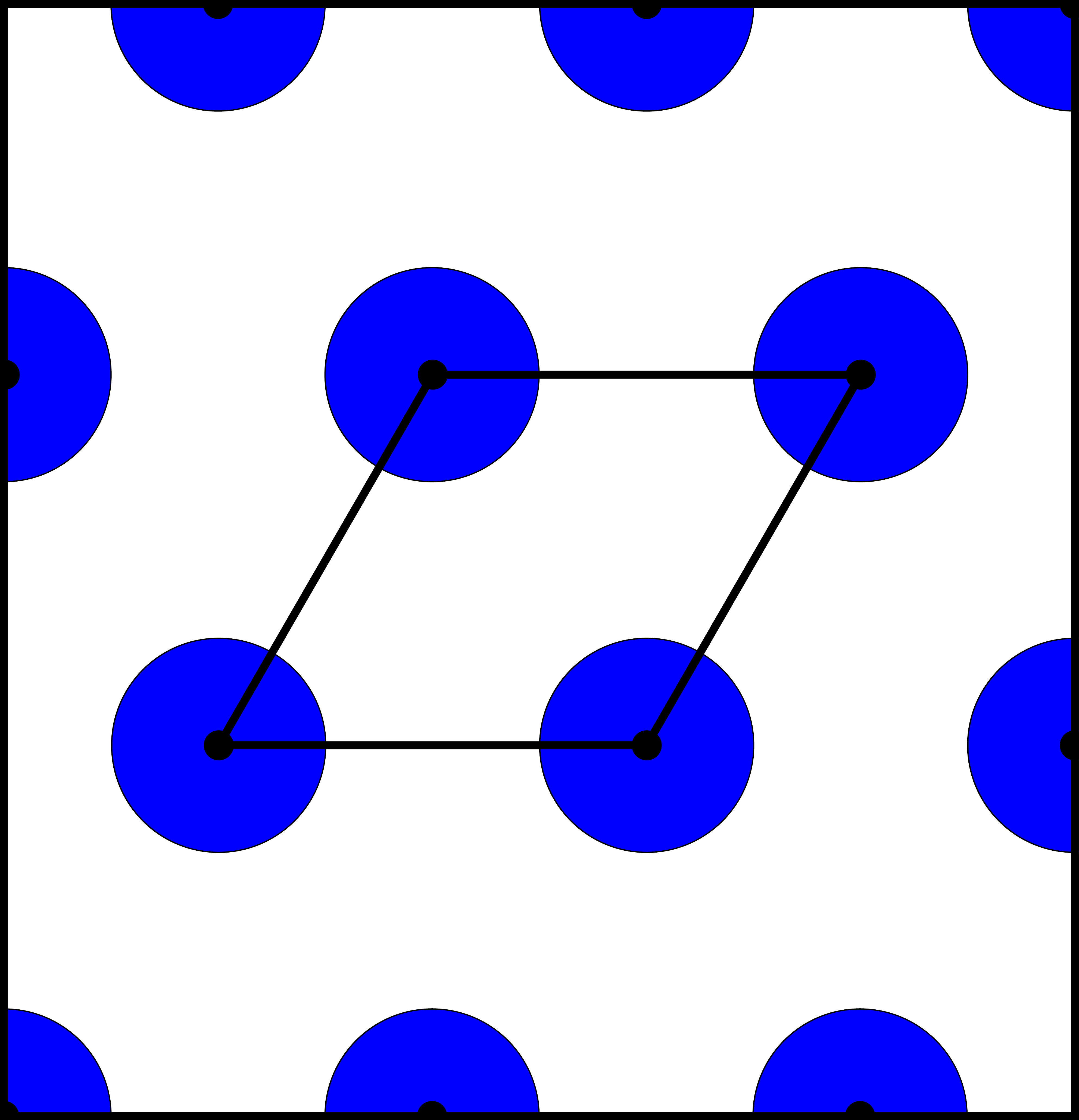}
  \hspace*{1mm}
  \includegraphics[height=\cheight]{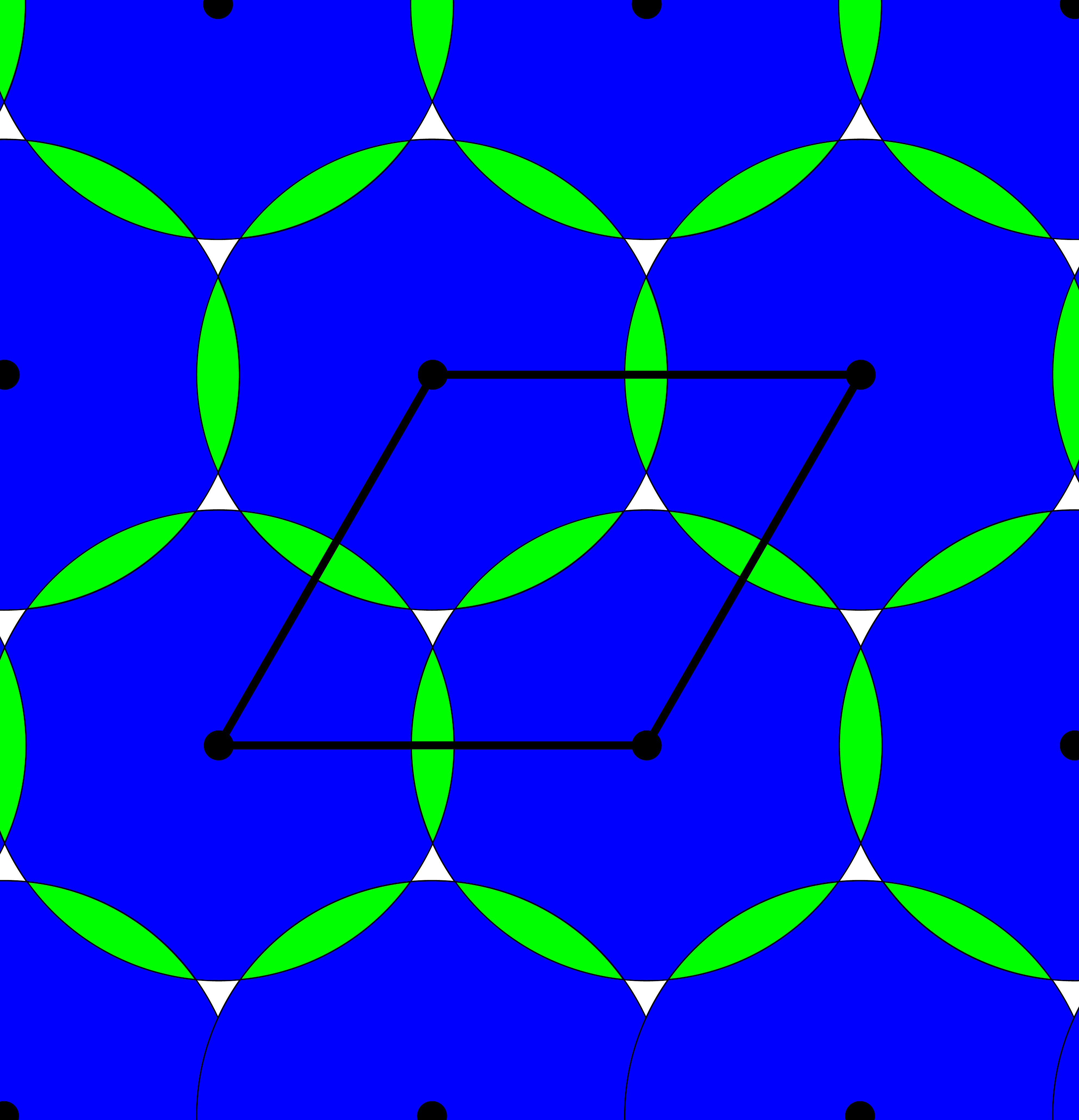}
  \medskip
  \includegraphics[height=\cheight]{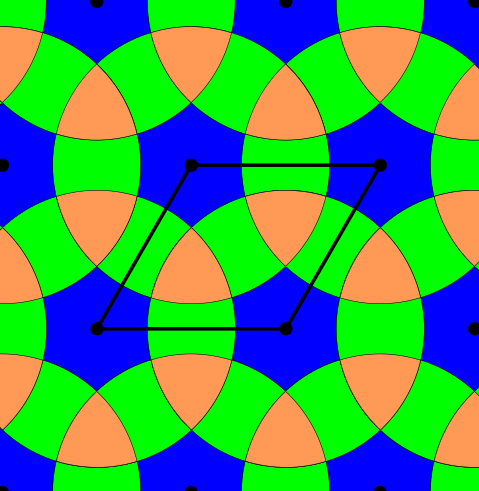}
  \hspace*{1mm}
  \includegraphics[height=\cheight]{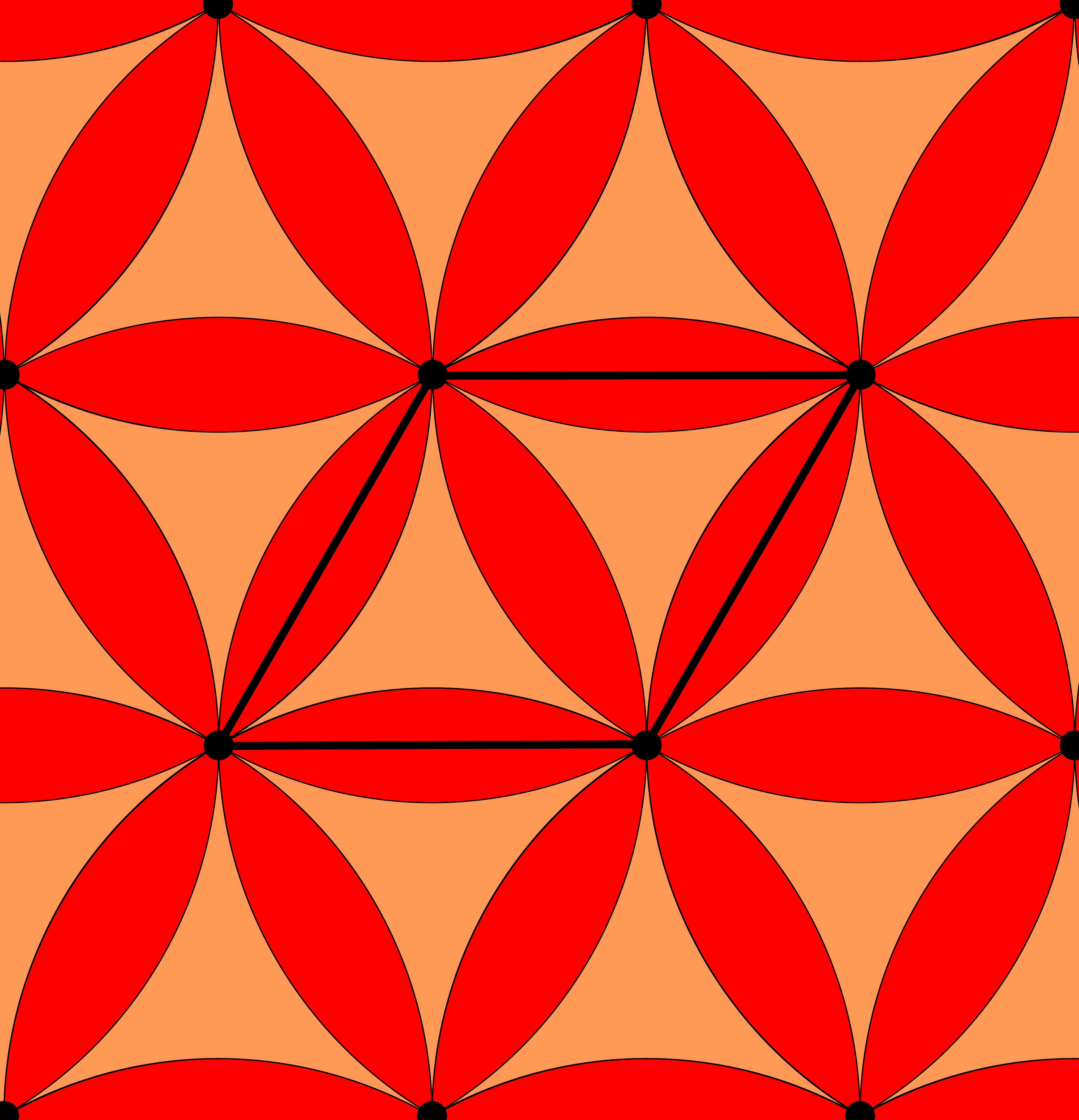}}
  \parbox{80mm}{
  \includegraphics[width=\dwidth]{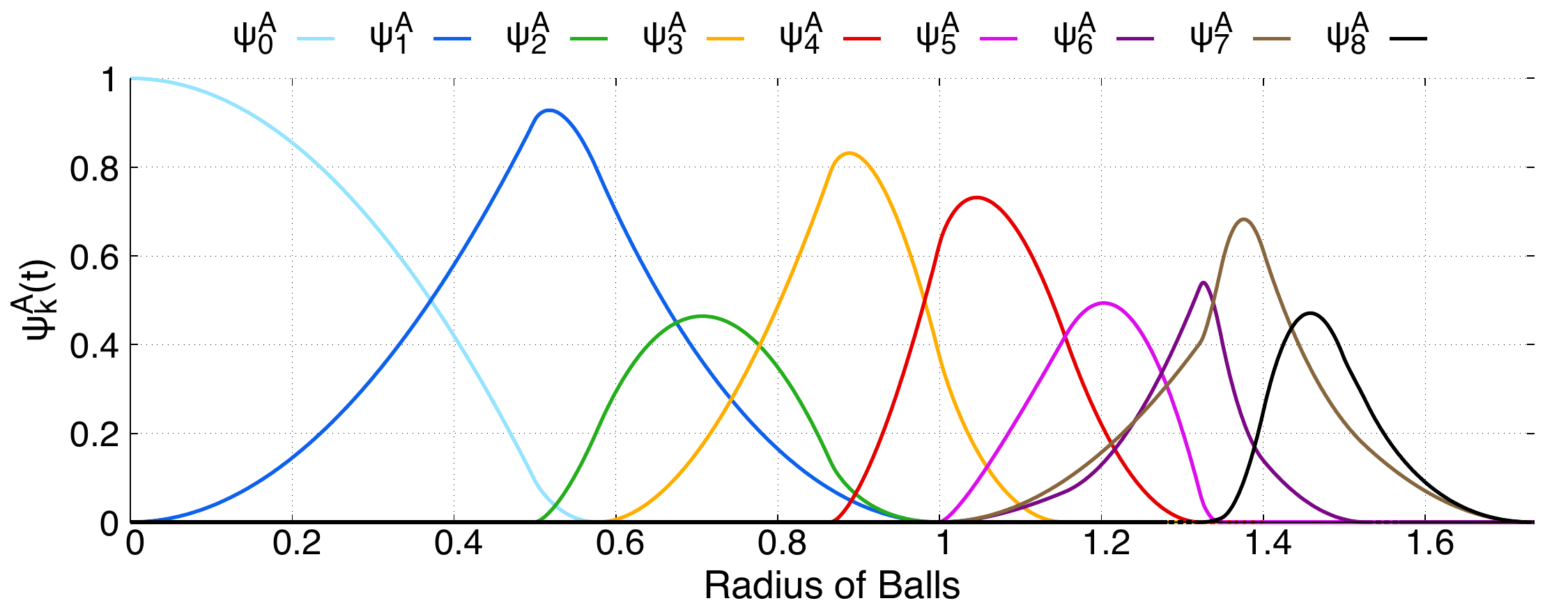}
  \includegraphics[width=\dwidth]{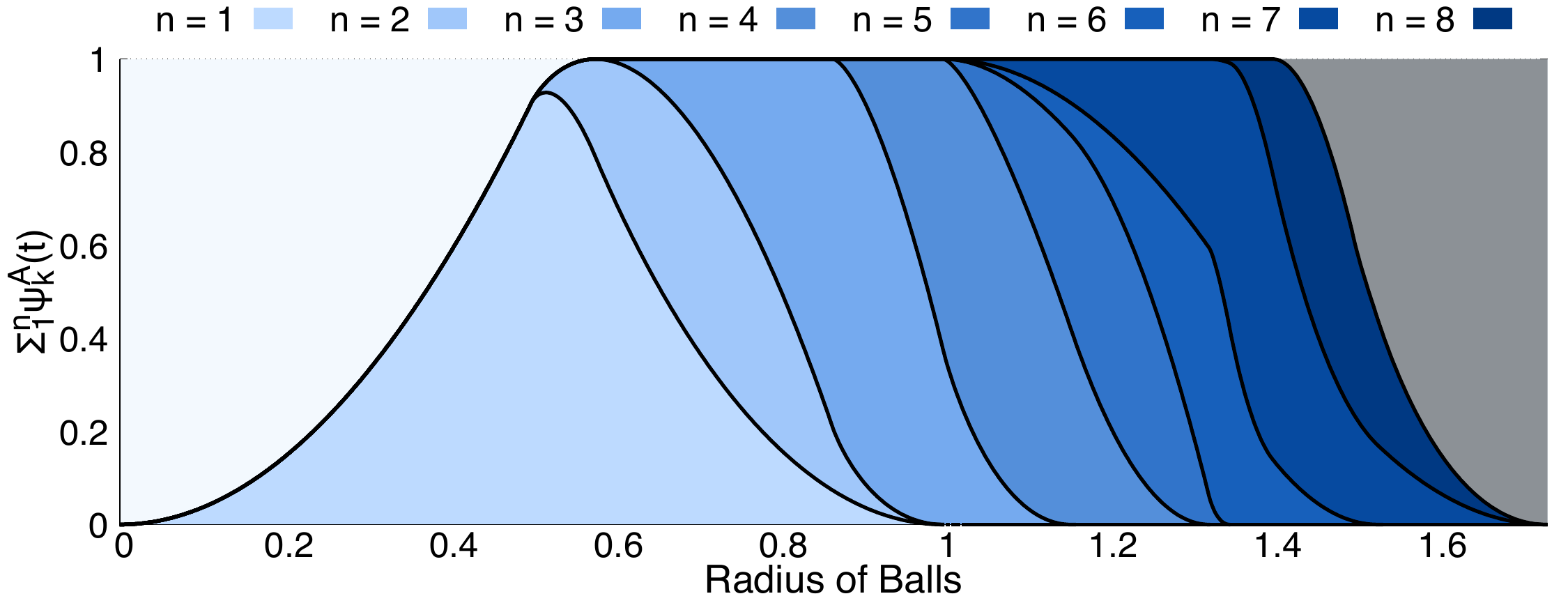}}
\vspace*{1mm}
  
  \parbox{\bwidth}{
  \includegraphics[height=\cheight]{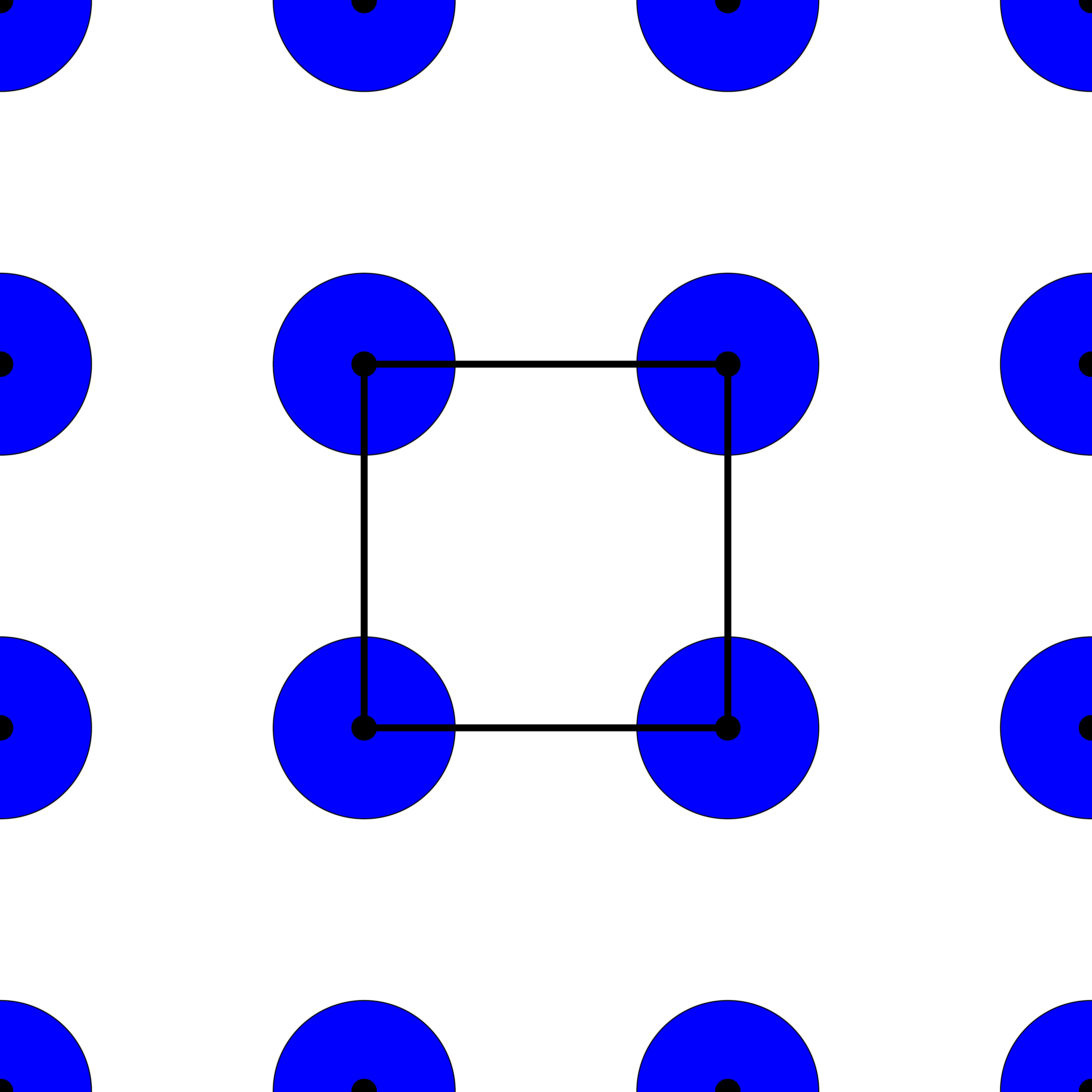}
  \hspace*{1mm}
  \includegraphics[height=\cheight]{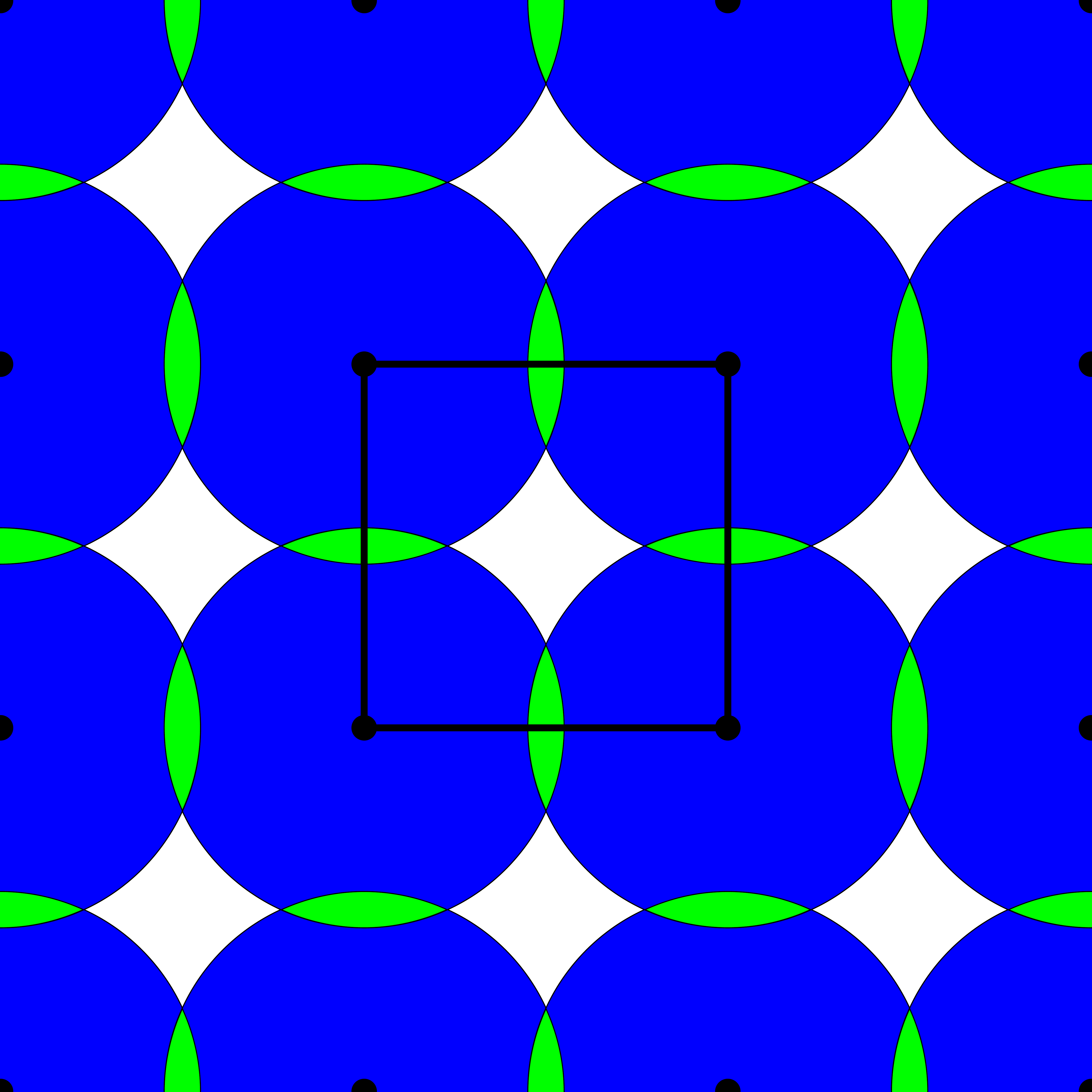}
  \medskip
  \includegraphics[height=\cheight]{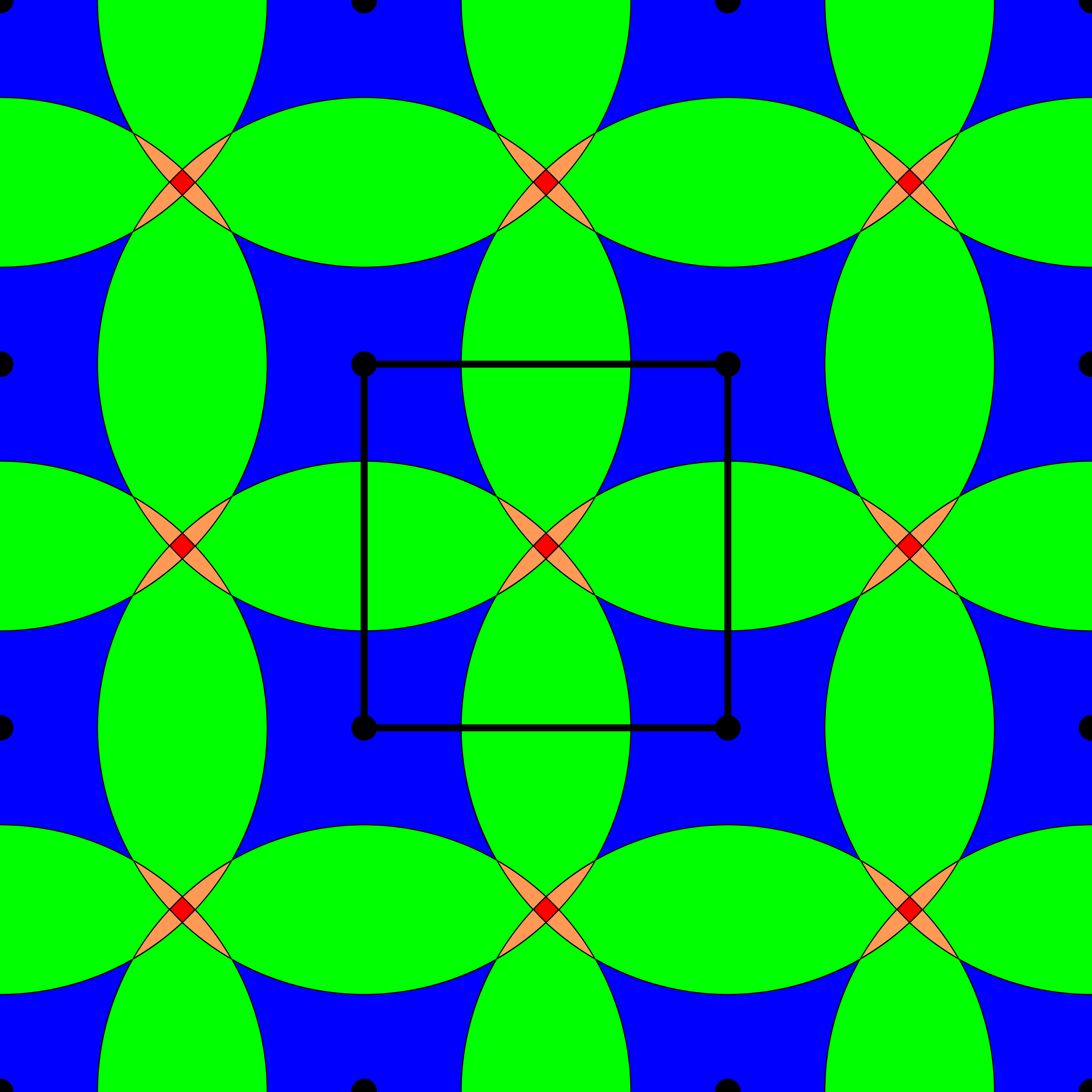}
  \hspace*{1mm}
  \includegraphics[height=\cheight]{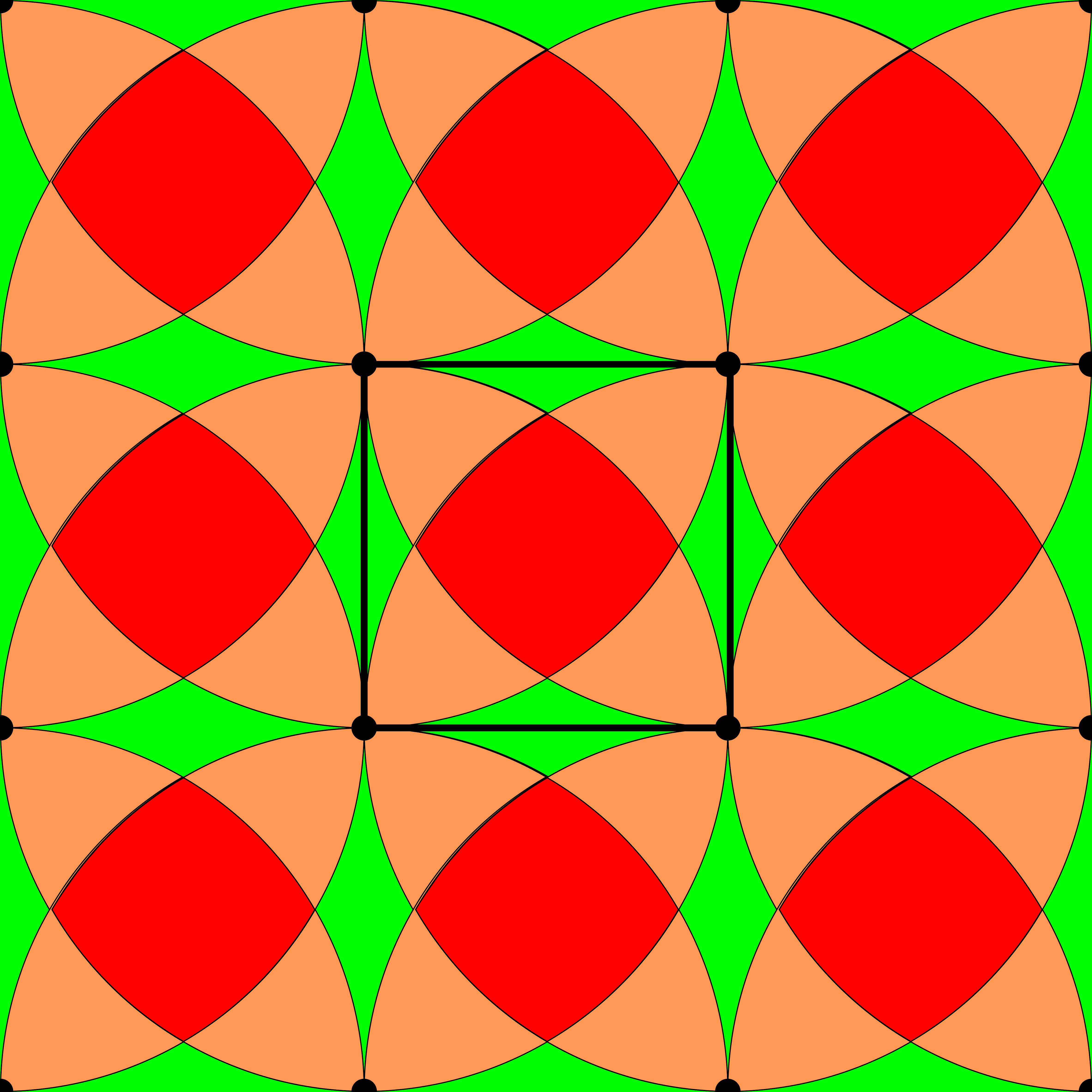}}
  \parbox{80mm}{
  \includegraphics[width=\dwidth]{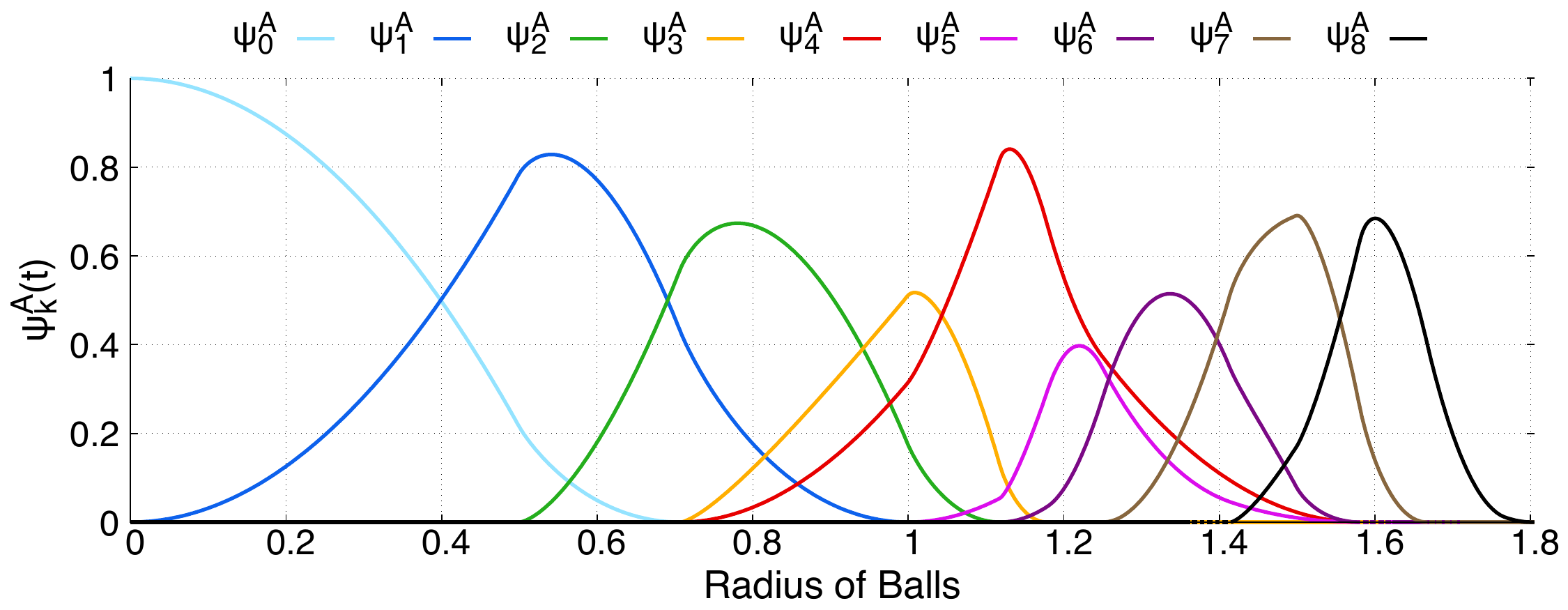}
  \includegraphics[width=\dwidth]{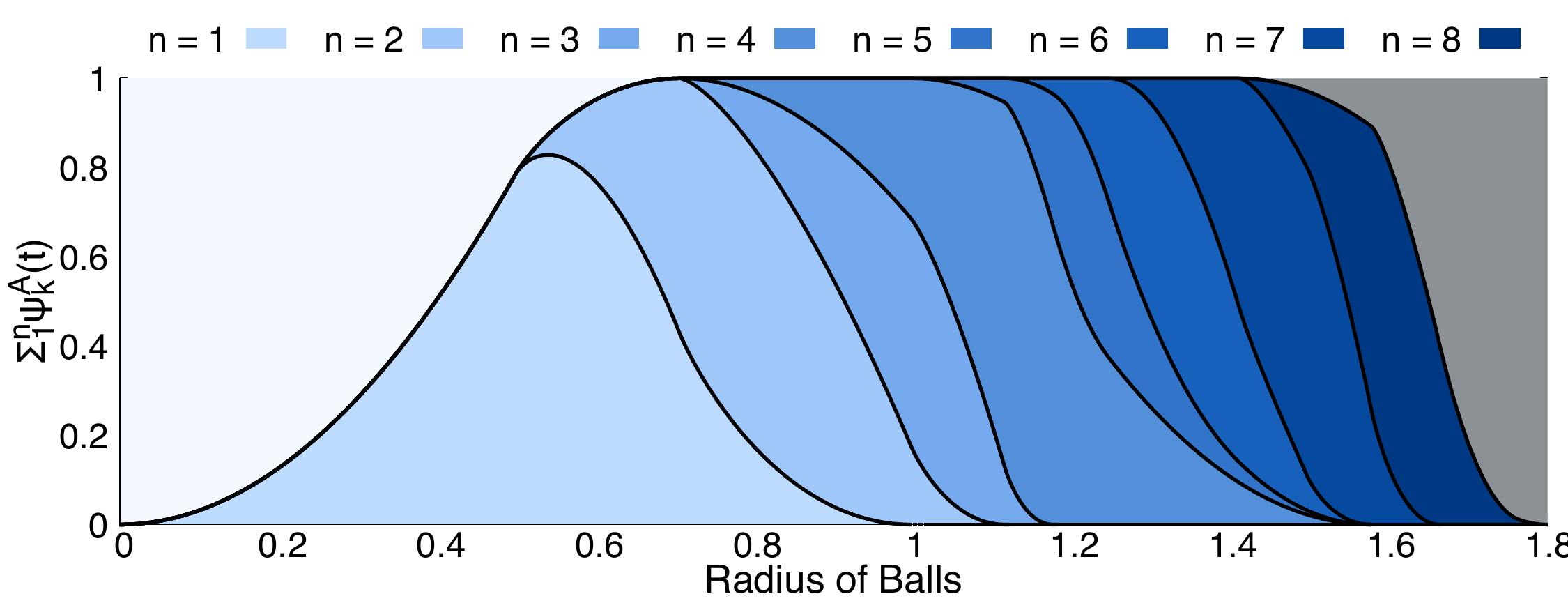}}
  \caption{\cite[Fig.~2]{edels2021}: illustration of Definition~\ref{dfn:densities} for the hexagonal and square lattices.
  \textbf{Left}: subregions $U_i(t)$ covered by $i$ disks for $t=0.25, 0.55, 0.75, 1.00$.
  \textbf{Right}: the graphs of the respective first nine density functions above the corresponding \emph{densigram}, in which the zeroth function can be seen upside-down and the remaining density functions are accumulated from left to right.}
  \label{fig:densities_hex+sq}
\end{figure}

The main proved results about density functions in \cite{edels2021} are the following.
\smallskip

\noindent
\textbf{Lemma~3.2}: 
isometry invariance of $\Psi(S)$ for all periodic sets $S\subset\R^3$.
\smallskip

\noindent
\textbf{Theorem~4.2}: 
continuity in the bottleneck distance for periodic sets $S\subset\R^3$.
\smallskip

\noindent
\textbf{Theorem~5.1}: 
completeness for \emph{generic} 3D sets defined in \cite[section~5.1]{edels2021}. 
\smallskip

\begin{figure}[ht]
\includegraphics[height=58mm]{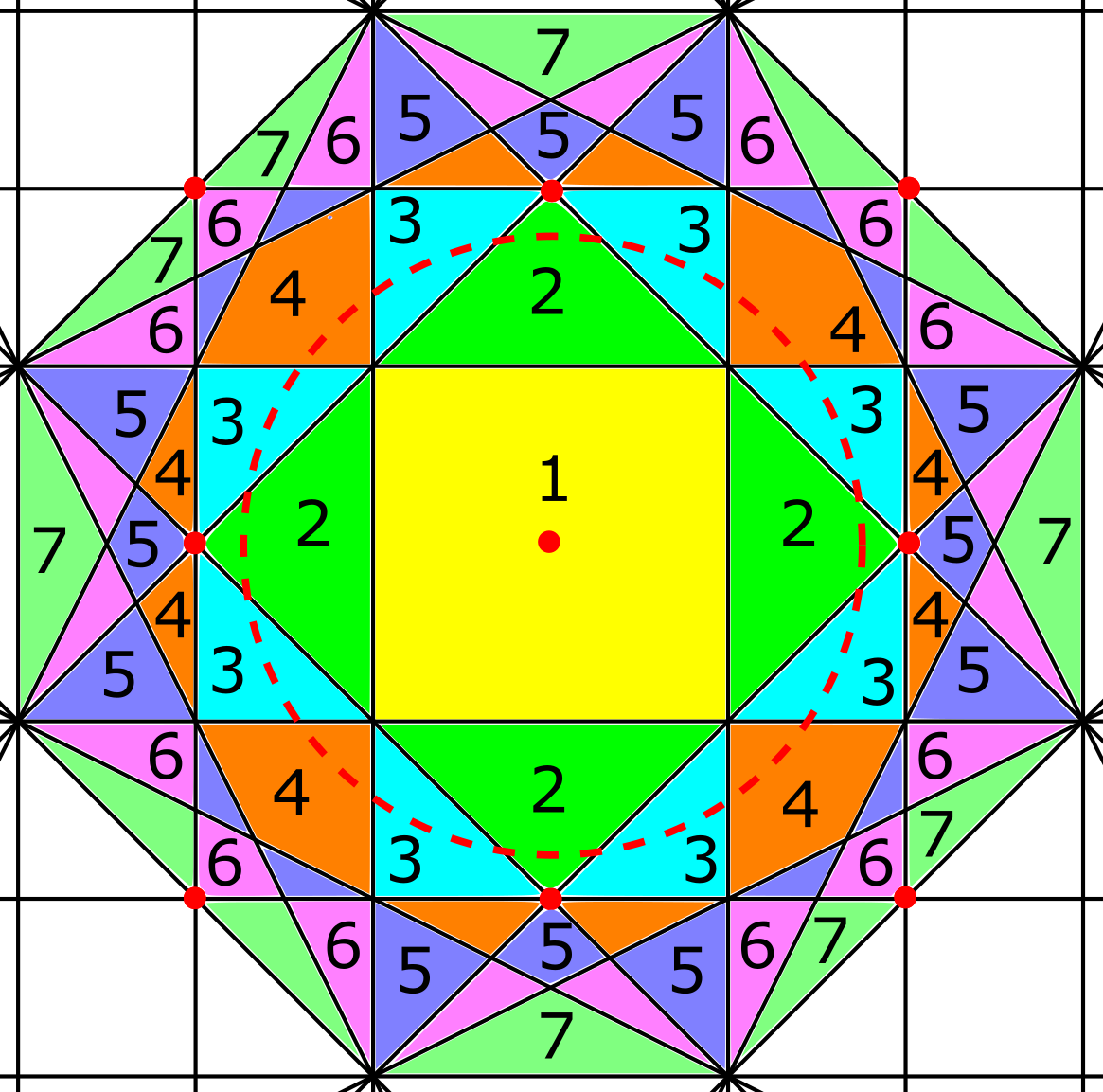}
\includegraphics[height=58mm]{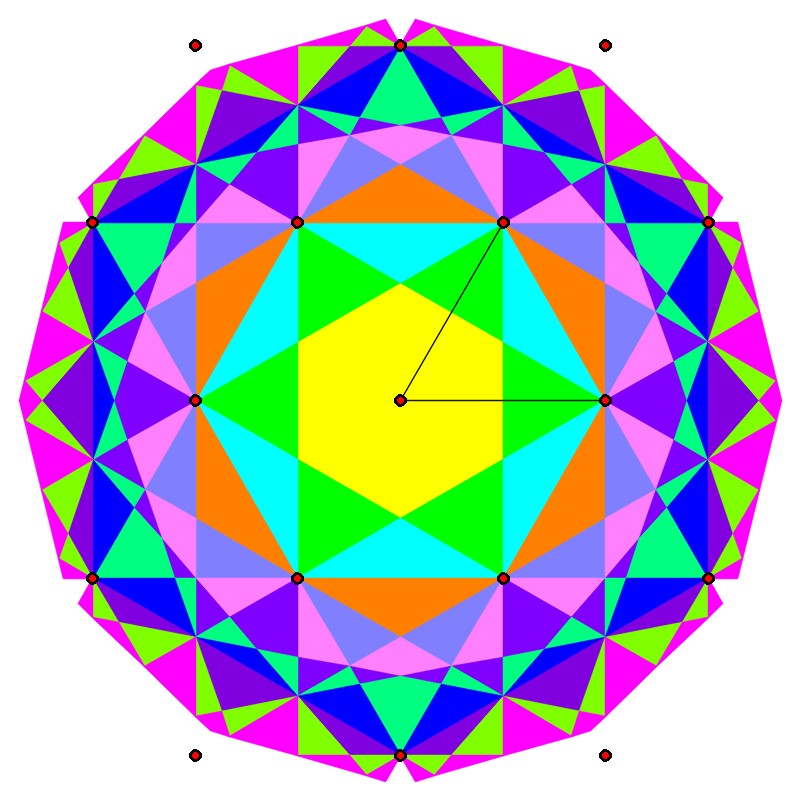}
\caption{\textbf{Left}: Brillouin zones of the square lattice.
\textbf{Right}: Brillouin zones of the hexagonal lattice.}
\label{fig:Brillouin_zones}      
\end{figure}

\noindent
\textbf{Theorem~6.1}: to compute $\psi_k(t)$ for a periodic set $S$, we intersect the $k$-th Brillouin zone $Z_i(S,p)$ with the ball $\bar B(p;t)$ for each point $p$ in a motif of $S$, see Fig.~\ref{fig:Brillouin_zones}.
\medskip

For a point $p$ in a periodic set $S\subset\R^n$, the $i$-th \emph{Brillouin zone} $Z_k(S,p)$ is the closure of the region consisting of all points $q\in\R^n$ that have $p$ as its $k$-th nearest neighbor in $S$.
For example, the first zone $Z_1(S,p)$ is the Voronoi diagram of the point $p$ in $S$.
In Fig.~\ref{fig:Brillouin_zones} all polygons of the same color belong to one Brillouin zone.
\medskip

Comparing density functions, we found that the crystal T2-$\de$ reported in \cite{pulido2017functional} was not deposited in the Cambridge Structural Database (CSD), which is the world's largest collection of over 1M synthesized crystals.
The AMD invariant has confirmed these comparisons.
After raising alert, T2-$\de$ is now deposited with the id SEMDIA.
\medskip

This story illustrates the fact that there were no invariant-based tools to continuously quantify a similarity between crystals.
The Python computation of $\AMD_k$ up to $k = 1000$ for 5679 crystals reported in \cite{pulido2017functional} took about 17 min on a standard desktop, while eight density functions required three days using the C++ code \cite{Densities_PS}.

\begin{exa}
\label{exa:densities1D}
Fig.~\ref{fig:densities1D} illustrates how density functions $\psi_k(t)$ are computed for the periodic set $S=\{0,\frac{1}{3},\frac{1}{2}\}+\Z$.
By Definition~\ref{dfn:densities} $\psi_0(t)$ is the fractional length within the period interval $[0,1]$ not covered by the 1D balls of radius $t$, which are the red intervals $[0,t]\cup[1-t,1]$, green dashed interval $[\frac{1}{3}-t,\frac{1}{3}+t]$ and blue dotted interval $[\frac{1}{2}-t,\frac{1}{2}+t]$.
The graph of $\psi_0(t)$ starts from the point $(0,1)$ at $t=0$.
Then $\psi_0(t)$ linearly drops to the point $(\frac{1}{12},\frac{1}{3})$ at $t=\frac{1}{12}$ when a half of the interval $[0,1]$ remains uncovered.
The next linear piece of $\psi_0(t)$ continues to the point $(\frac{1}{6},\frac{1}{6})$ at $t=\frac{1}{6}$ when only $[\frac{2}{3},\frac{5}{6}]$ is uncovered.
The graph of $\psi_0(t)$ finally returns to the $t$-axis at the point $(\frac{1}{4},0)$ and remains there for $t\geq \frac{1}{4}$.
The piecewise linear behavior of $\psi_0(t)$ can be briefly described via the four \emph{corner} points $(0,1)$, $(\frac{1}{12},\frac{1}{3})$, $(\frac{1}{6},\frac{1}{6})$, $(\frac{1}{4},0)$.
\medskip

The 1st density function $\psi_1(t)$ can be obtained as a sum of three \emph{trapezoid} functions $\eta_R$, $\eta_G$, $\eta_B$, each measuring the length of an interval covered by a single ball (interval of one color).
The red intervals $[0,t]\cup[1-t,1]$ grow until $t=\frac{1}{6}$ when they touch the green interval $[\frac{1}{6},\frac{1}{2}]$.
So the length $\eta_R(t)$ of this interval linearly grows from the origin $(0,0)$ to the corner point $(\frac{1}{6},\frac{1}{3})$.
For $t\in[\frac{1}{6},\frac{1}{4}]$, the left red interval is shrinking at the same rate due to the overlapping green interval, while the right red interval continues to grow until $t=\frac{1}{4}$, when it touches the blue interval $[\frac{1}{4},\frac{3}{4}]$.  
Hence the graph of $\eta_R(t)$ remains constant up to the corner point $(\frac{1}{4},\frac{1}{3})$.
After that $\eta_R(t)$ linearly returns to the $t$-axis at $t=\frac{5}{12}$.
Hence the trapezoid function $\eta_R$ has the piecewise linear graph through the corner points $(0,0)$, $(\frac{1}{6},\frac{1}{3})$, $(\frac{1}{4},\frac{1}{3})$, $(\frac{5}{12},\frac{1}{0})$. 
\medskip

\begin{figure}[ht]
\includegraphics[width=\textwidth]{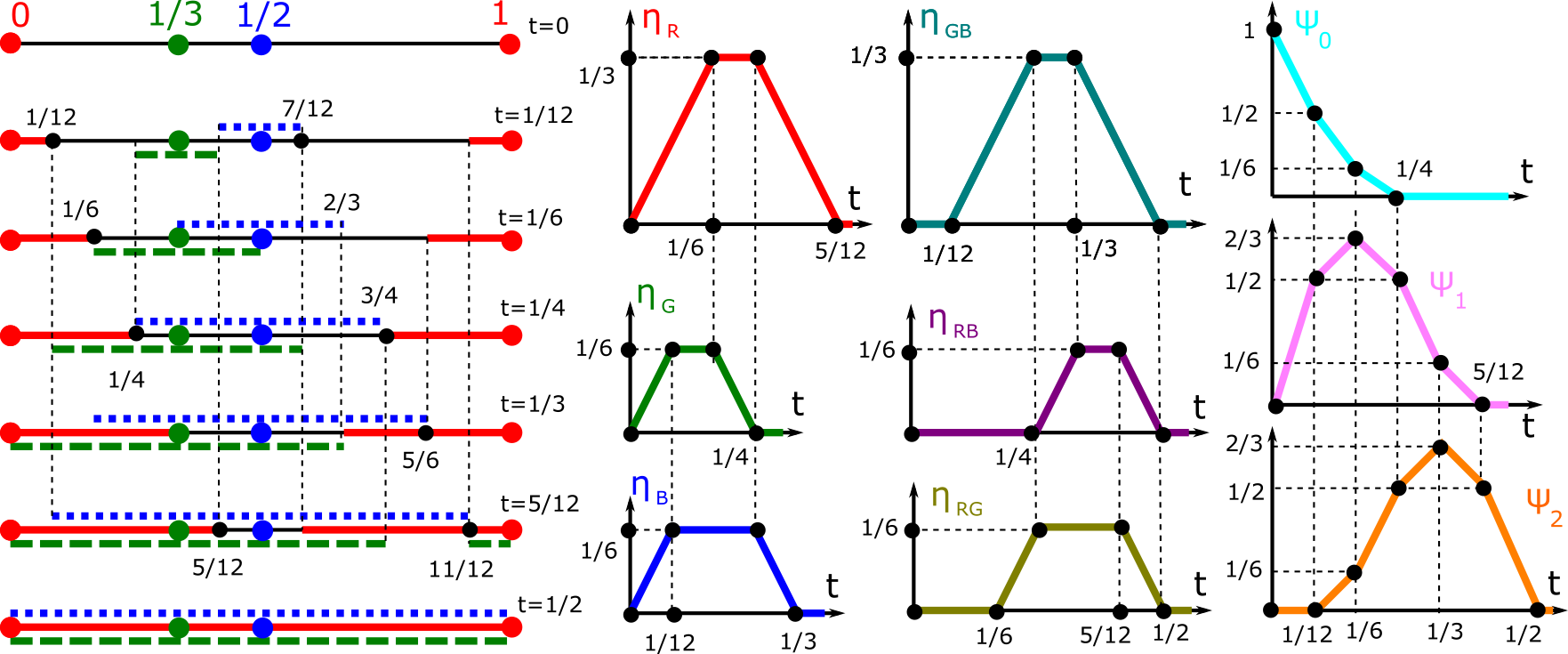}
\caption{\textbf{Left}: periodic set $S=\{0,\frac{1}{3},\frac{1}{2}\}+\Z$. The growing balls around the red point $0\equiv 1$, green point $\frac{1}{3}$ and blue point $\frac{1}{2}$ are shown in the same color for various radii $t$.
\textbf{Middles}: the trapezoid functons $\eta$ are explained in Example~\ref{exa:densities1D}.
\textbf{Right}: the density functions $\psi_i(S)$ from Definition~\ref{dfn:densities}.}
\label{fig:densities1D}      
\end{figure}

The 2nd density function $\psi_2(t)$ can be obtained as a sum of three \emph{trapezoid} functions $\eta_{GB},\eta_{RG},\eta_{RB}$, each measuring the length of a double interval intersection.
For the green interval $[\frac{1}{3}-t,\frac{1}{3}+t]$ and the blue interval $[\frac{1}{2}-t,\frac{1}{2}+t]$, the graph of the trapezoid function $\eta_{GB}(t)$ is piecewise linear and starts at the point $(\frac{1}{12},0)$, where the intervals touch.
The green-blue intersection interval $[\frac{1}{2}-t,\frac{1}{3}+t]$ grows until $t=\frac{1}{4}$, when $[\frac{1}{4},\frac{7}{12}]$ touches the red interval on the left.
At the same time $\eta_{GB}(t)$ is linearly growing to the point $(\frac{1}{4},\frac{1}{3})$.
For $t\in[\frac{1}{4},\frac{1}{3}]$, the green-blue intersection interval becomes shorter on the left, but grows at the same rate on the right until $[\frac{1}{3},\frac{2}{3}]$ touches the red interval $[\frac{2}{3},1]$.
Then $\eta_{GB}(t)$ remains constant up to the point $(\frac{1}{3},\frac{1}{3})$.
For $t\in[\frac{1}{3},\frac{1}{2}]$ the green-blue intersection interval is shortening from both sides.
Finally, the graph of $\eta_{GB}(t)$ returns to the $t$-axis at $(\frac{1}{2},0)$, see Fig.~\ref{fig:densities1D}.
\bs
\end{exa}

Though the density functions may look more informative than distance-based invariants such as AMD, the following result explicitly describes all density functions of a 1D periodic set $S\subset\R$ in terms of distances between closest neighbors.

\begin{thm}[density functions in 1D]
\label{thm:densities1D}
Any periodic point set $S\subset\R$ can be scaled to period 1 so that $S=\{p_1,\dots,p_m\}+\Z$.
Set $d_i=p_{i+1}-p_i\in(0,1)$, where $i=1,\dots,m$ and $p_{m+1}=p_1+1$. 
Put the distances in the increasing order $d_{[1]}\leq d_{[2]}\leq\dots\leq d_{[m]}$.
Any density function $\psi_k:[0,+\infty)\to[0,1]$ of $S$ is piecewise linear and passes through several \emph{corners} $(a_j,b_j)$ so that 
$\psi_k$ is linear between any successive corners.
\medskip

\noindent
\textbf{(\ref{thm:densities1D}a)}
The 0-th density function $\psi_0$ has the (unordered) corners: $(0,1)$ and also \\ 
$(\frac{1}{2}d_{[i]}, 1-\sum\limits_{j=1}^{i-1} d_{[j]}-(m-i+1)d_{[i]})$ for $i=1,\dots,m$, e.g. the last corner is $(\frac{1}{2}d_{[m]},0)$.
If any corner points are repeated, e.g. when $d_{[i-1]}=d_{[i]}$, they are collapsed into one. 
So $\psi_0(t)$ is uniquely determined by the ordered distances $d_{[1]}\leq d_{[2]}\leq\dots\leq d_{[m]}$.
\medskip

\noindent
Below we will consider any index $i=1,\dots,m$ (of a point $p_i$ or a distance $d_i$) modulo $m$ so that $m+1\equiv 1$.
Any interval $[p_i-t,p_i+t]$ is projected modulo $\Z$ to $[0,1]$.
\medskip

\noindent
\textbf{(\ref{thm:densities1D}b)}
The density function $\psi_1$ is the sum of $m$ \emph{trapezoid} functions $\eta_{1i}$ with the corner points $(0,0)$, $(\frac{d_{i-1}}{2}, d)$, $(\frac{d_{i}}{2},d)$, $(\frac{d_{i-1}+d_{i}}{2},0)$, where $d=\min\{d_{i-1},d_{i}\}$, $d_{0}=d_m$, $i=1,\dots,m$. 
If $d_{i-1}=d_{i}$, the two corner points are collapsed into one. 
Hence $\psi_1(t)$ is uniquely determined by the unordered set of unordered pairs $(d_{i-1},d_i)$, $i=1,\dots,m$.
\medskip

\noindent
\textbf{(\ref{thm:densities1D}c)}
For $1<k\leq m$, the density function $\psi_k$ is the sum of $m$ \emph{trapezoid} functions $\eta_{ki}$ with the corner points 
$(\frac{s}{2},0)$, 
$(\frac{d_{i-1}+s}{2},d)$, 
$(\frac{s+d_{i+k-1}}{2},d)$, 
$(\frac{d_{i-1}+s+d_{i+k-1}}{2},0)$, where 
$d=\min\{d_{i-1},d_{i+k-1}\}$, $s=\sum\limits_{j=i}^{i+k-2}d_j$,
$i=1,\dots,m$.
Then $\psi_k$ is determined by the unordered set of triples $(d_{i-1},s,d_{i+k-1})$ whose first and last entries are swappable.
\medskip

\noindent
\textbf{(\ref{thm:densities1D}d)}
The density functions satisfy the \emph{periodicity} 
$\psi_{k+m}(t+\frac{1}{2})=\psi_{k}(t)$ for any $k\geq 0$, $t\geq 0$, and the \emph{symmetry} $\psi_{m-k}(\frac{1}{2}-t)=\psi_k(t)$ for $k=1,\dots,[\frac{m}{2}]$ and $t\in[0,\frac{1}{2}]$.
\medskip

\noindent
\textbf{(\ref{thm:densities1D}e)}
Let $S,Q\subset\R$ be periodic sets whose motifs have at most $m$ points.
For $k\geq 1$, one can draw the graph of the $k$-th density function $\psi_k[S]$ in time $O(m^2)$.
One can also check in time $O(m^3)$ if the density fingerprints coincide: $\Psi(S)=\Psi(Q)$.
\bs 
\end{thm}
\begin{proof}
\textbf{(\ref{thm:densities1D}a)}
The function $\psi_0(t)$ measures the total length of subintervals in $[0,1]$ that are not covered by growing intervals $[p_i-t,p_i+t]$, $i=1,\dots,m$. 
Hence $\psi_0(t)$ linearly decreases on $t$ from the initial value $\psi_0(0)=1$ except for $m$ critical values of $t$ where one of the intervals $[p_i,p_{i+1}]$ between successive points become completely covered and can not longer shrink.
These critical radii $t$ are ordered according to the distances $d_{[1]}\leq d_{[2]}\leq\dots\leq d_{[m]}$.
The first critical radius is $t=\frac{1}{2}d_{[1]}$, when the shortest interval $[p_i,p_{i+1}]$ of the length $d_{[1]}$ is covered by the balls centered at $p_i,p_{i+1}$.
At this moment, all $m$ intervals cover the subregion of the length $md_{[1]}$.
Then the graph of $\psi_0(t)$ has the first corner points $(0,1)$ and $(\frac{1}{2}d_{[1]},1-md_{[1]})$. 
\smallskip

The second critical radius is $t=\frac{1}{2}d_{[2]}$, when the covered subregion has the length $d_{[1]}+(m-1)d_{[2]}$, i.e. the next corner point is $(\frac{1}{2}d_{[2]},1-d_{[1]}-(m-1)d_{[2]})$. 
If $d_{[1]}=d_{[2]}$, then both corner points coincide, so $\psi_0(t)$ will continue from the joint corner point.
\smallskip

The above pattern generalizes to the $i$-th critical radius $t=\frac{1}{2}d_{[i]}$, when the covered subregion has the length $\sum\limits_{j=1}^{i-1}d_{[j]}$ (for the already covered intervals) plus $(m-i+1)d_{[i]}$ (for the still growing intervals).
For the final critical radius $t=\frac{1}{2}d_{[m]}$, the whole interval $[0,1]$ is covered, because $\sum\limits_{j=1}^{m}d_{[j]}=1$, so the final corner point is $(\frac{1}{2}d_{[m]},0)$.
\smallskip

In Example~\ref{exa:densities1D}  for $S=\{0,\frac{1}{3},\frac{1}{2}\}+\Z$, the ordered distances $d_{[1]}=\frac{1}{6}<d_{[2]}=\frac{1}{3}<d_{[3]}=\frac{1}{2}$ give $\psi_0$ with the corner points $(0,1)$, $(\frac{1}{12},\frac{1}{2})$, $(\frac{1}{6},\frac{1}{6})$, $(\frac{1}{4},0)$ as in Fig.~\ref{fig:densities1D}.
\medskip

\noindent
\textbf{(\ref{thm:densities1D}b)}
The 1st density function $\psi_1(t)$ measures the total length of subregions covered by a single interval $[p_i-t,p_i+t]$.
Hence $\psi_1(t)$ splits into the sum of the functions $\eta_{1i}$, each equal to the length of the subinterval of $[p_i-t,p_i+t]$ not covered by other intervals.
Each $\eta_{1i}$ starts from $\eta_{1i}(0)=0$ and linearly grows up to $\eta_{1i}(\frac{1}{2}d)=d$, where $d=\min\{d_{i-1},d_{i}\}$, when the interval $[p_i-t,p_i+t]$ of the length $2t=d$ touches the growing interval centered at the closest of its neighbors $p_{i\pm 1}$.
\smallskip

If (say) $d_{i-1}<d_i$, then the subinterval covered only by $[p_i-t,p_i+t]$ is shrinking on the left and is growing at the same rate on the right until it touches the growing interval centered at the right neighbor.
During this period, when $t$ is between $\frac{1}{2}d_{i-1}$ and $\frac{1}{2}d_i$, the function $\eta_{1i}(t)=d$ remains constant.
If $d_{i-1}=d_i$, this horizontal piece collapses to one point in the graph of $\eta_{1i}(t)$.
For $t\geq\max\{d_{i-1},d_{i}\}$, the subinterval covered only by $[p_i-t,p_i+t]$ is shrinking on both sides until the intervals centered at $p_{i\pm 1}$ meet at a mid-point between them for $t=\frac{d_{i-1}+d_i}{2}$. 
So the graph of $\eta_{1i}$ has a trapezoid form with the corner points $(0,0)$, $(\frac{d_{i-1}}{2}, d)$, $(\frac{d_{i}}{2},d)$, $(\frac{d_{i-1}+d_{i}}{2},0)$.
\smallskip

In Example~\ref{exa:densities1D} for $S=\{0,\frac{1}{3},\frac{1}{2}\}+\Z$, the distances $d_{1}=\frac{1}{3}$, $d_{2}=\frac{1}{6}$, $d_{3}=\frac{1}{2}=d_0$ give $\eta_{11}=\eta_{R}$ with the corner points $(0,0)$, $(\frac{1}{4},\frac{1}{3})$, $(\frac{1}{6},\frac{1}{3})$, $(\frac{5}{12},0)$ as in Fig.~\ref{fig:densities1D}.
\medskip

\noindent
\textbf{(\ref{thm:densities1D}c)}
For $k>1$, the $k$-th density function $\psi_k(t)$ measures the total length of $k$-fold intersections among $m$ intervals $[p_i-t,p_i+t]$, $i=1,\dots,m$.
Such a $k$-fold intersection appears only when two intervals $[p_i-t,p_i+t]$ and $[p_{i+k-1}-t,p_{i+k-1}+t]$ overlap, because their intersection is covered by $k$ intervals centered at $k$ points $p_i<p_{i+1}<\dots<p_{i+k-1}$.
Since only $k$ successive intervals can contribute to $k$-fold intersections, $\psi_k(t)$ splits into the sum of the functions $\eta_{ki}$, each equal to the length of the subinterval of $[p_i-t,p_{i+k-1}+t]$ covered by exactly $k$ intervals of the form $[p_j-t,p_j+t]$, $j=1,\dots,m$.
The function $\eta_{ki}(t)$ remains 0 until the radius $t=\frac{1}{2}\sum\limits_{j=i}^{i+k-2}d_j$, because $2t$ is the length between the points $p_i<p_{i+k-1}$.
Then $\eta_{ki}(t)$ is linearly growing until the $k$-fold intersection touches one of the intervals centered at the points $p_{i-1},p_{i+k}$, which are left and right neighbors of $p_i,p_{i+k-1}$, respectively.
\medskip

If (say) $d_{i-1}<d_{i+k-1}$, this critical radius is $t=\frac{1}{2}\sum\limits_{j=i-1}^{i+k-2}d_j=\frac{d_{i-1}+s}{2}$.
The function $\eta_{ki}(t)$ measures the length of the $k$-fold intersection $[p_{i+k-1}-t,p_i+t]$, so $\eta_{ki}(t)=(p_i+t)-(p_{i+k-1}-t)=2t-(p_{i+k-1}-p_i)=(d_{i-1}+s)-s=d_{i-1}$.
Then the $k$-fold intersection is shrinking on the left and is growing at the same rate on the right until it touches the growing interval centered at the right neighbor $p_{i+k}$.
During this time, when $t$ is between $\frac{1}{2}\sum\limits_{j=i-1}^{i+k-2}d_j$ and $\frac{1}{2}\sum\limits_{j=i}^{i+k-1}d_j$, the function $\eta_{ki}(t)$ remains equal to $d_{i-1}$.
\medskip

If $d_{i-1}>d_{i+k-1}$, the last argument should include the smaller distance $d_{i+k-1}$ instead of $d_{i-1}$.
Hence we will use below $d=\min\{d_{i-1},d_{i+k-1}\}$ to cover both cases. 
If $d_{i-1}=d_i$, this horizontal piece collapses to one point in the graph of $\eta_{ki}(t)$.
The $k$-fold intersection within $[p_i,p_{i+k-1}]$ disappears when the intervals centered at $p_{i-1},p_{i+k}$ have the radius $t$ equal to the half-distance $\frac{1}{2}\sum\limits_{j=i-1}^{i+k-1}d_j$ between $p_{i-1},p_{i+k}$.
\medskip

Then the trapezoid function $\eta_{ki}(t)$ has the expected four corner points expressed as $(\frac{s}{2},0)$, $(\frac{d_{i-1}+s}{2},d)$, $(\frac{s+d_{i+k-1}}{2},d)$, $(\frac{d_{i-1}+s+d_{i+k-1}}{2},0)$ for $s=\sum\limits_{j=i}^{i+k-2}d_j$ and $d=\min\{d_{i-1},d_{i+k-1}\}$.
These corners are uniquely determined by the triple $(d_{i-1},s,d_{i+k-1})$, where the components $d_{i-1},d_{i+k-1}$ can be swapped. 
\medskip

Fig.~\ref{fig:SQ15density4} shows more specific examples of trapezoid functions $\eta(d_{i-1},s,d_{i+k-1})$.

\begin{figure}[ht]
\includegraphics[width=\textwidth]{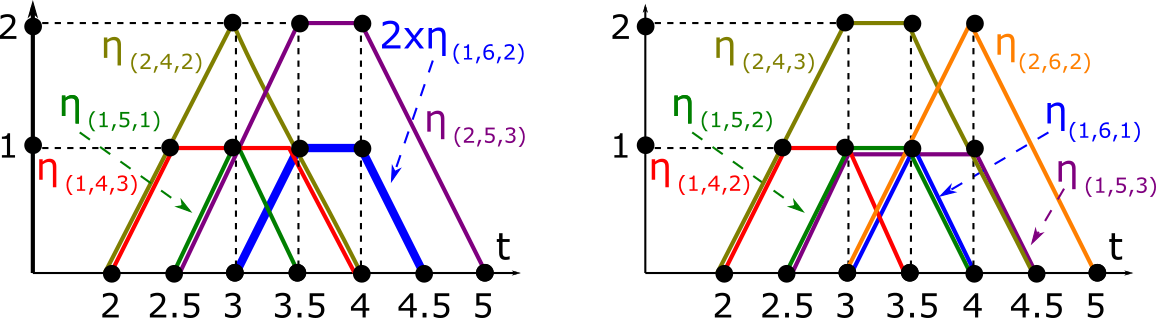}
\caption{The 4th-density function $\psi_4[S_{15}]$ includes the six trapezoid functions on the left, which are replaced by other six trapezoid functions in $\psi_4[Q_{15}]$ on the right, compare the last rows of Tables~\ref{tab:S15d} and~\ref{tab:Q15d}.
However, the sums of these six functions are equal, which can be checked at corner points: both sums of six functions have $\eta(2.5)=2$, $\eta(3)=5$, $\eta(3.5)=6$, $\eta(4)=4$, $\eta(4.5)=1$.
Hence $S_{15},Q_{15}$ in Fig.~\ref{fig:SQ15} have identical density functions $\psi_k$ for all $k\geq 0$, see details in Example~\ref{exa:SQ15densities}.} 
\label{fig:SQ15density4}      
\end{figure}

The above argument is similar to the proof of $(\ref{thm:densities1D}b)$, which can be considered as the partial case of (\ref{thm:densities1D}c) for $k=1$ if we replace all empty sums by 0.
In Example~\ref{exa:densities1D} for $S=\{0,\frac{1}{3},\frac{1}{2}\}+\Z$, we have $d_{1}=\frac{1}{3}$, $d_{2}=\frac{1}{6}$, $d_{3}=\frac{1}{2}=d_0$.
For $k=2$, $i=2$, we get $d_{i-1}=d_1=\frac{1}{3}$, $d_{i+k-1}=d_3=\frac{1}{2}$, i.e. $d=\min\{d_1,d_3\}=\frac{1}{3}$, $s=d_{2}=\frac{1}{6}$.
Then $\eta_{22}=\eta_{GB}$ has the corner points $(\frac{1}{12},0)$, $(\frac{1}{4},\frac{1}{3})$, $(\frac{1}{3},\frac{1}{3})$, $(\frac{1}{2},0)$ as in Fig.~\ref{fig:densities1D}.
\medskip

\noindent
\textbf{(\ref{thm:densities1D}d)}
To prove the symmetry $\psi_{m-k}(\frac{1}{2}-t)=\psi_k(t)$, we establish a bijection between the triples of $\psi_{m-k}$ and $\psi_k$ from (\ref{thm:densities1D}c).
Take a triple $(d_{i-1},s,d_{i+k-1})$ of $\psi_k$, where $s=\sum\limits_{j=i}^{i+k-2}d_j$ is the sum of $k-1$ distances from $d_{i-1}$ to $d_{i+k-1}$ in the increasing (cyclic) order of distance indices.
Under $t\mapsto\frac{1}{2}-t$, the corner points of trapezoid function $\eta_{k,i}$ map to
$(\frac{1-s}{2},0)$, 
$(\frac{1-s-d_{i-1}}{2},d)$, 
$(\frac{1-s-d_{i+k-1}}{2},d)$, 
$(\frac{1-d_{i-1}-s-d_{i+k-1}}{2},0)$.
\medskip

Notice that $\bar s=1-d_{i-1}-s-d_{i+k-1}$ is the sum of the $m-k-1$ intermediate distances from $d_{i+k-1}$ to $d_{i-1}$ in the increasing (cyclic) order of distance indices.
The above corner points can be re-written as
$(\frac{d_{i-1}+\bar s+d_{i+k-1}}{2},0)$, 
$(\frac{\bar s+d_{i+k-1}}{2},d)$, 
$(\frac{\bar s+d_{i-1}}{2},d)$, 
$(\frac{\bar s}{2},0)$.
The resulting points are re-ordered corners of the trapezoid function $\eta_{m-k,i+k}$.
Hence $\eta_{k,i}(\frac{1}{2}-t)=\eta_{m-k,i+k}(t)$.
Taking the sum over all indices $i=1,\dots,m$, we get the symmetry $\psi_{k}(\frac{1}{2}-t)=\eta_{m-k}(t)$.
Fig.~\ref{fig:densities1D} shows the symmetry  
$\psi_1(t)=\psi_2(\frac{1}{2}-t)$.
\medskip

To prove the periodicity, we compare the  functions $\psi_k$ and $\psi_{k-m}$ for $k>m$.
Any $k$-fold intersection should involve intervals centered at $k>m$ successive points of the infinite set $S\subset\R$.
Then we can find a period interval $[t,t+1]$ covering $m$ of these points.
By collapsing this interval to a single point, the $k$-fold intersection becomes $(k-m)$-fold, but its fractional length within any period interval of length 1 remains the same.
Since the radius $t$ is twice smaller than the length of the corresponding interval, the above collapse gives us
$\psi_k(t)=\psi_{k-m}(t-\frac{1}{2})$.
So the graph of $\psi_k$ is obtained from the graph of $\psi_{k-m}$ by the shift to the right (to larger radii) by $\frac{1}{2}$. 
\medskip

\noindent
\textbf{(\ref{thm:densities1D}e)}
To draw the graph of $\psi_k[S]$ or evaluate the $k$-th density function $\psi_k(t)$ at any $t$, we first use the symmetry and periodicity from (\ref{thm:densities1D}d) to reduce $k$ to the range $0,1,\dots,[\frac{m}{2}]$.
In time $O(m\log m)$ we put the points of a motif in the increasing (cyclic) order $p_1,\dots,p_m$ within a period interval scaled to $[0,1]$ for convenience.
In time $O(m)$ we compute the distances $d_i=p_{i+1}-p$ between successive points.
\medskip

For $k=0$, we put the distances in the increasing order $d_{[1]}\leq\dots\leq d_{[m]}$ in time $O(m\log m)$.
By (\ref{thm:densities1D}a), in time $O(m^2)$ we write down the $O(m)$ corner points whose horizontal coordinates are the critical radii where $\psi_0(t)$ can change its linear slope. 
We evaluate $\psi_0$ at every critical radius $t$ by summing up the values of $m$ trapezoid functions, which needs $O(m^2)$ time.
It remains to plot the points at all $O(m)$ critical radii and connect successive points by straight lines, so the total time in $O(m^2)$.
\medskip

For any fixed $k=1,\dots,[\frac{m}{2}]$, in time $O(m^2)$ we write down all $O(m)$ corner points from (\ref{thm:densities1D}c), which leads to the graph of $\psi_k(t)$ similarly to the above argument.
\medskip

To decide if $\Psi[S]=\Psi[Q]$, by (\ref{thm:densities1D}d) it suffices to check whether $O(m)$ density functions coincide: $\psi_k[S]=\psi_k[S]$ for $k=0,1,\dots,[\frac{m}{2}]$.
To check if two piecewise linear functions coincide it suffices to check a potential equality between their values at all $O(m)$ critical radii $t$ from the corner points in (\ref{thm:densities1D}ac) .
Since these values were found in time $O(m^2)$ above, the total time over $k=0,1,\dots,[\frac{m}{2}]$ is $O(m^3)$. 
\end{proof}

\begin{figure}[h!]
\includegraphics[width=\textwidth]{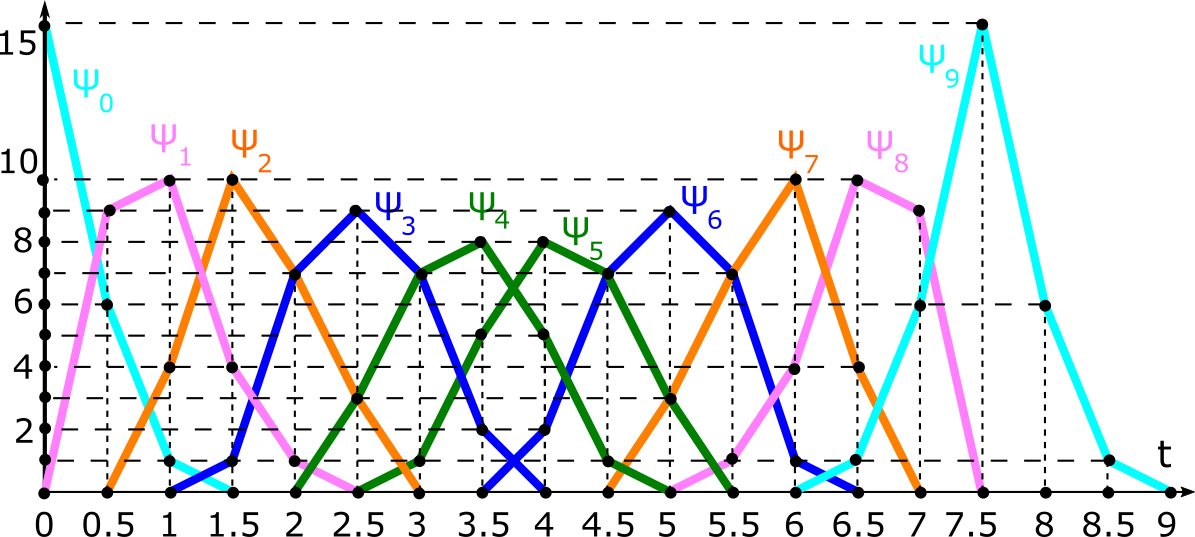}
\caption{$S_{15},Q_{15}$ in Fig.~\ref{fig:SQ15} have identical $\psi_k$, $k\geq 0$.
Both axes are scaled by factor 15.
Theorem~\ref{thm:densities1D}d implies
the symmetry $\psi_k(\frac{15}{2}-t)=\psi_{9-k}(t)$, $t\in[0,\frac{15}{2}]$, and periodicity $\psi_9(t+\frac{15}{2})=\psi_0(t)$, $t\geq 0$.}
\label{fig:SQ15densities}      
\end{figure}

\begin{exa}[identical density functions of $S_{15}$ and $Q_{15}$]
\label{exa:SQ15densities}
The beginning of section 5 in \cite{edels2021} claimed that the periodic 1D sets $S_{15}$ and $Q_{15}$ in Fig.~\ref{fig:SQ15} are undistinguishable by density functions $\psi_k$, which was experimentally checked up to $k=40$.
\medskip

Now Theorem~\ref{thm:densities1D} will help to theoretically prove that $\Psi(S_{15})=\Psi(Q_{15})$.
To avoid fractions, we keep the period 15 of the sets $S_{15},Q_{15}$, because all quantities in Theorem~\ref{thm:densities1D} can be scaled up by factor 15. 
To conclude that $\psi_0[S_{15}]=\psi_0[Q_{15}]$, by 
Theorem~\ref{thm:densities1D}a we check that $S_{15},Q_{15}$ have the same set of the ordered distances between successive points.
Indeed, Tables~\ref{tab:S15d} and~\ref{tab:Q15d} have identical rows 3.
\medskip

To conclude that $\psi_1[S_{15}]=\psi_1[Q_{15}]$, by Theorem~\ref{thm:densities1D}b we check that $S_{15},Q_{15}$ have the same unordered set of unordered pairs $(d_{i-1},d_i)$ of distances between successive points.
Indeed, Tables~\ref{tab:S15d} and~\ref{tab:Q15d} have identical rows 5, where all pairs are \emph{lexicograpically} ordered, i.e. $(a,b)<(c,d)$ if $a<b$ or $a=b$ and $c<d$.
\medskip

To conclude that $\psi_k[S_{15}]=\psi_k[Q_{15}]$ for $k=2,3,4$, 
 we compare the triples $(d_{i-1},\mathbf{s},d_{i+k-1})$ from Theorem~\ref{thm:densities1D}c for $S_{15},Q_{15}$.
Tables~\ref{tab:S15d} and~\ref{tab:Q15d} have identical rows 7 and 9, where the triples are ordered as follows.
If needed, we swap $d_{i-1},d_{i+k-1}$ to make sure that the first entry is not larger than the last.
Then we order by the middle bold number $\mathbf s$.
Finally, we lexicographically order the triples that have the same $s$.
\medskip

Final rows 11 of Tables~\ref{tab:S15d} and~\ref{tab:Q15d} look different for $k=4$.
More exactly, the rows share three triples (1,{\bf 4},2), (1,{\bf 5},2), (1,{\bf 6},4), but the remaining six triples are different.
However, the density function $\psi_4$ is the \emph{sum} of nine trapezoid functions.
\medskip

Fig.~\ref{fig:SQ15density4} shows that these sums are equal for $S_{15},Q_{15}$.
Hence the periodic sets $S_{15},Q_{15}$ have identical density functions $\psi_k$ for $k=0,1,2,3,4$, hence for all $k$ by the symmetry and periodicity from Theorem~\ref{thm:densities1D}d.
Fig.~\ref{fig:SQ15densities} shows $\psi_k$, $k=0,1,\dots,9$.
\bs
\end{exa}

\vspace*{-2mm}

\noindent
\begin{table}[h!]
\caption{\textbf{Row 1}: points $p_i$ from the set $S_{15}$ in Fig.~\ref{fig:SQ15}.
\textbf{Row 2}: the distances $d_i$ between successive points of $S_{15}$. 
\textbf{Row 3}: the distances are put in the increasing order.
\textbf{Row 4}: the unordered set of these pairs determines the density function $\psi_1$ by Theorem~\ref{thm:densities1D}b. 
\textbf{Row 5}: the pairs are lexicographically ordered for comparison with row 5 in Table~\ref{tab:Q15d}. 
Further rows are explained in Example~\ref{exa:SQ15densities}.
}
\label{tab:S15d}
\begin{tabular}{C{25mm}|C{9mm}C{9mm}C{9mm}C{9mm}C{9mm}C{9mm}C{9mm}C{9mm}C{9mm}}
\noalign{\smallskip}\svhline\noalign{\smallskip}
$p_i$                      & 0 & 1 & 3 & 4 & 5 & 7 & 9 & 10 & 12 \\
\noalign{\smallskip}\svhline\noalign{\smallskip}
$d_i=p_{i+1}-p_i$ & 1 & 2 & 1 & 1 & 2 & 2 & 1 &  2 & 3 \\
\noalign{\smallskip}\noalign{\smallskip}
ordered $d_{[i]}$ & 1 & 1 & 1 & 1 & 2 & 2 & 2 &  2 & 3 \\
\noalign{\smallskip}\svhline\noalign{\smallskip}
$(d_{i-1},d_i)$  & (3,1) & (1,2) & (2,1) & (1,1) & (1,2) & (2,2) & (2,1) & (1,2) & (2,3) \\
\noalign{\smallskip}\noalign{\smallskip}
ord. $(d_{i-1},d_i)$  & (1,1) & (1,2) & (1,2) & (1,2) & (1,2) & (1,2) & (1,3) & (2,2) & (2,3) \\
\noalign{\smallskip}\svhline\noalign{\smallskip}
$(d_{i-1},\mathbf{d_i},d_{i+1})$  & (3,{\bf 1},2) & (1,{\bf 2},1) & (2,{\bf 1},1) & (1,{\bf 1},2) & (1,{\bf 2},2) & (2,{\bf 2},1) & (2,{\bf 1},2) & (1,{\bf 2},3) & (2,{\bf 3},1) \\
\noalign{\smallskip}\noalign{\smallskip}
ord. $(d_{i-1},\mathbf{d_i},d_{i+1})$ & (1,{\bf 1},2)  & (1,{\bf 1},2) & (2,{\bf 1},2) & (2,{\bf 1},3)  & (1,{\bf 2},1) & (1,{\bf 2},2) & (1,{\bf 2},2) & (1,{\bf 2},3) & (1,{\bf 3},2) \\
\noalign{\smallskip}\svhline\noalign{\smallskip}
$(d_{i-1},\mathbf{s},d_{i+2})$  & (3,{\bf 3},1) & (1,{\bf 3},1) & (2,{\bf 2},2) & (1,{\bf 3},2) & (1,{\bf 4},1) & (2,{\bf 3},2) & (2,{\bf 3},3) & (1,{\bf 5},1) & (2,{\bf 4},2) \\
\noalign{\smallskip}\noalign{\smallskip}
ord. $(d_{i-1},\mathbf{s},d_{i+2})$ & (2,{\bf 2},2) & (1,{\bf 3},1) & (1,{\bf 3},2) & (1,{\bf 3},3)  & (2,{\bf 3},2)  & (2,{\bf 3},3) & (1,{\bf 4},1) & (2,{\bf 4},2) & (1,{\bf 5},1) \\
\noalign{\smallskip}\svhline\noalign{\smallskip}
$(d_{i-1},\mathbf{s},d_{i+3})$  & (3,{\bf 4},1) & (1,{\bf 4},2) & (2,{\bf 4},2) & (1,{\bf 5},1) & (1,{\bf 5},2) & (2,{\bf 5},3) & (2,{\bf 6},1) & (1,{\bf 6},2) & (2,{\bf 6},1) \\
\noalign{\smallskip}\noalign{\smallskip}
ord. $(d_{i-1},\mathbf{s},d_{i+3})$  & (1,{\bf 4},2) & (1,{\bf 4},3) & (2,{\bf 4},2) & (1,{\bf 5},1) & (1,{\bf 5},2) & (2,{\bf 5},3) & (1,{\bf 6},2) & (1,{\bf 6},2) & (1,{\bf 6},2) 
\end{tabular}
\end{table}

\noindent
\begin{table}[h!]
\caption{\textbf{Row 1}: points $p_i$ from the set $Q_{15}$ in Fig.~\ref{fig:SQ15}.
\textbf{Row 2}: the distances $d_i$ between successive points of $Q_{15}$. 
\textbf{Row 3}: the distances are put in the increasing order.
\textbf{Row 4}: the unordered set of these pairs determines the density function $\psi_1$ by Theorem~\ref{thm:densities1D}b. 
\textbf{Row 5}: the pairs are lexicographically ordered for easier comparison with row 5 in Table~\ref{tab:S15d}. 
Further rows are explained in Example~\ref{exa:SQ15densities}.
}
\label{tab:Q15d}
\begin{tabular}{C{25mm}|C{9mm}C{9mm}C{9mm}C{9mm}C{9mm}C{9mm}C{9mm}C{9mm}C{9mm}}
\noalign{\smallskip}\svhline\noalign{\smallskip}
$p_i$                      & 0 & 1 & 3 & 4 & 6 & 8 & 9 & 12 & 14 \\
\noalign{\smallskip}\svhline\noalign{\smallskip}
$d_i=p_{i+1}-p_i$ & 1 & 2 & 1 & 2 & 2 & 1 & 3 &  2 & 1 \\
\noalign{\smallskip}\noalign{\smallskip}
ordered $d_{[i]}$ & 1 & 1 & 1 & 1 & 2 & 2 & 2 &  2 & 3 \\
\noalign{\smallskip}\svhline\noalign{\smallskip}
$(d_{i-1},d_i)$  & (1,1) & (1,2) & (2,1) & (1,2) & (2,2) & (2,1) & (1,3) & (3,2) & (2,1) \\
\noalign{\smallskip}\noalign{\smallskip}
ordered $(d_{i-1},d_i)$  & (1,1) & (1,2) & (1,2) & (1,2) & (1,2) & (1,2) & (1,3) & (2,2) & (2,3) \\
\noalign{\smallskip}\svhline\noalign{\smallskip}
$(d_{i-1},\mathbf{d_i},d_{i+1})$  & (1,{\bf 1},2) & (1,{\bf 2},1) & (2,{\bf 1},2) & (1,{\bf 2},2) & (2,{\bf 2},1) & (2,{\bf 1},3) & (1,{\bf 3},2) & (3,{\bf 2},1) & (2,{\bf 1},1) \\
\noalign{\smallskip}\noalign{\smallskip}
ord. $(d_{i-1},\mathbf{d_i},d_{i+1})$ & (1,{\bf 1},2)  & (1,{\bf 1},2) & (2,{\bf 1},2) & (2,{\bf 1},3)  & (1,{\bf 2},1) & (1,{\bf 2},2) & (1,{\bf 2},2) & (1,{\bf 2},3) & (1,{\bf 3},2) \\
\noalign{\smallskip}\svhline\noalign{\smallskip}
$(d_{i-1},\mathbf{s},d_{i+2})$  & (1,{\bf 3},1) & (1,{\bf 3},2) & (2,{\bf 3},2) & (1,{\bf 4},1) & (2,{\bf 3},3) & (2,{\bf 4},2) & (1,{\bf 5},1) & (3,{\bf 3},1) & (2,{\bf 2},2) \\
\noalign{\smallskip}\noalign{\smallskip}
ord. $(d_{i-1},\mathbf{s},d_{i+2})$ & (2,{\bf 2},2) & (1,{\bf 3},1) & (1,{\bf 3},2) & (1,{\bf 3},3)  & (2,{\bf 3},2)  & (2,{\bf 3},3) & (1,{\bf 4},1) & (2,{\bf 4},2) & (1,{\bf 5},1) \\
\noalign{\smallskip}\svhline\noalign{\smallskip}
$(d_{i-1},\mathbf{s},d_{i+3})$  & (1,{\bf 4},2) & (1,{\bf 5},2) & (2,{\bf 5},1) & (1,{\bf 5},3) & (2,{\bf 6},2) & (2,{\bf 6},1) & (1,{\bf 6},1) & (3,{\bf 4},2) & (2,{\bf 4},1) \\
\noalign{\smallskip}\noalign{\smallskip}
ord. $(d_{i-1},\mathbf{s},d_{i+3})$  & (1,{\bf 4},2) & (1,{\bf 4},2) & (2,{\bf 4},3) & (1,{\bf 5},2) & (1,{\bf 5},2) & (1,{\bf 5},3)  & (1,{\bf 6},1) & (1,{\bf 6},2) & (2,{\bf 6},2) \\
\end{tabular}
\end{table} 

\section{Isosets are complete isometry invariants of periodic point sets}
\label{sec:isoset_complete}

This section describes the results from \cite{anosova2021isometry} and new Lemma~\ref{lem:upper_bounds} to simplify algorithms in section~\ref{sec:algorithms}.
First we remind auxiliary concepts from Dolbilin's papers \cite{dolbilin2019regular}, \cite{bouniaev2017regular}.
Then we introduce the isotree $\IT(S)$ to visualize
the invariant \emph{isoset} in Definition~\ref{dfn:isoset}.

\begin{dfn}[$m$-regular sets]
\label{dfn:m-regular}
For any point $p$ in a periodic set $S\subset\R^n$, the \emph{global cluster} $C(S,p)$ is the infinite set of vectors $q-p$ for all points $q\in S$.
The set $S\subset\R^n$ is called \emph{1-regular} if all global clusters of $S$ are isometric, so for any points $p,q\in S$, there is an isometry $f:C(S,p)\to C(S,q)$ such that $f(p)=q$. 
A periodic set $S$ is called \emph{$m$-regular} if all global clusters of $S$ form exactly $m>1$ isometry classes.
\bs
\end{dfn}

For any point $p\in S$, its global cluster is a view of $S$ from the position of $p$, so $C(S;p)$ represents all stars in the infinite universe $S$ viewed from our planet Earth located at $p$.
\medskip

Any lattice is 1-regular, because all its global clusters are related by translations.
Though the global clusters $C(S,p)$ and $C(S,q)$ at any different points $p,q\in S$ seem to contain the same set $S$, they can be different even modulo translations.
The global clusters are infinite, hence distinguishing them up to isometry is not easier than original sets.
However, regularity can be checked in terms of local clusters below.

\begin{dfn}[local $\al$-clusters $C(S,p;\al)$ and symmetry groups $\sym(S,p;\al)$]
\label{dfn:local_cluster}
For a point $p$ in a crystal $S\subset\R^n$ and any radius $\al\geq 0$, the local \emph{$\al$-cluster} $C(S,p;\al)$ is the set of all vectors $q-p$ such that $q\in S$ and $|q-p|\leq\al$.
An isometry $f\in\iso(\R^n)$ between local clusters should match their centers.
The \emph{symmetry} group $\sym(S,p;\al)$ consists of all \emph{self-isometries} of the $\al$-cluster $C(S,p;\al)$ that fix the center $a$.
\bs
\end{dfn}

\begin{figure}[h!]
\includegraphics[height=55mm]{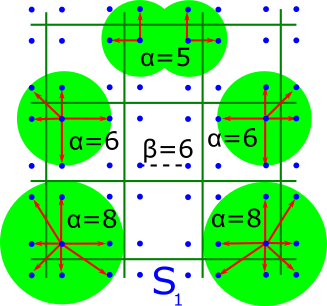}
\hspace*{2mm}
\includegraphics[height=55mm]{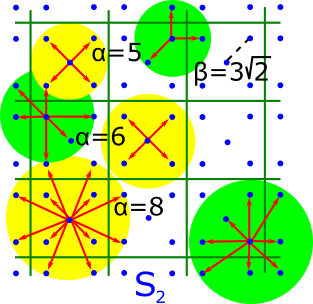}
\caption{
\textbf{Left}: $S_1$ has the four points $(2,2),(2,8),(8,2),(8,8)$ in the square unit cell $[0,10)^2$, so $S_1$ isn't a lattice, but is 1-regular by Definition~\ref{dfn:m-regular}, the bridge length is $\be(S_1)=6$.
All local $\al$-clusters are isometric, shown by red arrows for $\al=5,6,8$, see Definition~\ref{dfn:local_cluster}.
\textbf{Right}: $S_2$ has the extra point $(5,5)$ in the center of $[0,10)^2$ and is 2-regular with $\be(S_2)=3\sqrt{2}$. Local $\al$-clusters have two isometry types. }
\label{fig:alpha-clusters}
\end{figure}

Both periodic sets in Fig.~\ref{fig:alpha-clusters} are not lattices.
The first picture in Fig.~\ref{fig:alpha-clusters} shows the regular set $S_1\subset\R^2$, where all points have isometric global clusters related by translations and rotations through $\frac{\pi}{2},\pi,\frac{3\pi}{2}$, so $S_1$ is not a lattice. 
The periodic set $S_2$ in the second picture has extra central points all unit square cells.
Local $\al$-clusters of these new central points and previous corner points differ for $\al\geq\be=3\sqrt{2}$. 
\medskip

If $\al>0$ is smaller than the minimum distance between any points,
 then any $\al$-cluster $C(S,p;\al)$ is the single-point set $\{p\}$ and its symmetry group $\Or(\R^n)$ consists of all isometries fixing the center $a$.
When the radius $\al$ is increasing, the $\al$-clusters $C(S,p;\al)$ become larger and can have fewer (not more) self-isometries, so the symmetry group $\sym(S,p;\al)$ can become smaller (not larger) and eventually stabilizes.
The regular set $S_1$ in Fig.~\ref{fig:alpha-clusters} for any point $p\in S_1$ has the symmetry group $\sym(S_1,p;\al)=\Or(\R^2)$ for $\al\in[0,4)$. 
The group $\sym(S_1,p;\al)$ stabilizes as $\Z_2$ for $\al\geq 4$ as soon as the local $\al$-cluster $C(S_1,p;\al)$ includes one more point.
\medskip

\begin{dfn}[bridge length $\be(S)$]
\label{dfn:bridge_length}
For a periodic point set $S\subset\R^n$, the \emph{bridge length} is a minimum $\be(S)>0$ such that any points $p,q\in S$ can be connected by a finite sequence $p_0=p,p_1,\dots,p_k=q$ such that 
the Euclidean distance $|p_{i-1} -p_{i}|\leq\be(S)$ for $i=1,\dots,k$.
\bs
\end{dfn}

The past research on Delone sets focused on criteria of $m$-regularity of a single set, see \cite[Theorem~1.3]{dolbilin1998multiregular}.
We extended these ideas to compare different periodic sets. 
The concept of an \emph{isotree} in Definition~\ref{dfn:isotree} is inspired by a dendrogram of hierarchical clustering, though points are partitioned according to isometry classes of local $\al$-clusters at different radii $\al$, not according to a distance threshold.

The past research on Delone sets focused on criteria of $m$-regularity of a single set, e.g. \cite[Theorem~1.3]{dolbilin1998multiregular}.
These ideas are now extended to compare different periodic sets. 
The \emph{isotree} in Definition~\ref{dfn:isotree} is inspired by a clustering dendrogram.
However, points of $S$ are partitioned according to isometry classes of $\al$-clusters at different $\al$, not by a distance threshold.

\begin{figure}[h]
\includegraphics[width=\textwidth]{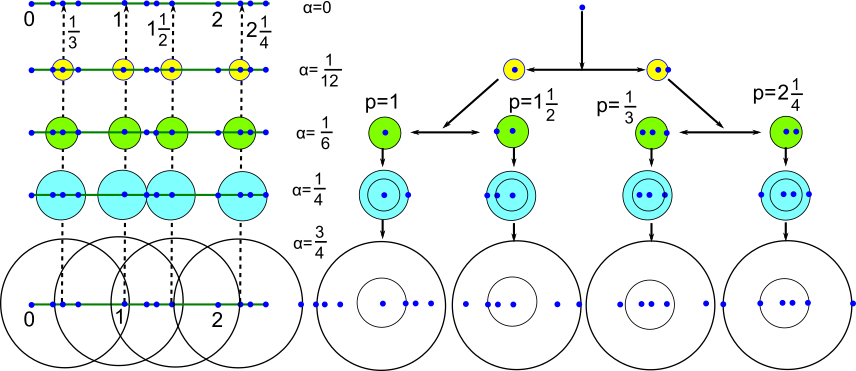}
\caption{\textbf{Left}: 1D periodic set $S_4=\{0,\frac{1}{4},\frac{1}{3},\frac{1}{2}\}+\Z$ is 4-regular by Definition~\ref{dfn:m-regular}.
\textbf{Right}: interval $\al$-clusters with radii $\al=0,\frac{1}{12},\frac{1}{6},\frac{1}{4},\frac{3}{4}$ represent points in the isotree $\IT(S_4)$ from Definiton~\ref{dfn:isotree}.
}
\label{fig:4-regular_set_isotree}
\end{figure}

\begin{dfn}[isotree $\IT(S)$ of $\al$-partitions]
\label{dfn:isotree}
Fix a periodic point set $S\subset\R^n$ and $\al\geq 0$.
Points $p,q\in S$ are \emph{$\al$-equivalent} if 
their $\al$-clusters $C(S,p;\al)$ and $C(S,q;\al)$ are isometric.
The \emph{isometry class} $[C(S,p;\al)]$ consists of all $\al$-clusters isometric to $C(S,p;\al)$.
The \emph{$\al$-partition} $P(S;\al)$ is the splitting of $S$ into $\al$-equivalence classes of points.
The size $|P(S;\al)|$ is the number of $\al$-equivalence classes.
When $\al$ is increasing, the $\al$-partition can be refined by subdividing $\al$-equivalence classes into subclasses.
If we represent each $\al$-equivalence class by a point, the resulting points form the \emph{isotree} $\IT(S)$ of all $\al$-partitions, see Fig.~\ref{fig:4-regular_set_isotree}.
\bs
\end{dfn}

The $\al$-clusters of the periodic set $S_4\subset\R$ in Fig.~\ref{fig:4-regular_set_isotree} are intervals in $\R$, but are shown as disks only for better visibility.
The isotree $\IT(S)$ is continuously parameterized by $\al\geq 0$ and can be considered as a metric tree.
Branched vertices of $\IT(S)$ correspond to the values of $\al$ when an $\al$-equivalence class is subdivided into subclasses for $\al'$ slightly larger than $\al$.
\medskip

The root vertex of $\IT(S)$ at $\al=0$ is the single class $S$, because any $C(S,p;0)$ consists only of its center $p$.
In Fig.~\ref{fig:4-regular_set_isotree} this class persists until $\al=\frac{1}{12}$, when all points $p\in S_4$ are partitioned into two classes: one represented by 1-point clusters $\{p\}$ and another represented by 2-point clusters $\{p,p+\frac{1}{12}\}$.
The set $S_4$ has four $\al$-equivalence classes for any $\al\geq\frac{1}{6}$.
For any point $p\in\Z\subset S_4$, the symmetry group $\sym(S_4,p;\al)=\Z_2$ is generated by the reflection in $p$ for $\al\in[0,\frac{1}{4})$.
For all $p\in S_4$, the symmetry group $\sym(S_4,p;\al)$ is trivial for $\al\geq\frac{1}{4}$. 
\medskip

When a radius $\al$ is increasing, $\al$-clusters $C(S,p;\al)$ include more points, hence are less likely to be isometric, so $|P(S;\al)|$ is a non-increasing function of $\al$. 
Any $\al$-equivalence class from $P(S;\al)$ may split into two or more classes, which cannot merge at any larger $\al'$.
\medskip

Lemma~\ref{lem:isotree} justifies that the isotree $\IT(S)$ can be visualized as a merge tree of $\al$-equivalence classes of points represented by their $\al$-clusters.

\begin{lem}[isotree properties]
\label{lem:isotree} 
The isotree $\IT(S)$  
has the following properties:
\medskip

\noindent
(\ref{lem:isotree}a)
for $\al=0$, the $\al$-partition $P(S;0)$ consists of one class;
\medskip

\noindent
(\ref{lem:isotree}b) 
if $\al<\al'$, then $\sym(S,p;\al')\subseteq\sym(S,p;\al)$ for any point $a\in S$;
\medskip

\noindent
(\ref{lem:isotree}c) 
if $\al<\al'$, the $\al'$-partition $P(S;\al')$ \emph{refines} $P(S;\al)$, so any $\al'$-equivalence class from $P(S;\al')$ is included into an $\al$-equivalence class from $P(S;\al)$.
\bs
\end{lem}
\begin{proof}
(\ref{lem:isotree}a)
If $\al\geq 0$ is smaller than the minimum distance $r$ between point of $S$, every cluster $C(S,p;\al)$ is the single-point set $\{p\}$.
All these single-point clusters are isometric to each other.
So $|P(S;\al)|=1$ for all small radii $\al<r$.
\medskip

\noindent
(\ref{lem:isotree}b)
For any point $p\in S$, the inclusion of clusters $C(S,p;\al)\subseteq C(S,p;\al')$ implies that any self-isometry of the larger cluster $C(S,p;\al')$ can be restricted to a self-isometry of the smaller cluster $C(S,p;\al)$.
So $\sym(S,p;\al')\subseteq\sym(S,p;\al)$.
\medskip

\noindent
(\ref{lem:isotree}c)
If points $p,q\in S$ are $\al'$-equivalent at the larger radius $\al'$, i.e. the clusters $C(S,p;\al')$ and $C(S,q;\al')$ are isometric, then $p,q$ are $\al$-equivalent at the smaller radius $\al$.
Hence any $\al'$-equivalence class in $S$ is a subset of an $\al$-equivalence class.
\end{proof}

If a point set $S$ is periodic, the $\al$-partitions of $S$ stabilize in the sense below.

\begin{dfn}[a stable radius]
\label{dfn:stable_radius}
Let a periodic point set $S\subset\R^n$ have an upper bound $\be$ of its bridge length $\be(S)$.
A radius $\al\geq\be$ is called \emph{stable} if these conditions hold:
\medskip

\noindent
(\ref{dfn:stable_radius}a) 
the $\al$-partition $P(S;\al)$ coincides with the $(\al-\be)$-partition $P(S;\al-\be)$ of $S$;
\medskip

\noindent
(\ref{dfn:stable_radius}b)
the symmetry groups stabilize: 
$\sym(S,p;\al)=\sym(S,p;\al-\be)$ for $p\in S$. 
\bs
\end{dfn}

Though (\ref{dfn:stable_radius}b) is stated for all $p\in S$, one can check only points from a finite motif $M$ of $S$.
A minimum $\al$ satisfying (\ref{dfn:stable_radius}ab) for the bridge length $\be(S)$ from Definition~\ref{dfn:bridge_length} can be called \emph{the minimum stable radius} and denoted by $\al(S)$.
All stable radii of $S$ form the interval $[\al(S),+\infty)$ by \cite[Lemma~13]{anosova2021isometry}.
The 1D set $S_4$ in Fig.~\ref{fig:4-regular_set_isotree} has $\be(S_4)=\frac{1}{2}$ and $\al(S)=\frac{3}{4}$, because the $\al$-partition and symmetry groups $\sym(S_4,p;\al)$ remain unchanged over $\frac{1}{4}\leq\al\leq\frac{3}{4}$.
\medskip

Due to Lemma~(\ref{lem:isotree}bc), conditions (\ref{dfn:stable_radius}ab) imply that the $\al'$-partitions $P(S;\al')$ and the symmetry groups $\sym(S,p;\al')$ remain the same for all $\al'\in[\al-\be,\al]$. 
\medskip

Condition (\ref{dfn:stable_radius}b) doesn't follow from condition (\ref{dfn:stable_radius}a) due to the following example. 
Let $\La$ be the 2D lattice with the basis $(1,0)$ and $(0,\be)$ for $\be>1$.
Then $\be$ is the bridge distance of $\La$. 
Condition (\ref{dfn:stable_radius}a) is satisfied for any $\al\geq 0$, because all points of any lattice are equivalent up to translations. However, condition (\ref{dfn:stable_radius}b) fails for any $\al<\be+1$.
Indeed, the $\al$-cluster of the origin $(0,0)$ contains five points $(0,0),(\pm 1,0),(0,\pm\be)$, whose symmetries are generated by the two reflections in the axes $x,y$, but the $(\al-\be)$-cluster of the origin consists of only $(0,0)$ and has the symmetry group $\Or(2)$. 
Condition (\ref{dfn:stable_radius}b) might imply condition (\ref{dfn:stable_radius}a), but in practice it makes sense to verify (\ref{dfn:stable_radius}b) only after checking much simpler condition (\ref{dfn:stable_radius}a). Both conditions are essentially used in the proofs of key Theorem~\ref{thm:isoset_complete}. 
\medskip

New Lemma~\ref{lem:upper_bounds} proves important upper bounds for the bridge length $\be(S)$ and the minimum stable radius $\al(S)$ to simplify algorithms in section~\ref{sec:algorithms}.

\begin{lem}[upper bounds]
\label{lem:upper_bounds}
Let $S\subset\R^n$ be a periodic point set whose unit cell $U$ has maximum edge-length $b$, diameter $d$.
Then $\be(S)\leq r$, $\al(S)\leq \be(S)+r$ for $r=\max\{b,\frac{d}{2}\}$.
For any lattice $\La\subset\R^n$, consider a basis whose longest vector has a minimum possible length $b$.
The bridge length is $\be(\La)=b$ and $\al(\La)\leq 2b$, which becomes equality for generic $\La$.
\bs
\end{lem}
\begin{proof}
For any point $p\in S$, shift a given unit cell $U$ so that $p$ becomes the origin of $\R^n$ and a corner of $U$.
Then any translations of $p$ along basis vectors of $U$ are within the maximum edge-length $b$ of $U$.
The center of $U$ is at most $\frac{d}{2}$ away from the corner $p$, where $d$ is the diameter (length of a longest diagonal) of $U$.
Then all points in a motif $M\subset U$ are at most $\frac{d}{2}$ away from one of the corners of $U$.
So any points of $S$ can be connected by a finite sequence whose successive points are at most $r=\max\{b,\frac{d}{2}\}$ away from each other.
\medskip

To prove that $\al(S)\leq\be(S)+r$ by Definition~\ref{dfn:stable_radius}, it suffices to show that the $\al$-partition $P(S;\al)$ and $\sym(S,p;\al)$ remain unchanged for any $\al\geq r$.
The stabilization of $\sym(S,p;\al)$ follows if for any $p\in S$ a self-isometry $f$ of $C(S,p;r)$ such that $f(p)=p$ can be extended to a global isometry $S\to S$ fixing $p$.
Consider a unit cell $U$ of $S$ with a corner at $p$ and straight-line edges that are directed from $p$ and form a basis $v_1,\dots,v_n$ of a lattice of $S$.  
Since $r=\max\{b,\frac{d}{2}\}$, the closed ball $\bar B(p;r)$ covers 
$p$ with all endpoints $p+v_i$ of the basis vectors, $i=1,\dots,n$, and also $\bar B(p;r)$ covers the shifted cell $U'=U-\frac{1}{2}\sum\limits_{i=1}^n c_i v_i$, which is centered at $p$ and contains a full motif $M$ of $S$.
Then we know the vectors $f(v_i)=f(p+v_i)-f(p)$ along edges of the parallelepiped $f(U')$ centered at $p$.
For any point $q\in M$, the formula $f(q+\sum\limits_{i=1}^n c_i v_i)=f(q)+\sum\limits_{i=1}^n c_i f(v_i)$ linearly extends $f$ to the global isometry $S\to S$ fixing $p$.  
\medskip

Indeed, choose an orthonormal basis of $\R^n$ and put the origin at $p$.
In this basis, the orthogonal map $f$ above can be represented by an orthogonal matrix $A$ satisfying $AA^T=I=A^T A$, where $I$ is the identity $n\times n$ matrix.
Since the above extension is linear, the extended map $f$ on the whole $\R^n$ is represented by the same orthogonal matrix $A$, so $f$ preserves distances.
Any point $q'\in S$ equals $f(q)$ for a unique point $q\in S$, because the inverse map $f^{-1}$ exists and has the matrix $A^{-1}=A^T$ in the orthonormal basis above.
\medskip

The stabilization of $P(S;\al)$ similarly follows if for any $p,q\in S$ an isometry $f:C(S,p;r)\to C(S,q;r)$ can be extended to a global isometry $S\to S$ mapping $p$ to $q$.
The above argument works for the similarly extended affine map $f$ composed of the translation by the vector $q-p$ and a linear map that has an orthogonal matrix $A$ in a suitable orthonormal basis at $q$.
\medskip

In the case of a lattice $\La$, let $\La$ contain the origin $0\in\R^n$ and have a basis $v_1,\dots,v_n$ whose longest vector $v_n$ has a minimum possible length $b$. 
The cluster $C(\La,0;b)$ contains $2n$ neighbors $\{\pm v_i,i=1,\dots,n\}$, which are at most $b$ away from $0$, so $\be(\La)\leq b$.
The bridge length $\be(\La)$ cannot be smaller than $b=|v_n|$, otherwise $v_n$ can be replaced by a shorter vector, so $\be(\La)=b$.
Any self-isometry of $C(\La,0;b)$ maps the basis $v_1,\dots,v_n$ to another basis at $0$ and linearly extends to a global self-isometry of $\La$ fixing the origin.
Hence the symmetry group $\sym(\La,0;\al)$ remains the same for all radii $\al\geq b$.
Then $\al(\La)\leq b+\be(\La)=2b$.
\medskip

For any smaller radius $\al<b$, the cluster $C(\La,0;\al)$ misses the longest vectors $\pm v_n$ and allows a self-symmetry fixed on the subspace spanned by all shorter basis vectors, hence $\sym(\La,0;\al)$  for $\al<b$ is larger than $\sym(\La,0;b)=\Z_2$ generated only by the central symmetry with respect to $0$ in general position.
Definition~\ref{dfn:stable_radius} implies that $\al(\La)=b+\be(\La)=2b$ for all lattices $\La$ whose longest basis vector is not orthogonal to all other basis vectors.
\end{proof}

Bouniaev and Dolbilin \cite{bouniaev2017regular} wrote conditions (\ref{dfn:stable_radius}ab) for $\rho,\rho+t$.
Since crystallographers use $\rho$ for the density and have many types of bond distances $t$, we replaced $t$ by the bridge distance $\be$ and replaced $\rho+t$ by $\al$, which is commonly used for similarly growing $\al$-shapes in Topological Data Analysis.
Any regular periodic set in $\R^3$ with a bridge distance $\be$ has a stable radius $\al=7\be$ or $\rho=6t$ in the past notations \cite{dolbilin2019regular}.
\medskip

Definition~\ref{dfn:isoset} introduces the invariant \emph{isoset}, whose completeness (or injectivity) in Isometry Classification Problem~\ref{pro:isometry_classification} is proved in Theorem~\ref{thm:isoset_complete}.
  
\begin{dfn}[isoset $I(S;\al)$ at a radius $\al$]
\label{dfn:isoset}
Let a periodic point set $S\subset\R^n$ have a motif $M$ of $m$ points.
Split all points $p\in M$ into $\al$-equivalence classes.
Each $\al$-equivalence class consisting of (say) $k$ points in $M$ can be associated with the \emph{isometry class} $\si=[C(S,p;\al)]$ of an $\al$-cluster centered at one of these $k$ points $p\in M$.
The \emph{weight} of $\si$ is $w=k/m$.
The \emph{isoset} $I(S;\al)$ is the unordered set of all isometry classes with weights $(\si;w)$ for $p\in M$.
\bs
\end{dfn}

All points $p$ of a lattice $\La\subset\R^n$ are $\al$-equivalent to each other for any radius $\al\geq 0$, because all $\al$-clusters $C(\La,p;\al)$ are isometrically equivalent to each other by translations. 
Hence the isoset $I(\La;\al)$ is one isometry class of weight 1 for $\al\geq 0$.
\medskip

All isometry classes $\si$ in $I(S;\al)$ are in a 1-1 correspondence with all $\al$-equivalence classes in the $\al$-partition $P(S;\al)$ from Definition~\ref{dfn:isotree}.
So $I(S;\al)$ without weights can be viewed as a set of points in the isotree $\IT(S)$ at the radius $\al$.
The size of the isoset $I(S;\al)$ equals the number $|P(S;\al)|$ of $\al$-equivalence classes in the $\al$-partition. 
Formally, $I(S;\al)$ depends on $\al$, because $\al$-clusters grow in $\al$.
To distinguish any periodic point sets $S,Q\subset\R^n$ up to isometry, we will compare their isosets at a common (maximum) stable radius $\al$ of $S,Q$.
\medskip

An equality $\si=\xi$ between isometry classes of clusters means that there is an isometry $f$ from a cluster in $\si$ to a cluster in $\xi$ such that $f$ respects the centers of the clusters.

\begin{thm}[isometry classification]
\label{thm:isoset_complete}
For any periodic point sets $S,Q\subset\R^n$, let $\al$ be a common stable radius satisfying Definition~\ref{dfn:stable_radius} for an upper bound $\be$ of $\be(S),\be(Q)$.
Then $S,Q$ are isometric if and only if there is a bijection $I(S;\al)\to I(Q;\al)$ respecting all weights.
\bs
\end{thm}

Theoretically a complete invariant of periodic points sets should include isosets $I(S;\al)$ for all sufficiently large radii $\al$.
However, when comparing two sets $S,Q$ up to isometry, it sufficies to build their isosets only at a common stable radius $\al$.
\medskip

The $\al$-equivalence and isoset in Definition~\ref{dfn:isoset} can be refined by labels of points such as chemical elements, which keeps Theorem~\ref{thm:isoset_complete} valid for labeled points.
Recall that isometries include reflections, however an orientation sign can be easily added to $\al$-clusters, hence we focus on the basic case of all isometries.

Lemmas~\ref{lem:local_extension} and~\ref{lem:global_extension} help to extend an isometry between local clusters to full periodic sets to prove the complete isometry classification in Theorem~\ref{thm:isoset_complete}.

\begin{lem}[local extension]
\label{lem:local_extension}
Let periodic point sets $S,Q\subset\R^n$ have bridge distances at most $\be$ and a common stable radius $\al$ such that $\al$-clusters $C(S,p;\al)$ and $C(Q,q;\al)$ are isometric for some $p\in S$, $q\in Q$.
Then any isometry $f:C(S,p;\al-\be)\to C(Q,q;\al-\be)$ extends to an isometry $C(S,p;\al)\to C(Q,q;\al)$. 
\bs
\end{lem}
\begin{proof}
Let $g:C(S,p;\al)\to C(Q,q;\al)$ be any isometry, which may not coincide with $f$ on the $(\al-\be)$-subcluster $C(S,p;\al-\be)$.
The composition $f^{-1}\circ g$ isometrically maps $C(S,p;\al-\be)$ to itself.
Hence $f^{-1}\circ g=h\in\sym(S,p;\al-\be)$ is a self-isometry.
Since the symmetry groups stabilize by condition~(\ref{dfn:stable_radius}b), the isometry $h$ maps the larger cluster $C(S,p;\al)$ to itself.
Then the initial isometry $f$ extends to the isometry $g\circ h^{-1}: C(S,p;\al)\to C(Q,q;\al)$.
\end{proof}

\begin{lem}[global extension]
\label{lem:global_extension}
For any periodic point sets $S,Q\subset\R^n$, let $\al$ be a common stable radius satisfying Definition~\ref{dfn:stable_radius} for an upper bound $\be$ of both $\be(S),\be(Q)$.
Assume that $I(S;\al)=I(Q;\al)$.
Fix a point $p\in S$.
Then any local isometry $f:C(S,p;\al)\to C(Q,f(p);\al)$  extends to a global isometry $S\to Q$.
\bs
\end{lem}
\begin{proof}
We shall prove that the image $f(q)$ of any point $q\in S$ belongs to $Q$, hence $f(S)\subset Q$.
Swapping the roles of $S$ and $Q$ will prove that $f^{-1}(Q)\subset S$, i.e. $f$ is a global isometry $S\to Q$. 
By Definition~\ref{dfn:bridge_length} the above points $p,p'\in S$ are connected by a sequence of points $q=p_0,p_1,\dots,p_m=p'\in S$ such that all distances $|p_{i-1}- p_{i}|\leq\be$ are bounded by any upper bound $\be$ of both $\be(S),\be(Q)$ for $i=1,\dots,m$.
\medskip

The cluster $C(S,p;\al)$ is the intersection $S\cap\bar B(p;\al)$.
The closed ball $\bar B(a;\al)$ contains the smaller ball $\bar B(p_1;\al-\be)$ around the closely located center $p_1$.
Indeed, since $|p-p_1|\leq\be$, the triangle inequality for the Euclidean distance implies that any point $a\in\bar B(p_1;\al)$ with $|p_1-a|\leq\al-\be$ satisfies $|p-a|\leq |p -p_1|+|p_1 -a|\leq\al$.
\medskip
 
Due to $I(S;\al)=I(Q;\al)$ the isometry class of $C(S,p_1;\al)$ coincides with an isometry class of $C(Q,q;\al)$ for some $q\in Q$, i.e. $C(S,p_1;\al)$ is isometric to $C(Q,q;\al)$.
Then the smaller clusters $C(S,p_1;\al-\be)$ and $C(Q,q;\al-\be)$ are isometric.
\medskip

By condition~(\ref{dfn:stable_radius}a), the splitting of $Q$ into $\al$-equivalence classes coincides with the splitting into $(\al-\be)$-equivalence classes. 
Take the $(\al-\be)$-equivalence class represented by the cluster $C(Q,q;\al-\be)$ centered at $q$.
This cluster includes the point $f(p_1)\in Q$, because $f$ restricts to the isometry $f:C(S,p_1;\al-\be)\to C(Q,f(p_1);\al-\be)$ and $C(S,p_1;\al-\be)$ was shown to be isometric to $C(Q,q;\al-\be)$.
The $\al$-equivalence class represented by $C(Q,q;\al)$ includes both points $q$ and $f(p_1)$.
The isometry class $[C(Q,q;\al)]=[C(S,p_1;\al)]$ can be represented by the cluster $C(Q,f(p_1);\al)$, which is now proved to be isometric to $C(S,p_1;\al)$.
\medskip

We apply Lemma~\ref{lem:local_extension} for $f$ restricted to $C(S,p_1;\al-\be)\to C(Q,f(p_1),\al-\be)$ and conclude that $f$ extends to an isometry $C(S,p_1;\al)\to C(Q,f(p_1);\al)$.
\medskip

Continue applying Lemma~\ref{lem:local_extension} to the clusters around the next center $p_2$ and so on until we conclude that the initial isometry $f$ maps the $\al$-cluster centered at $p_m=p'\in S$ to an isometric cluster within $Q$, so $f(p')\in Q$ as required.
\end{proof}

\begin{lem}[all stable radii of a periodic point set]
\label{lem:stable_radius}
If $\al$ is a stable radius of a periodic point set $S\subset\R^n$, then so is any  larger radius $\al'>\al$.
Then all stable radii form the interval $[\al(S),+\infty)$, where $\al(S)$ is the minimum stable radius of $S$.
\bs
\end{lem}
\begin{proof} 
Due to Lemma~(\ref{lem:isotree}bc), conditions (\ref{dfn:stable_radius}ab) imply that the $\al'$-partition $P(S;\al')$ and the symmetry groups $\sym(S,p;\al')$ remain the same for all $\al'\in[\al-\be,\al]$. 
We need to show that they remain the same for any larger $\al'>\al$.
\medskip

Below we will apply Lemma~\ref{lem:global_extension} for the same set $S=Q$ and $\be=\be(S)$.
Let points $p,q\in S$ be $\al$-equivalent, i.e. there is an isometry $f:C(S,p;\al)\to C(S,q;\al)$.
Then $f$ extends to a global self-isometry $S\to S$ such that $f(p)=q$.
Then all larger $\al'$-clusters of $p,q$ are isometric, so $p,q$ are $\al'$-equivalent  and $P(S;\al)=P(S,\al')$.
\medskip

Similarly, any self-isometry of $C(S,p;\al)$ extends to a global self-isometry, so the symmetry group $\sym(S,p;\al')$ for any $\al'>\al$ is isomorphic to $\sym(S,p;\al')$. 
\end{proof}

Condition (\ref{dfn:stable_radius}b) doesn't follow from condition (\ref{dfn:stable_radius}a) due to the following example. 
Let $\La$ be the 2D lattice with the basis $(1,0)$ and $(0,\be)$ for $\be>1$.
Then $\be$ is the bridge length of $\La$. 
Condition (\ref{dfn:stable_radius}a) is satisfied for any $\al\geq 0$, because all points of any lattice are equivalent up to translations. However, condition (\ref{dfn:stable_radius}b) fails for any $\al<\be+1$.
Indeed, the $\al$-cluster of the origin $(0,0)$ contains five points $(0,0),(\pm 1,0),(0,\pm\be)$, whose symmetries are generated by the two reflections in the axes $x,y$, but the $(\al-\be)$-cluster of the origin consists of only $(0,0)$ and has the symmetry group $\Or(\R^2)$. 
Condition (\ref{dfn:stable_radius}b) might imply condition (\ref{dfn:stable_radius}a), but in practice it makes sense to verify (\ref{dfn:stable_radius}b) only after checking much simpler condition (\ref{dfn:stable_radius}a). Both conditions are essentially used in the proof of Isometry Classification Theorem~\ref{thm:isoset_complete}. 

\begin{proof}[Theorem~\ref{thm:isoset_complete}]
The part \emph{only if} $\Rightarrow$ follows by restricting any given global isometry $f:S\to Q$ between the infinite sets of points to the local $\al$-clusters $C(S,p;\al)\to C(Q,f(p);\al)$ for any point $p$ in a motif $M$ of $S$.
Hence the isometry class $[C(S,p;\al)]$ is considered equivalent to the class $[C(Q,f(p);\al)]$, which can be represented by the $\al$-cluster $C(Q,q;\al)$ centered at a point $q$ in a motif of $Q$.
Since $f$ is a bijection and the point $p\in M$ was arbitrary, we get a bijection between isometry classes with weights in $I(S;\al)=I(Q;\al)$.
\medskip

The part \emph{if} $\Leftarrow$.
Fix a point $p\in S$. 
The $\al$-cluster $C(S,p;\al)$ represents a class with a weight $(\si,w)\in I(S;\al)$.
Due to $I(S;\al)=I(Q;\al)$, there is an isometry $f:C(S,p;\al)\to C(Q,f(p);\al)$ to a cluster from an equal class $(\si,w)\in I(Q;\al)$. 
By Lemma~\ref{lem:global_extension} the local isometry $f$ extends to a global isometry $S\to Q$. 
\end{proof}


\section{Lipschitz continuity of isosets under point perturbations}
\label{sec:isoset_continuous}

The key new result is the continuity of the isoset $I(S;\al)$ in Theorem~\ref{thm:continuity} for the Earth Mover's Distance (EMD) from Definition~\ref{dfn:EMD}, which needs Definition~~\ref{dfn:cluster_distance}.
\medskip

For a center $a\in\R^n$ and a radius $\ep$, the {\em open} ball is $B(a;\ep)=\{b\in\R^n \vl | b- a|<\ep\}$, the {\em closed} ball is $\bar B(a;\ep)=\{b\in\R^n \vl | b- a|\leq\ep\}$.
For any $C\subset\R^n$, the Minkowski sum $C+\bar B(0;\ep)=\{ a+ b \vl a\in C, b\in \bar B(0;\ep)\}$ is the \emph{$\ep$-offset} of $C$.

\begin{dfn}[distances $d_H$, $d_R$, $d_C$]
\label{dfn:cluster_distance}
For any finite sets $C,D\subset\R^n$, the directed \emph{Hausdorff} distance $d_H(C,D)=\max\limits_{p\in C}\min\limits_{q\in D}|p-q|$ is the minimum $\ep\geq 0$ such that $C\subseteq D+\bar B(0;\ep)$.
The \emph{rotationally-invariant} distance $d_R(C,D)=\min\limits_{f\in\Or(\R^n)}d_H(f(C),D)$ is minimized over all orthogonal maps $f\in\Or(\R^n)$. 
For any periodic point sets $S,Q\subset\R^n$, let isometry classes $\si\in I(S;\al)$ and $\xi\in I(Q;\al)$ be represented by clusters $C(S,p;\al)$ and $C(Q,q;\al)$, respectively.
The \emph{boundary-tolerant cluster distance} $d_C(\si,\xi)$ is the minimum $\ep\in[0,\al]$ such that 
\medskip

\noindent
(\ref{dfn:cluster_distance}a) 
$d_R(\; C(S,p;\al-\ep)-p,\; C(Q,q;\al)-q \;)\leq\ep$;
\medskip

\noindent
(\ref{dfn:cluster_distance}b) 
$d_R(\; C(Q,q;\al-\ep)-q,\; C(S,p;\al)-p \;)\leq\ep$.
\bs
\end{dfn}

\begin{figure}[h!]
\includegraphics[width=\linewidth]{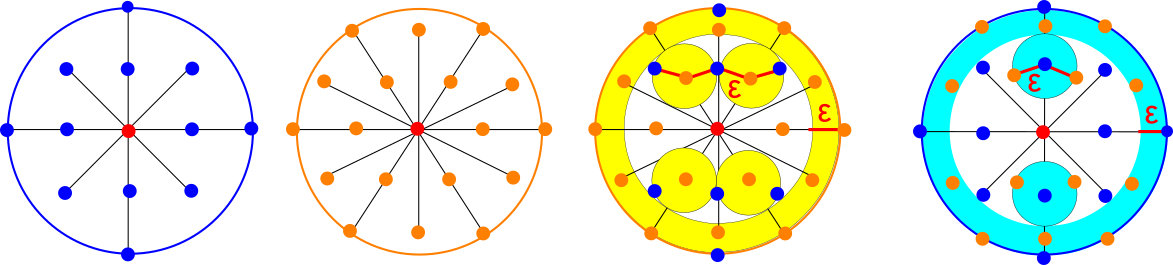}
\caption{
\textbf{1st and 2nd}: clusters $C(\La,0;2)$ of the square and hexagonal lattices with the minimum interpoint distance 1 at the stable radius $\al=2$.
\textbf{3rd}: the minimal $\ep$-offset of the boundary $\bd\bar B(0;2)$ and hexagonal cluster with $\ep=\sqrt{(1-\frac{\sqrt{3}}{2})^2+(\frac{1}{2})^2}=\sqrt{2-\sqrt{3}}\approx 0.52$ covers the square cluster.
\textbf{4th}: the minimal $\ep$-offset of the boundary $\bd\bar B(0;2)$ and square cluster covers the hexagonal cluster.
 }
\label{fig:square_vs_hexagon}
\end{figure}

The notation $C(S,p;\al-\ep)-p$ means that the cluster $C(S,p;\al-\ep)$ is translated by the vector $-p$ so that its central point $p$ becomes the origin $0$ of $\R^n$.
Conditions (\ref{dfn:cluster_distance}ab) can be re-stated for the full clusters at the radius $\al$ if we also use the $\ep$-offset of the boundary $\bd B(0;\al)$ covering all points that are $\ep$-close to this boundary as shown in Fig.~\ref{fig:square_vs_hexagon}.
\medskip

Since an isometry class consists of all local clusters that are isometric to each other, the distance $d_C(\si,\xi)$ in Definition~\ref{dfn:cluster_distance} is independent of representative clusters $C(S,p;\al)$, $C(Q,q;\al)$. 
Isometries between local clusters respect their centers, hence form a subgroup of the compact group $\Or(\R^n)$, so the minimum value of $\ep$ in Definition~\ref{dfn:cluster_distance} is always attained. 

The isoset $I(\La;\al)$ of any lattice $\La\subset\R^n$ containing the origin $0$ consists of a single isometry class $[C(\La,0;\al)]$.
For the square (hexagonal) lattice with minimum distance 1 between points in Fig.~\ref{fig:square_vs_hexagon}, the cluster $C(\La,0;\al)$ consists of only 0 for $\al<1$ and then includes four (six) nearest neighbors of 0 for $\al\geq 1$.
Hence $\sym(\La,0;\al)$ stabilizes as the symmetry group of the square (regular hexagon) for $\al\geq 1$.
Both lattices have the minimum stable radius $\al(\La)=2$ and $\be(\La)=1$ by Lemma~\ref{lem:upper_bounds}.
Fig.~\ref{fig:square_vs_hexagon} shows how to compute  $d_C=\sqrt{2-\sqrt{3}}$ as the distance between points $(0,1)$ and $(\frac{1}{2},\frac{\sqrt{3}}{2})$. 
In the last picture, $(\pm 1,\pm\sqrt{3})$ and $(\pm\frac{3}{2},\pm\frac{\sqrt{3}}{2})$ are covered by the $\ep$-offsets of the boundary and don't need to be covered by blue $\ep$-disks by Definition~\ref{dfn:cluster_distance}.
No rotation can make $\ep$ smaller while keeping the mutual coverings above.

\begin{lem}
\label{lem:cluster_distance_metric}
The cluster distance $d_C$ from Definition~\ref{dfn:cluster_distance} satisfies the metric axioms: 
\medskip

\noindent
(\ref{lem:cluster_distance_metric}a)
$d_C(\si,\xi)=0$ if and only if $\si=\xi$ as isometry classes of $\al$-clusters;
\medskip

\noindent
(\ref{lem:cluster_distance_metric}b)
symmetry : $d_C(\si,\xi)=d_C(\xi,\si)$ for any isometry classes of $\al$-clusters;
\medskip

\noindent
(\ref{lem:cluster_distance_metric}c)
triangle inequality : $d_C(\si,\zeta)\leq d_C(\si,\xi)+d_C(\xi,\zeta)$ for any classes $\si,\xi,\zeta$.
\bs
\end{lem}
\begin{proof}
(\ref{lem:cluster_distance_metric}a)
If $d_C(\si,\xi)=0$, then $\ep=0$ in Definition~\ref{dfn:cluster_distance}.
In this case the shifted cluster representatives $C(S,p;\al)-p$ and $C(Q,q;\al)-q$ of these classes are related by an orthogonal map $f\in\Or(\R^n)$.
Then there is an isometry $C(S,p;\al)\to C(Q,q;\al)$, so $\si=\xi$.
\medskip

\noindent
(\ref{lem:cluster_distance_metric}b)
Definition~\ref{dfn:cluster_distance} of $d_C$ is symmetric with respect to the isometry classes of $\al$-clusters.
\medskip

\noindent
(\ref{lem:cluster_distance_metric}c)
An equivalent form of Definition~\ref{dfn:cluster_distance} says that $d_C(\si,\xi)$ is a minimum value of $\ep$ across all isometries $f,\ti f\in\iso(\R^n)$ such that $f(p)=q$, $\ti f(q)=p$, and
\medskip

\noindent
$f(C(S,p;\al-\ep))\subset C(Q,q;\al)+\bar B(0,\ep)$ and
$\ti f(C(Q,q;\al-\ep))\subset C(S,p;\al)+\bar B(0,\ep)$. 
\medskip

Let cluster representatives $C(S,p;\al)$, $C(Q,q;\al)$, $C(T,u;\al)$ of classes $\si,\xi,\zeta$ and corresponding pairs of isometries $(f,\ti f)$ and $(g,\ti g)$ minimize the cluster distances $d_C(\si,\xi)$ and $d_C(\xi,\zeta)$, respectively, from Definition~\ref{dfn:cluster_distance}. 
The inclusions
$$f(C(S,p;\al-d_C(\si,\xi)))\subset C(Q,q;\al)+\bar B(0,d_C(\si,\xi))\text{ and}$$
$$g(C(Q,q;\al-d_C(\xi,\zeta)))\subset C(T,u;\al)+\bar B(0,d_C(\xi,\zeta))\text{ imply that}$$ 
$$g\circ f(C(S,p;\al-d_C(\si,\xi)-d_C(\xi,\zeta)))\subset g(C(Q,q;\al-d_C(\xi,\zeta))+\bar B(0,d_C(\si,\xi))$$
$$\subset C(T,u;\al)+\bar B(0,d_C(\xi,\zeta))+\bar B(0,d_C(\si,\xi))
=C(T,u;\al)+\bar B(0,d_C(\si,\xi)+d_C(\xi,\zeta)).$$
Since similar inclusions hold for the dual versions $\ti f,\ti g$ of isometries as in (\ref{dfn:cluster_distance}), 
the distance $d_C(\si,\zeta)$ cannot be smaller than $d_C(\si,\xi)+d_C(\xi,\zeta)$.
\end{proof}

Non-isometric periodic sets $S,Q$, for example the perturbations of the square lattice in Fig.~\ref{fig:lattice_perturbations}, can have isosets consisting of different numbers of isometry classes of clusters.
A similarity between such distributions of different sizes can be measured by the distance below.

\begin{dfn}[Earth Mover's Distance on isosets]
\label{dfn:EMD}
For any periodic point sets $S,Q\subset\R^n$ with a common stable radius $\al$, let their isosets be $I(S;\al)=\{(\si_i,w_i)\}$ and $I(Q;\al)=\{(\xi_j,v_j)\}$, where $i=1,\dots,m(S)$ and $j=1,\dots,m(Q)$.
The \emph{Earth Mover's Distance} \cite{rubner2000earth} 
$\EMD( I(S;\al), I(Q;\al) )=\sum\limits_{i=1}^{m(S)} \sum\limits_{j=1}^{m(Q)} f_{ij} d_C(\si_i,\xi_j)$ is minimized over 
$f_{ij}\in[0,1]$ subject to 
$\sum\limits_{j=1}^{m(Q)} f_{ij}\leq w_i$ for $i=1,\dots,m(S)$,
$\sum\limits_{i=1}^{m(S)} f_{ij}\leq v_j$ for $j=1,\dots,m(Q)$ and 
$\sum\limits_{i=1}^{m(S)}\sum\limits_{j=1}^{m(Q)} f_{ij}=1$.
\bs
\end{dfn}

Since EMD satisfies all metric axioms \cite[Appendix]{rubner2000earth},
Definition~\ref{dfn:EMD} introduces the first metric for periodic point sets $d(S,Q)=\EMD( I(S;\al), I(Q;\al) )$, which can be 0 for a common stable radius $\al$ of $S,Q$ only if $I(S;\al)=I(Q;\al)$, so
$S,Q$ are isometric by Theorem~\ref{thm:isoset_complete}.
\medskip

Lemma~\ref{lem:common_lattice} is needed to prove Theorem~\ref{thm:continuity}.
Slightly different versions of Lemma~\ref{lem:common_lattice} are \cite[Lemma 4.1]{edels2021}, \cite[Lemma~7]{widdowson2020average}.
The proof below is more detailed for any dimension $n\geq 1$.

\begin{lem}[common lattice]
\label{lem:common_lattice}
Let periodic point sets $S,Q\subset\R^n$ have bottleneck distance $d_B(S,Q)<r(Q)$, where $r(Q)$ is the packing radius.
Then $S,Q$ have a common lattice $\La$ with a unit cell $U$ such that $S=\La+(U\cap S)$ and $Q=\La+(U\cap Q)$.
\bs
\end{lem}
\begin{proof}
Let $S=\La(S)+(U(S)\cap S)$ and $Q=\La(Q)+(U(Q)\cap Q)$, where 
$U(S),U(Q)$ are initial unit cells of $S,Q$ and the lattices $\La(S),\La(Q)$ of $S,Q$ contain the origin.
\medskip

By shifting all points of $S,Q$ (but not their lattices), we guarantee that $S$ contains the origin $0$ of $\R^n$.
Assume by contradiction that the given periodic point sets $S,Q$ have no common lattice.
Then there is a vector $p\in\La(S)$ whose all integer multiples $kp\not\in\La(Q)$ for $k\in\Z-0$.
Any such multiple $kp$ can be translated by a vector $v(k)\in\La(Q)$ to the initial unit cell $U(Q)$ so that $q(k)=kp-v(k)\in U(Q)$.
\medskip

Since $U(Q)$ contains infinitely many points $q(k)$,
one can find a pair $q(i),q(j)$ at a distance less than $\de=r(Q)-d_B(S,Q)>0$.
The formula $q(k)\equiv kp\pmod{\La(Q)}$ implies that 
 $q(i+k(j-i)) \equiv (i+k(j-i))p\pmod{\La(Q)} \equiv
 q(i) + k(q(j)-q(i))\pmod{\La(Q)}$.
If the point $q(i) + k(q(j)-q(i))$ belongs to $U(Q)$, we get the equality $q(i+k(j-i))=q(i) + k(q(j)-q(i))$.
All these points over $k\in\Z$ lie on a straight line within $U(Q)$ and have the distance $|q(j)-q(i)|<\de$ between successive points.
\medskip

The closed balls with radius $d_B(S,Q)$ and centers at points in $Q$ are at least $2\de$ away from each other.
Then one of the points $q(i+k(j-i))$ is more than $d_B(S,Q)$ away from $Q$. 
Hence the point $(i+k(j-i))p\in S$ also has a distance more than $d_B(S,Q)$ from any point of $Q$, which contradicts the definition
of the bottleneck distance.  
\end{proof}

\begin{thm}[continuity of isosets under perturbations]
\label{thm:continuity}
Let periodic point sets $S,Q\subset\R^n$ have bottleneck distance $d_B(S,Q)<r(Q)$, where the packing radius $r(Q)$ is the minimum half-distance between any points of $Q$. 
Then the isosets $I(S;\al)$ and $I(Q;\al)$ are close in the Earth Mover's Distance: 
$\EMD( I(S;\al), I(Q;\al) )\leq 2d_B(S,Q)$ for any radius $\al\geq 0$. 
\bs
\end{thm}
\begin{proof}
By Lemma~\ref{lem:common_lattice} the given periodic point sets $S,Q$ have a common unit cell $U$.
Let $g:S\to Q$ be a bijection from Definition~\ref{dfn:bottleneck_distance} such that $|p-g(p)|\leq\ep=d_B(S,Q)$ for all points $p\in S$.
Since the bottleneck distance $\ep<r(Q)$ is small, for any point $p\in S$, its bijective image $g(p)$ is a unique $\ep$-close point of $Q$ and vice versa.
\medskip

Hence we can assume that the common unit cell $U\subset\R^n$ contains the same number (say, $m$) points from $S$ and $Q$.
The bijection $g$ will induce flows $f_{ij}\in[0,1]$ from Definition~\ref{dfn:EMD} between weighted isometry classes from the isosets $I(S;\al)$ and $I(Q;\al)$.
\medskip

First we expand the initial $m(S)$ isometry classes $(\si_i,w_i)\in I(S;\al)$
to $m$ isometry classes (with equal weights $\frac{1}{m}$) represented by clusters $C(S,p;\al)$ for $m$ points $p\in S\cap U$.
If the $i$-th initial isometry class had a weight $w_i=\frac{k_i}{m}$, $i=1,\dots,m(S)$, the expanded isoset contains $k_i$ equal isometry classes of weight $\frac{1}{m}$.
For example, the 1-regular set $S_1$ in Fig.~\ref{fig:alpha-clusters} initially has the isoset consisting of a single class $[C(S_1,p;\al)]$, which is expanded to four identical classes of weight $\frac{1}{4}$ for the four points in the motif of $S_1$.
The isoset $I(Q;\al)$ is similarly expanded to the set of $m$ isometry classes of weight  $\frac{1}{m}$, possibly with repetitions. 
\medskip

The bijection $g$ between points $S\cap U\to Q\cap U$ induces the bijection between the expanded sets of $m$ isometry classes above.
Each correspondence $\si_l\mapsto\xi_l$ in this bijection can be visualized as a horizontal arrow with the flow $f_{ll}=\frac{1}{m}$ for $l=1,\dots,m$, so $\sum\limits_{l=1}^m f_{ll}=1$.
\medskip

To show that the Earth Mover's Distance (EMD) between any initial isoset and its expansion is 0, we collapse all identical isometry classes in the expanded isosets, but keep the arrows with the flows above.
Only if both tail and head of two (or more) arrows are identical, we collapse these arrows into one arrow that gets the total weight.
\medskip

All equal weights $\frac{1}{m}$ correctly add up at heads and tails of final arrows to the initial weights $w_i,v_j$ of isometry classes.
So the total sum of flows is $\sum\limits_{i=1}^{m(S)} \sum\limits_{j=1}^{m(Q)} f_{ij}=1$ as required by Definition~\ref{dfn:EMD}.
Hence it suffices to consider the EMD only between the expanded isosets.
\medskip
 
It remains to estimate the cluster distance between isometry classes $\si_l,\xi_l$ whose centers $p$ and $g(p)$ are $\ep$-close within the common unit cell $U$.
For any fixed point $p\in S\cap U$, shift $S$ by the vector $g(p)-p$.
This shift makes $p\in S$ and $g(p)\in Q$ identical and keeps all pairs of points $q,g(q)$ for $q\in C(S,p;\al)$ within $2\ep$ of each other.
Using the identity map $f\in\Or(\R^n)$ in Definition~\ref{dfn:cluster_distance}, we conclude that the cluster distance is $d_C([C(S,p;\al)],[C(Q,g(p);\al)])\leq 2\ep$.
Then $\EMD(I(S;\al),I(Q;\al))\leq
\sum\limits_{l=1}^{m} f_{ll} d_C([C(S,p;\al)],[C(Q,g(p);\al)])\leq
2\ep\sum\limits_{l=1}^{m} f_{ll}=
2\ep$.
\end{proof}

\section{Polynomial time algorithms for isosets and their metric}
\label{sec:algorithms}

This section proves the key results: polynomial time algorithms for computing the complete invariant isoset (Theorem~\ref{thm:compute_isoset}), comparing isosets (Theorem~\ref{thm:compare_isosets}) and approximating 
Earth Mover's Distance on isosets (Corollary~\ref{cor:isoset_distance}).  
Let $V_n=\dfrac{\pi^{n/2}}{\Ga(\frac{n}{2}+1)}$ be the volume of the unit ball in $\R^n$, where the Gamma function $\Ga$ has $\Ga(k)=(k-1)!$ and $\Ga(\frac{k}{2}+1)=\sqrt{\pi}(k-\frac{1}{2})(k-\frac{3}{2})\cdots\frac{1}{2}$ for any integer $k\geq 1$.
The diameter of a unit cell $U$ of a periodic set $S\subset\R^n$ is
$d=\sup\limits_{p,q\in U}|p-q|$.
Set $\nu(S,\al,n)=\dfrac{(\al+d)^n V_n}{\vol[U]}$.
Stirling's approximation 
implies that the volume $V_n$ approaches $\dfrac{1}{\sqrt{\pi}n}\left(\dfrac{2\pi e}{n}\right)^{n/2}$ as $n\to+\infty$, hence $\nu(S,\al,n)\to 0$ as $n\to+\infty$ for fixed $S,\al$.
All complexities below assume the real Random-Access Machine (RAM) model and a fixed dimension $n$.

\begin{lem}[computing a local cluster]
\label{lem:compute_cluster}
Let a periodic point set $S\subset\R^n$ have $m$ points in a unit cell $U$ of diameter $d$.
For any $\al\geq 0$ and a point $p\in M=S\cap U$, the $\al$-cluster $C(S,p;\al)$ has at most $k=\nu(S,\al,n)m$ points and can be found in time $O(k)$.
\bs
\end{lem}
\begin{proof}
To find all points in $C(S,p;\al)$, we will extend $U$ by iteratively adding adjacent cells around $U$.
For any new shifted cell $U+v$ with $v\in\La$, we check if any translated points $M+v$ are within the closed ball $\bar B(p;\al)$ of radius $\al$.
The upper union $\bar U=\bigcup \{(U+v) :  v\in\La, (U+v)\cap \bar B(p;\al)\neq\emptyset\}$
consists of $\dfrac{\vol[\bar U]}{\vol[U]}$ cells and is contained in the larger ball $B(p;\al+d)$, because any shifted cell $U+v$ within $\bar U$ has the diameter $d$ and intersects $B(p;\al)$.
Since each $U+v$ contains $m$ points of $S$, we check at most 
$m\dfrac{\vol[\bar U]}{\vol[U]}$ points.
So
$|C(S,p;\al)|\leq 
m\dfrac{\vol[\bar U]}{\vol[U]}\leq 
m\dfrac{\vol[B(p;\al+d)]}{\vol[U]}=
m\dfrac{(\al+d)^n V_n}{\vol[U]}=\nu(S,\al,n)m$.
\end{proof}


We measure the size of a periodic set as the number $m$ of motif points.
Theorems~\ref{thm:compute_isoset} and~\ref{thm:compare_isosets} analyze the time with respect to $m$, while hidden constants can depend on $S$.

\begin{thm}[compute an isoset]
\label{thm:compute_isoset}
For any periodic point set $S\subset\R^n$ given by a motif $M$ of $m$ points in a unit cell $U$ of diameter $d$, the isoset $I(S;\al)$ at a stable radius $\al$ can be found in time $O(m^2k^{n-2}\log k)$, where $k=\nu m$, $\nu=\dfrac{(\al+d)^n V_n}{\vol[U]}$, $V_n$ is the unit ball volume.
\bs
\end{thm}
\begin{proof}
Lemma~\ref{lem:upper_bounds} gives the easy stable radius $\al=\max\{2b,d\}$, where $b$ is the longest edge-length of the unit cell $U$ and $d$ is the length of a longest diagonal of $U$.
Lemma~\ref{lem:compute_cluster} computes $\al$-clusters of all $m$ points $p\in M$ in time $O(km)$.
The algorithm from \cite[Theorem~1]{alt1988congruence} checks if two finite sets of $k$ points are isometric in time $O(k^{n-2}\log k)$ for $n\geq 3$.
The isoset $I(S;\al)$ with weights is obtained after identifying all isometric clusters through $O(m^2)$ comparisons.
The total time is $O(m^2k^{n-2}\log k)$.
For $n=2$, the time is $O(m^2k\log k)$ due to \cite{atallah1984checking}. 
\end{proof}

For simplicity, all further complexities will hide 
the factor $\nu(S,\al,n)=\dfrac{(\al+d)^n V_n}{\vol[U]}$. 

\begin{thm}[compare isosets]
\label{thm:compare_isosets}
For any periodic point sets $S,Q\subset\R^n$ with at most $m$ points in their motifs, one can decide if $S,Q$ are isometric in time $O(m^n\log m)$, $n\geq 3$.
\bs
\end{thm}
\begin{proof}
Theorem~\ref{thm:compute_isoset} finds isosets $I(S;\al),I(Q;\al)$ with a common stable radius in time $O(m^2k^{n-2}\log k)$, where each cluster consists of $k=\nu(S,\al,n)m=O(m)$ points by Lemma~\ref{lem:compute_cluster}.
By \cite[Theorem~1]{alt1988congruence} any two classes from $I(S;\al),I(Q;\al)$ can be compared in time $O(k^{n-2}\log k)$ or $O(m^{n-2}\log m)$.
Finally, $O(m^2)$ comparisons are enough to decide if $I(S;\al)=I(Q;\al)$.  
\end{proof}

In dimension $n=4$, the complexity in Theorem~\ref{thm:compare_isosets} reduces to $O(m^3\log m)$ due to \cite{kim2016congruence}.

\begin{lem}[max-min formula for the distance $d_C$]
\label{lem:max-min_formula}
Let $C,D\subset\R^n$ be any finite sets.
Order all points $p_1\dots,p_k\in C$ by their distance to the origin so that $|p_1|\leq\dots\leq|p_k|$.
For any fixed radius $\al\geq|p_k|$, the minimum $\ep\in[0,\al]$ satisfying condition~(\ref{dfn:cluster_distance}a) $d_R(C\cap\bar B(0;\al-\ep),D)\leq\ep$ from Definition~\ref{dfn:cluster_distance} of $d_C$ equals $d_M(C,D)=\max\limits_{i=1,\dots,k}\min\{\;\al-|p_i|,\; d_R(\{p_1,\dots,p_i\}, D) \;\}$.
\bs
\end{lem}
\begin{proof}
Let $\ep\in[0,\al]$ be a minimum value satisfying $d_R(C\cap\bar B(0;\al-\ep),D)\leq\ep$.
Then after a suitable orthogonal map $f\in\Or(\R^n)$ all points of $f(C\cap\bar B(0;\al-\ep))$ are covered by the $\ep$-offset $D+\bar B(0;\ep)$.
Let $j\in\{1,\dots,k\}$ be the largest index so that $|p_j|\leq\al-\ep$.
Then $C\cap\bar B(0;\al-\ep)=\{p_1,\dots,p_j\}$ and $d_R(\{p_1,\dots,p_i\}, D)\leq d_R(C\cap\bar B(0;\al-\ep),D)\leq\ep$ for all $i=1,\dots,j$.
By the above choice of $j$, if $j<i\leq k$, then $\al-|p_i|<\ep$ for all $i>j$.
Then $\min\{\; \al-|p_i|,\; d_R(\{p_1,\dots,p_i\}, D)\; \}\leq\ep$ for all $i=1,\dots,k$, so $d_M\leq\ep$.
\medskip

Conversely, $\ep\leq d_M$ will follow from $d_R(C\cap\bar B(0;\al-d_M),D)\leq d_M$.
Indeed, let $i\in\{1,\dots,k\}$ be the largest index so that $|p_i|\leq\al-d_M$.
Since $\al-|p_i|\geq d_M$ and 
$\min\{\; \al-|p_i|,\; d_R(\{p_1,\dots,p_i\}, D)\; \}\leq d_M$, we get $d_R(\{p_1,\dots,p_i\}, D)\leq d_M$. 
Due to $C\cap\bar B(0;\al-d_M)=\{p_1,\dots,p_i\}$, we get 
$d_R(C\cap\bar B(0;\al-d_M),D)\leq d_M$.
\end{proof}

Lemma~\ref{lem:rot-inv_distance} extends \cite[section 2.3]{goodrich1999approximate} from the case of $n=3$ to any dimension $n>1$. 

\begin{lem}[approximate $d_R$]
\label{lem:rot-inv_distance}
For any sets $C,D\subset\R^n$ of maximum $k$ points, the directed rotationally-invariant distance $d_R(C,D)$ is approximated within a  factor $\eta=2(n-1)(1+\de)$ for any $\de>0$ in time $O(c_{\de}k^n\log k)$, where $c_{\de}\leq n\lceil 1+6n/\de\rceil^n$ is independent of $k$.
\bs
\end{lem}
\begin{proof}
Let $p_1\in C$ be a point that has a maximum Euclidean distance to the origin $0\in\R^n$.
If there are several points at the same maximum distance, choose any of them.
We can make similar random choices below.
For any $1<i<n$, let $p_i\in C$ be a point that has a maximum perpendicular distance to the linear subspace spanned by the vectors $p_1,\dots,p_{i-1}$.
\medskip

For any point $q\in D$, let a map $f_1[q]\in\Or(\R^n)$ move $p_1$ to the straight line through $0,q$.
For any $1<i<n$, let a map $f_i[q]\in\Or(\R^n)$ fix the linear subspace spanned by $0,p_1,\dots,p_{i-1}$ and move $p_i$ to the subspace spanned by $0,p_1,\dots,p_{i-1},q$.
The required approximation will be computed as $d_a=\min\limits_{q_1,\dots,q_{n-1}\in D}
d_H(C',D)$, where $C'=f_{n-1}[q_{n-1}]\circ\cdots\circ f_1[q_1](C)$.
\medskip

Indeed, let $f\in\Or(\R^n)$ be an optimal map minimizing $d_R(C,D)$ as $d_o=d_H(f(C),D)$.
For simplicity, assume that $f$ is the identity, else any $p\in C$ should be replaced by $f(p)$ below. 
We will find points $q_1,\dots,q_{n-1}\in D$ such that $d_o\leq \eta d_H(C',D)$ for the set $C'$ above.  
Associate any point $p\in C$ via a map $g$ to its $d_o$-neighbor  $q=g(p)\in D$.
For $q_1=g(p_1)\in D$, let a map $f_1[q_1]$ move $p_1$ to the straight line through $0$ and $q_1$, which is $d_o$-close to $p_1$.
Since $p_1$ is a furthest point of $C$ from $0$, the map $f_1[q_1]$ moves any point of $C$ by at most $d_o$.
\medskip

For any $1<i<n$, set $q_i=g(p_i)\in D$.
Let a map $f_i[q_i]$ move $p_i$ to the subspace spanned by $0,p_1,\dots,p_{i-1}$ and $q_i$, which is $d_o$-close to $p_i$.
Since $p_i$ and its image $f_i[q_i](p_i)$ are $d_o$-close to $q_i$,
due to the triangle inequality, $f_i[q_i]$ moves $p_i$ by at most $2d_o$.
Since $p_i$ is a furthest point of $C$ from the subspace spanned by $0,p_1,\dots,p_{i-1}$, any point of $C$ moves under $f_i[q_i]$ by at most $d_o$.
The composition of $n-1$ maps $f_{n-1}[q_{n-1}]\circ\cdots\circ f_1[q_1](C)$ will be considered in the approximation $d_a$ above and moves any point of $C$ by at most $(2n-3)d_o$.
Since any point $p\in C$ is $d_o$-close to its associated $g(p)\in D$, the final factor is $\eta=(2n-2)(1+\de)$. 
\medskip

It remains to justify the time.
The points $p_1,\dots,p_{n-1}\in C$ are found in time $O(kn)$.
The optimal algorithm from \cite{arya1998optimal} preprocesses the set $D\subset\R^n$ of $k$ points in time $O(nk\log k)$ and for any point $p\in C'$ finds its $(1+\de)$-approximate nearest neighbor in $D$ in time $O(c_{\de}\log k)$.  
Hence $d_H(C',D)=\max\limits_{p\in C'}\min\limits_{q\in D}|p-q|$ can be $(1+\de)$-approximated in time $O(c_{\de}k\log k)$. 
The minimization over $q_1,\dots,q_{n-1}\in D$ gives the total time $O(c_{\de}k^n\log k)$ as required.
\end{proof}

\begin{thm}[approximate $d_C$]
\label{thm:cluster_distance}
Let periodic points sets $S,Q\subset\R^n$ have isometry classes $\si,\xi$ represented by clusters $C,D$ of maximum $k$ points.
The cluster distance $d_C(\si,\xi)$ can be approximated within a factor $\eta=2(n-1)(1+\de)$ for any $\de>0$ in time $O(c_{\de}k^{n+1}\log k)$.  
\bs
\end{thm}
\begin{proof}
The cluster distance $d_C$ from Definition~\ref{dfn:cluster_distance}
is $\max\{d_M(C,D),d_M(D,C)\}$, see $d_M$ in Lemma~\ref{lem:max-min_formula}.
The distance $d_M$ needs the maximization of $i=1,\dots,k$ and the approximation of $d_R$ from Lemma~\ref{lem:rot-inv_distance} in time $O(c_{\de}k^{n}\log k)$.
The total time is $O(c_{\de}k^{n+1}\log k)$.
\end{proof}

\begin{cor}[approximating EMD on isosets]
\label{cor:isoset_distance}
Let $S,Q\subset\R^n$ be any periodic point sets with at most $m$ points in their motifs.
Then $\EMD(I(S;\al),I(Q;\al))$ can be approximated within a factor
$\eta=2(n-1)(1+\de)$ for any $\de>0$ in time 
 in time $O(m^{n+3}\log m)$.
\bs
\end{cor}
\begin{proof}
Since $S,Q$ have at most $m$ points in their motifs, their isosets at any radius $\al$ have at most $m$ isometry classes.
By Theorem~\ref{thm:cluster_distance} the cluster distance $d_C(\si,\xi)$ between any two classes $\si\in I(S;\al)$ and $\xi\in I(Q;\al)$ can be approximated within a factor $\eta$ in time $O(c_{\de}k^{n+1}\log k)$.
Since Definition~\ref{dfn:EMD} uses normalized distributions, $\eta$ emerges as a multiplicative upper bound in $\EMD(I(S;\al),I(Q;\al))$.
After computing $O(m^2)$ pairwise distances between $\al$-clusters, the exact EMD is found in time $O(m^3\log m)$ \cite{orlin1993faster}.
If we substitute $k=\nu m$ from Lemma~\ref{lem:compute_cluster} and  keep only the most important input size $m$, the total time is $O(m^{n+3}\log m)$.
\end{proof}

\section{Periodic topology of textile structures up to periodic isotopy}
\label{sec:textiles}

This section introduces \emph{Periodic Topology}, which studies 2-periodic structures up to continuous (non-isometric) deformations.
These textiles \cite{grishanov2009topological} were studied in the past almost exclusively up to isotopies preserving a unit cell.
However real textiles are naturally equivalent up to a more flexible periodic isotopy in Definition~\ref{dfn:textile}.

\begin{dfn}[periodic isotopy of textiles]
\label{dfn:textile}
A \emph{textile} is an embedding of (possibly infinitely many) lines or circles $K\subset\R^2\times[0,1]$ preserved under translations by basis vectors in $\R^2$.
A \emph{periodic isotopy} is any continuous family of textiles. 
\bs
\end{dfn}

\begin{figure}[h!]
\includegraphics[width=\linewidth]{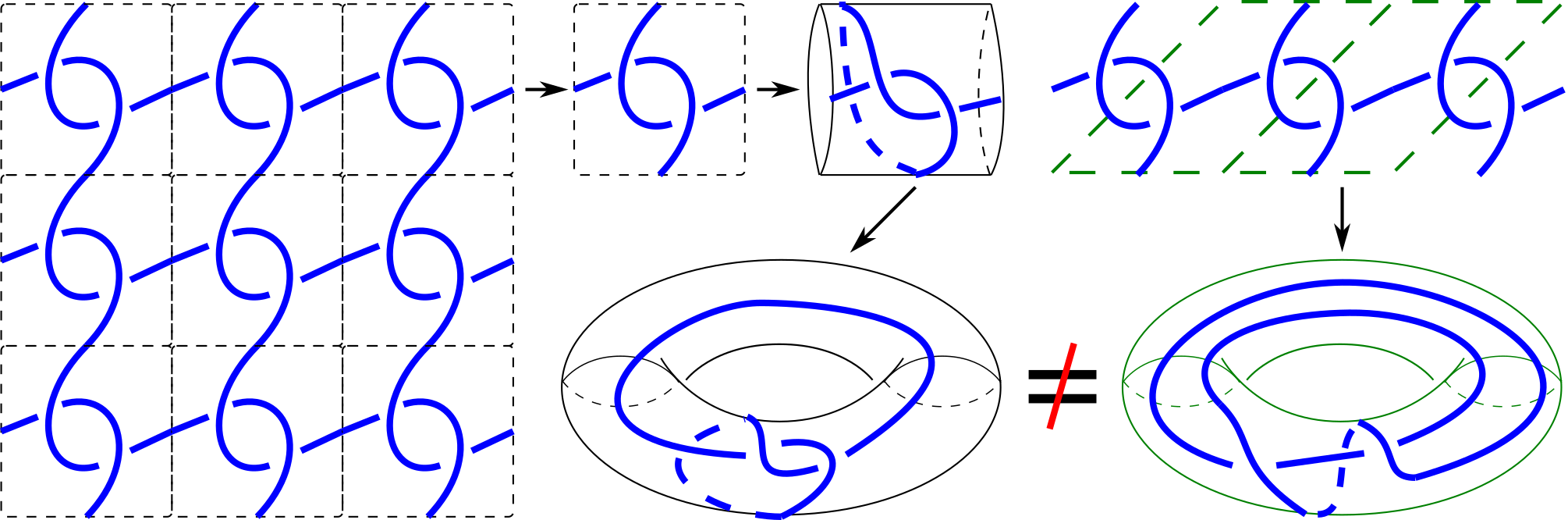}
\caption{A textile with a square cell can be mapped to a knot (one closed curve) in a thickened torus.
Choosing a different unit cell produces another link (of two closed curves) in a thickened torus.}
\label{fig:textile_different_cells}
\end{figure}

The above definition may seem as a straightforward extension of classical knot theory.
However, almost all past studies of periodic knots assumed periodicity with respect to fixed basis vectors $ v_1, v_2$.
Then the equivalence is restricted to an isotopy (continuous deformation) \cite{akimova2020classification} within a fixed thickened torus $T=S^1\times S^1\times[0,1]$.  
\medskip

Similarly to periodic point sets in Definition~\ref{dfn:crystal}, any textile can have infinitely many bases $ v_1, v_2$ that span different unit cells containing periodic patterns.
Such a unit cell or a periodic pattern is even more ambiguous for textiles than for periodic point sets.
Indeed, a periodic textile is a continuous object, not a discrete set of points.
Hence unit cells can be chosen from a continuous family, not from a discrete lattice.
\medskip

Fig.~\ref{fig:textile_different_cells} shows that another basis for the same textile produces a non-isotopic link in a thickened torus $T$. 
A classification of textiles up to periodic isotopy seems harder than up to isotopies in a fixed $T$, because our desired invariants should be preserved by a substantially larger family of periodic isotopies without a fixed basis.
\medskip

\begin{pro}[textile classification]
\label{pro:textile_classification}
Find an algorithm to distinguish textiles up to periodic isotopy, at least up to a certain complexity.
The simpler untangling question is to detect if a textile is periodically isotopic to a textile without crossings. 
\bs
\end{pro}

The only past attempt to construct invariants of periodic isotopy was based on the multivariable Alexander polynomial of classical links \cite{morton2009doubly} and has lead to a classification of about 10 textiles in Fig.~\ref{fig:textile_patterns} through manual computations.
\medskip

\newcommand{\theight}{11mm}
\begin{figure}[h]
\includegraphics[height=\theight]{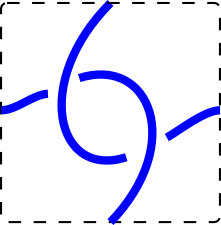}
\hspace*{1pt}
\includegraphics[height=\theight]{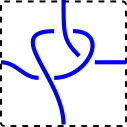}
\hspace*{1pt}
\includegraphics[height=\theight]{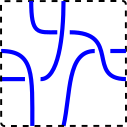}
\hspace*{1pt}
\includegraphics[height=\theight]{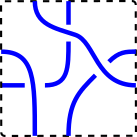}
\hspace*{1pt}
\includegraphics[height=\theight]{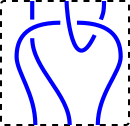}
\hspace*{1pt}
\includegraphics[height=\theight]{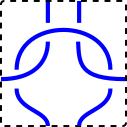}
\hspace*{1pt}
\includegraphics[height=\theight]{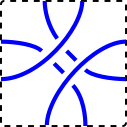}
\hspace*{1pt}
\includegraphics[height=\theight]{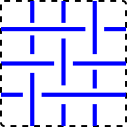}
\hspace*{1pt}
\includegraphics[height=\theight]{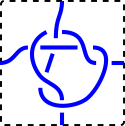}
\caption{All textile structures that were distinguished modulo periodic isotopies in 2009 \cite{grishanov2009topological, morton2009doubly}. }
\label{fig:textile_patterns}
\end{figure} 

The more recent approach \cite{bright2020encoding} has suggested an automatic enumeration of all textiles by adapting Gauss codes of classical links to square diagrams of textiles.
These combinatorial codes are based on the earlier classical case of links in $\R^3$ \cite{kurlin2008gauss}.

\section{Discussion: a summary of first results and further problems}
\label{sec:discussion}

This paper has introduced the new research area of \emph{Periodic Geometry and Topology} motivated by real-life applications to periodic crystals and textile structures. 
\medskip

Periodic Geometry studies crystals as periodic sets of points, possibly with labels, up to \emph{isometries}, because most solid crystalline materials are rigid bodies.
\medskip

In the last year, Problem~\ref{pro:isometry_classification} has been attacked from several directions: AMD invariants in section~\ref{sec:AMD}, density functions in section~\ref{sec:densities} and isosets in section~\ref{sec:isoset_complete}.
However, final condition (\ref{pro:isometry_classification}f) of an explicit continuous parameterization might need a new easier invariant.
Such an invariant should finally enable a guided exploration of a continuous space of crystals instead of the current random sampling.
\medskip

Experimental results of large crystal datasets are presented in \cite[section~8]{widdowson2020average} and \cite[section~7]{edels2021}.
Further problems on periodic crystals include (1) understanding energy barriers around local minima, (2) finding potential phase transitions as optimal paths between local minima of the energy, (3) optimizing a search for all deepest minima, which represent the most stable crystals for a given chemical composition.
\medskip

Periodic Geometry can be extended beyond Problem~\ref{pro:isometry_classification}, for example in the theory of dense packings.
Biological cells pack not as hard balls or other solid bodies, but more like slightly squashed balls.
For such soft packings with overlaps, one can maximize the probability that a  random point belongs to a single ball.
Among 2D lattices, the hexagonal lattice achieves the maximum probability \cite{edelsbrunner2015relaxed} of about $0.928$, which is higher than the classical packing density of about $0.907$ for hard disks.
\medskip

Similar packing problems can be stated for the new density functions $\psi_k$.
For instance, what lattice maximizes the global maximum of a fixed function $\psi_k$ or minimizes the overall maximum of all density functions?
Can we characterize periodic sets that whose every $\psi_k$ has a single local maximum as in Fig.~\ref{fig:SQ15densities}? 
\medskip

In the light of Theorem~\ref{thm:densities1D} we conjecture that, for any periodic set $S\subset\R^n$, all density functions $\psi_k$ can be obtained from distance-based isometry invariants of $S$.
\medskip

Periodic Topology studies textiles, which are periodic in two directions.
However, 3-periodic continuous curves or graphs naturally appear in physical simulations \cite{evans2015ideal} and crystalline networks \cite{power2020isotopy}, hence can be also studied up to periodic isotopy. 
\medskip

A \emph{periodic isotopy} is the most natural model for macroscopic deformations of textile clothes by keeping their microscopic periodicity, but was largely ignored in the past.
Problem~\ref{pro:textile_classification} is widely open.
A couple of first steps are in \cite{morton2009doubly,bright2020encoding}. 
\medskip

In conclusion, here are the most important technical contributions of this paper. 
\medskip

\noindent
$\bullet$
Examples~\ref{exa:SQ15}, \ref{exa:SQ15densities} show that the AMD can be stronger than the density functions.
\medskip

\noindent
$\bullet$
Theorem~\ref{thm:densities1D} completely describes the density functions of   periodic sets in $\R$.
\medskip

\noindent
$\bullet$
Theorem~\ref{thm:continuity} proves that the complete invariant isoset is continuous under point perturbations for a suitably adapted Earth Mover's Distance on isosets. 
\medskip

\noindent
$\bullet$
Theorems~\ref{thm:compute_isoset},~\ref{thm:compare_isosets},~\ref{cor:isoset_distance}
provide polynomial time algorithms for comparing periodic point sets by complete invariant isosets, which are fast enough for real crystals.
\medskip

We are open to collaboration on any potential problems in Periodic Geometry and Topology, and thank all reviewers in advance for their time and helpful suggestions.

\begin{acknowledgement}
We thank all co-authors of the cited papers, our colleagues in the Material Innovation Factory, the EPSRC for the $\pounds$3.5M grant `Application-driven Topological Data Analysis' (ref EP/R018472/1) and all reviewers in advance for their valuable time and helpful suggestions
\end{acknowledgement}

\bibliographystyle{spmpsci}
\bibliography{introduction_PGT}

\end{document}